%% file: TESI.tex
\pdfoutput=1
\pdfoutput=1
\pdfoutput=1
\pdfoutput=1
\pdfoutput=1
\documentclass[12pt,a4paper]{book}
\usepackage[latin2]{inputenc}
\usepackage[english]{babel}
\usepackage{amsmath}
\usepackage{amsfonts}
\usepackage{amssymb}
\usepackage{graphicx}
\usepackage{float}
\usepackage{wrapfig}
\usepackage{notoccite}
\usepackage[square,numbers]{natbib} 
\usepackage{graphicx}
\usepackage{epsfig}
\usepackage{rotate}
\usepackage{rotating}
\usepackage{minitoc} 
\usepackage{times} 
\usepackage{url}
\usepackage{subcaption}
\usepackage{rotating}
\usepackage{geometry}
\usepackage{pdflscape}
\usepackage{fancyhdr} 
\title{An agent-based model for preservation of digital objects}
\author{Jacopo Pellegrino}

\begin{document}
\newgeometry{margin=1cm}
\thispagestyle{empty}

\begin{figure}[!t]
\begin{center}
\includegraphics[width=2.0cm]{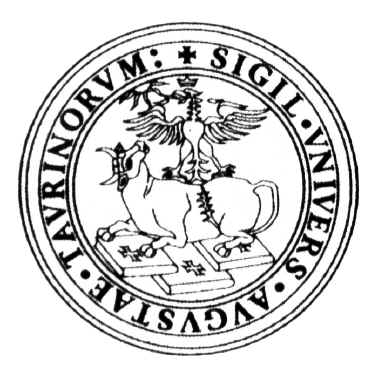}
\end{center}
\end{figure}

\begin{center}
  {\Large \bf Universit\`a degli Studi di Torino} 
\end{center}
\begin{center}
{\Large \bf Dipartimento di Fisica} 
\end{center}

\vspace{0.8cm} 
\begin{center}
{\Large \bf Laurea Magistrale in Fisica delle Tecnologie Avanzate} 
\end{center} 

\vspace{1.5cm} 


\begin{center}
  {\Large \bf A MULTI-AGENT BASED DIGITAL\\ 
\vspace{0.2cm}
 PRESERVATION MODEL 
} \\
\end{center}

\vspace{3.0cm}

\begin{table}[!h]	
	\begin{center}
    \begin{tabular}{p{0.47\textwidth} p{0.1\textwidth} p{0.27\textwidth}}	
	 Tesi presentata da: Jacopo Pellegrino & & \\
  	    \end{tabular}
    	\end{center}	
%
%
	\begin{center}
    \begin{tabular}{p{0.4\textwidth} p{0.4\textwidth}}
	 & \\ 	
 	 \hspace{2.5cm}  Relatore:  &  \hspace{1cm} Correlatore: \\ 
	 & \\    
     \hspace{2.5cm} Prof. M. Maggiora &  \hspace{1cm} Prof. W. Allasia   \\
     \hspace{2.5cm} \footnotesize{Dipartimento di Fisica} & \hspace{1cm} \footnotesize{Innovation Department} \\
     \hspace{2.5cm} \footnotesize{Universit\`a di Torino} & \hspace{1cm} \footnotesize{EURIX} \\
    \end{tabular}
    	\end{center}	
%
%
	\begin{center}
	\begin{tabular}{p{0.27\textwidth} p{0.3\textwidth} p{0.27\textwidth}}	
	& & \\
	Controrelatori: & & \\
	& &\\
	 Dean and Prof. S. Ross  & Prof. P. Terna & Prof. D. Teruggi\\
	\footnotesize{Faculty of Information}  & \footnotesize{Dip. di scienze economico-sociali } & \footnotesize{Head of Research}\\	
	\footnotesize{University of Toronto}  & \footnotesize{e matematico-statistiche 
	
	 Universit\`a di Torino} & \footnotesize{Institut National de l'Audiovisuel, Paris }\\
	\end{tabular}
	\end{center}	

\end{table}
 		
\normalsize

\vspace{1.0cm} 
\begin{center}
  \textbf{Anni Accademici:}  2012/2013 2013/2014 \\
\end{center}

\restoregeometry

\clearpage

\raggedbottom

\parindent=0pt
\parskip=8pt
\setlength{\evensidemargin}{-24pt}
\setlength{\oddsidemargin}{0pt}
\setlength{\marginparsep}{0.0in}
\setlength{\marginparwidth}{0.0in}
\marginparpush=0pt


	
\newpage
\thispagestyle{empty}
\mbox{}		

\newpage
\thispagestyle{empty}
\mbox{}		
	
\begin{table}[!h]	
	\begin{center}
    \begin{tabular}{p{0.3\textwidth} p{0.3\textwidth} p{0.3\textwidth}}
	 & & \textit{Ai miei genitori, che da sempre credono in me e mi sostengono in quello che faccio.}\\
	& & \\	
	& & \textit{To my parents who always believed in me and support me in what I do.}\\
    \end{tabular}
	\end{center}	
\end{table}

\newpage
\thispagestyle{empty}
\mbox{}			
	
\clearpage
\thispagestyle{empty}
\chapter*{Acknowledgements}
It is not easy for me to show my gratitude to those people who supported me through the long, hard path that led me to the completion of this work. I take advantage of the following lines to thank the most important of these, starting from my supervisors Prof. Marco Maggiora and Prof. Walter Allasia who gave me the chance to prove myself with such an interesting and innovative work. I'm grateful for their attention and their support through this long period. Special thanks also to EURIX where I spent the most part of my time for these research.

I would like to thank Prof. Seamus Ross, Prof. Pietro Terna and Prof. Daniel Teruggi who were so kind and helpful to be my external reviewers, that has been very important for me.

Last but not least, particular thanks go to my closest friends, my mother Irene and my father Enzo who have always supported me, especially in this year full of difficult moments, but also of great satisfaction.

\begin{table}[!h]	
	\begin{center}
    \begin{tabular}{p{0.3\textwidth} p{0.3\textwidth} p{0.3\textwidth}}
	 & & \\	 
	 
	 & & \textit{Jacopo Pellegrino}\\
	
    \end{tabular}
	\end{center}	
\end{table}
\thispagestyle{empty}
\newpage
\thispagestyle{empty}
\mbox{}		
	
\clearpage

\chapter*{Abstract}
The thesis describes an agent-based model aimed to simulate those processes in which a digital object faces the risk of obsolescence, a migration process has to be performed and the most appropriate file format has to be adopted. Agents have been designed in order to monitor and control the local system where they reside and its environment. They are able to become aware of obsolescent formats based on global parameters such as their diffusion. They communicate as well with each other to find out the most suitable preservation action to be performed. Agents request suggestions that are evaluated and propagated according to a weighting based on the level of trust assigned to both the agents who identified the problem and proposed the solution. In the current research, the definition of the trust level has been chosen based on the cultural and geographical distances, the expertise of the involved agents and the file format numerosity. The level of trust between two agents is automatically updated after every interaction by the mean of a feedback mechanism profiting of an inter agent communication based on stigmergy. Summing up, the thesis demonstrates  how a multi-agent system can either perform an autonomous preservation action or suggest a list of best candidate solutions to the user. It benefits the management of several kinds of digital archive, especially those with limited resources specifically dedicated to digital preservation, such as small personal collections and many public institutions.

\thispagestyle{empty}
\newpage
\thispagestyle{empty}
\mbox{}	
	
\chapter*{Abstract}
La tesi descrive un modello ad agenti che si propone di simulare quei processi in cui un oggetto digitale va incontro al rischio di obsolescenza, un processo di migrazione \`e \-ne\-ces\-sa\-rio e il formato pi\`u adatto deve essere adottato. Gli agenti sono stati implementati con lo scopo di monitorare e controllare il sistema locale e l'ambiente in cui risiedono. Sono in grado di identificare formati obsolescenti basandosi su parametri globali come la loro diffusione. Comunicano inoltre tra loro per individuare l'azione pi\`u adatta da intraprendere per preservare i contenuti. Gli agenti richiedono suggerimenti che \-ven\-go\-no pesati e propagati secondo pesi basati sul livello di fiducia associato sia all'agente che ha individuato il problema che a quello che ha proposto una soluzione. In questa attivit\`a di ricerca, la definizione del livello di fiducia si basa, per scelta, sulla distanza geografica e culturale, la competenza in ambito di digital preservation degli agenti coinvolti e sulla diffusione del formato. Il livello di fiducia tra due agenti viene aggiornato automaticamente a seguito di ogni interazione attraverso un meccanismo di feedback che sfrutta la comunicazione tra gli agenti basata sulla stigmergia. Riassumendo, la tesi dimostra come un sistema ad agenti possa sia eseguire azioni autonome di preservation, sia suggerire la soluzione pi\`u appropriata all'utente. Ne trae beneficio la gestione di vari tipi di archivi digitali, in particolar modo quelli con risorse dedicate alla digital preservation limitate, come piccoli archivi personali e molte istituzioni pubbliche.
	
\thispagestyle{empty}
\newpage
\thispagestyle{empty}
\mbox{}	
	
	
\tableofcontents


\newpage
\thispagestyle{empty}
\mbox{}	

\chapter{Introduction}

\label{intro}
Digital preservation is becoming a fundamental issue for those who have to protect and maintain usable any kind of cultural heritage, for example a personal collection of photos and videos as well as large archives of digital books and web pages in a modern library.

Each digital object faces the risk of obsolescence since new formats are arising and have often a short life time. New applications to open and edit those objects are released which may not be able to read the older formats. There are essentially two different ways to deal with this serious problem. The first one is the migration process that consists in converting the digital object into a new and more recent format. The other one is the \-e\-mu\-la\-tion where the original environment, in which the object was originally opened, is purposely recreated to render an old digital object. \cite{planets}

Migration and the related risks and difficulties is the preservation solution we are going to take into account in the research described in this thesis. This kind of \-ope\-ra\-tion is difficult to reverse and could permanently compromise the file, \-ma\-king it unusable with any software. This means that migration should be performed under particular conditions. The user has to evaluate the risk for the file to become obsolete and compare it to the risk of damage during migration. To this end the questions to answer before performing a migration are: how high the risk of obsolescence is, which will be the new format, why should this format be chosen and how the migration will be performed. 

A key point is how to decide when a digital object is becoming obsolete. To answer this question several aspects must be taken into account. Here emerges the need for some kind of distributed and dynamic intelligence able of acquiring information from the environment,  for example if a new format has been released by a research group, and capable of autonomous actions after a reasoning process. Digital preservation cannot be viewed as a local or personal issue any more, it becomes instead a collective and distributed concept \cite{allasia2012} \cite{skinner2010guide}. In this approach information becomes global and decisions can be taken after an evaluation based on global statistics.
An agent based model could be appropriate to implement this kind of distributed intelligence that should emerge from the agent society. The agents will acquire, evaluate and share a certain set of information in order to understand how serious the risk of obsolescence is and which is the best action to perform. An important step is to verify the stability of this kind of system. Before proceeding it must be ensured that the system is able to convey the load of necessary information and that is characterized by a sort of inertia  which avoids chaotic evolution.    

\section{Scenario}
In this section we provide an example of a possible scenario in which the \-ob\-so\-le\-scen\-ce issue comes out. Let us consider an user trying to perform the rendering of a certain digital object, such as a video or an audio file, on his own personal computer. This user may encounter a very common problem: the rendering cannot be performed. It could be due to several reasons, but this situation might be a cue of obsolescence and it must be addressed with the most appropriated approach. By the way, asserting that a digital object is under risk of obsolescence only because a certain operating system (included all the needed applications) cannot render it any more could lead to a wrong view of the problem. 

As underlined in the introduction, the concept of obsolescence must be \-con\-si\-de\-red as global. The rendering failure encountered by the user may be related, for example, to an old version of the software needed to open and render the digital object. In such a case the digital object cannot be considered obsolete at all.
The impossibility to render a file could still be an useful information if we consider it under a wider perspective. Let us assume that an agent, which is monitoring the user's system, is warned that a failure took place while attempting to render a file. The agent has the task to figure out if the information acquired is a real index of obsolescence of the object or not. It will interact with the environment and the other agents within it, in order to understand if other users encountered the same problem when trying to render the same kind of object. If so, the information acquires a certain weight related to the number of agents that received a similar warning. It is possible to see how a local information becomes relevant when it is shared among a group of agents.

The agents must be able to know what is happening to and into the environment at any time. They will acquire and share information about the objects they are monitoring in order to give the warning a proper relevance. Only after such a stage they will then eventually warn the user about the risk of obsolescence of the object and suggest how to deal with it. In case no other users encountered the same problem the solution could simply be an update of the application needed to render the file. Otherwise a migration process may be suggested. 

\section{Structure of the Thesis}

We dedicate Chapter \ref{capPres} to a brief introduction to the concept of digital preservation. First of all we underline the most common issue related to the preservation of digital objects, then we focus on the concept of ``trust" as far as digital objects and repositories are concerned. After that we introduce metadata describing how important is the role they play in digital preservation and providing examples of the most common tools for their extraction. The remaining part of that chapter concerns the preservation planning and the strategies to prevent digital \-ob\-so\-le\-scen\-ce. The last section contains a brief list of some well known and adopted digital preservation platforms.

Chapter \ref{capAgents} is fully dedicated to agents and multi-agent systems. First of all the \-de\-fi\-ni\-tion of ``agent" is provided followed by the main features of such a computer system. After that we discuss the concept of environment, its several types and the interaction of the agents with it. We then introduce intelligent agents \-cha\-rac\-te\-r\-ized by their utility function and extend our discussion to the case in which many agents share the same environment that is a multi-agent system. In the last part of the chapter we focus on the communication process which is a fundamental aspect in such a system.

The following Chapter \ref{capImple} certainly is the core of this work. Here we describe the \-de\-vel\-op\-ment environment and then provide a detailed description of the framework structure. The main parts of the model are taken into account: global, entities and experiment. The global part contains the description of those variables or behaviours that are common to all the agents in the model. All the agent species are introduced in the entities section. We end this chapter with the description of the experiment part in which the user can define the variables to observe in the simulations. 

Once that the framework has been implemented some experiments has been performed to verify its behaviour and its flexibility. All these experiments constitute the content of Chapter \ref{using_the_model}. Each section is dedicated to a specific simulation that aims to test a particular model configuration and observe the framework response. 

Chapter \ref{statAnalysis} has a similar structure to Chapter \ref{using_the_model}. Here we provide a statistical \-a\-na\-ly\-sis of the data acquired by means of the simulations run. We also discuss the results of the analysis which lead us to very interesting conclusions as we shall see both in Chapter \ref{statAnalysis} and \ref{capConcl} that is the conclusive part of this thesis.

\chapter{Preservation of Digital Objects}
\label{capPres}
Since ancient times, the mankind has tried to create objects for representing \-si\-tu\-a\-tions or thoughts from the real world. The earlier attempts are cave paintings, after that clay tablets were used for writing and then books. Though all these items belong to different periods they show the aim to keep a content accessible for the future. Everyone can watch the paintings or read an old book and the contents are immediately shared.

The same cannot be said for digital objects. Unlike books or paintings a digital object needs a software environment to render it. Each technological environment is subjected to continuous changes and improvements due to the need to quickly store, transmit and find digital contents. The term \emph{technological obsolescence} \cite{gladney} indicates the process that makes a software or hardware environment unable to reproduce a digital content. Another tendency named \emph{media deterioration} regards specifically the digital object itself \cite{gladney}, in this case the digital content is somehow altered or corrupted to an extent that even error correction algorithms cannot allow a recovery. In both cases the user cannot access the digital content which thus becomes obsolete. The situation so far described is usually referred as \emph{digital obsolescence} \cite{wiki:digpres} \cite{skinner2010guide}.

Since the amount of digital objects is facing an ongoing growth in the last years, the need for a solution addressing digital obsolescence is a critical challenge. As an example the 2009 Consumer Electronics Storage Report \cite{2009CEreport} affirmed that the digital content in an average US home may reach a size of 12 terabytes. The growth is depicted in Figure \ref{growth}:

\begin{figure}[h!]
\centering
\includegraphics[width=11cm]{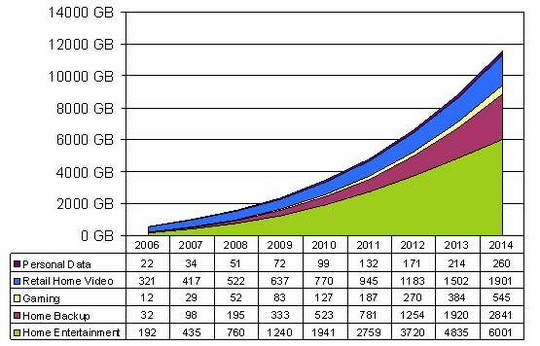}
\caption{Digital content in an average US home \cite{2009CEreport}.}
\label{growth}
\end{figure}

With the concept of \emph{digital preservation} we define the possible strategies and actions performed to maintain a digital object updated, therefore keeping its content accessible \cite{wiki:digpres}. Moreover, as described in \cite{webb2003} and \cite{gladney}, the task of preserving a digital object must be considered from several perspectives: first, it can be \-cha\-rac\-te\-ri\-zed as a physical phenomenon, since it is related to time; it is a set of logical encodings and it may represent an abstract concept for a human being; finally, a digital object is a set of essential elements that must be protected. 

\section{The Concept of ``Trust" for Digital Objects}
\subsection{The OAIS Reference Model}
\label{OAIS}
In the previous paragraphs we cited how quickly the amount of digital data is growing for private users. Thus we can guess which can be the scenario when considering public entities such as national libraries, universities or government archives. \-Di\-gi\-tal obsolescence is an urgent and critical issue for digital repositories. One of the first efforts to contrast digital obsolescence has been made in 1994 by the Research Libraries Group \cite{wiki:rlg} and the Commission of Preservation and Access. They formed the Task Force on Archiving of Digital Information that two years later published one of the most important documents about digital preservation: \emph{``Preserving digital information. Report of the task force on archiving of digital information"} \cite{taskforce}.\\ 

Based on this report, a reference model named OAIS (Open Archival Information System) \cite{oaisbriefing} was developed. It discusses the concept of long-term digital preservation and aims to point out the various stages of the life cycle of a digital object and of the related preservation process \cite{borghoff2006long}. It is a theoretical reference model for the organization of both conventional and digital archives. To a closer look the OAIS model appears to be composed of an information model and of a process model.

The information model describes the relationship between the Data Object and the Information Object. Data are understood and interpreted as information with the help of a Knowledge Base and a Representation Information that together allow to map data into meaningful information. Inside the archive the information is described through the Information Package. It consists of the Descriptive Information, for the description of the content, which is related to the Packaging Information that, in contrast, contains the Content Information for the interpretation and the Preservation Description Information for the preservation process \cite{borghoff2006long}.

The process model involves the relationship between a Producer that sends \-do\-cu\-ments to the archive, and the Management that specifies the purposes of the archive and distributes \-do\-cu\-ments to a Consumer. Each of these entities has its own Information Packages, that is shared with the others. Once the information is sent by the user as a SIP (Submission Information Package) the archive receives and maps it into a AIP (Archive Information Package). The stored data are then managed by the archive which operates to preserve them. When access to data is required a DIP (Dissemination Information Package) is created and made \-avai\-la\-ble for consumers \cite{borghoff2006long}.  

When a digital object is structured according to the AIP it is called a Trustworthy Digital Object (TDO) \cite{gladney}. It is interesting to underline the difference that emerges between what is \emph{trusted} and what is \emph{trustworthy}: as explained in \cite{gladney}, an object may have some features that make it  trustworthy but this does not guarantee the object to be trusted. Whether the object deserves to be trusted or not may be a choice of the user.

\subsection{Trusted Digital Repositories}

The OAIS model has been a guideline for another document that was published in 2002: ``Trusted Digital Repositories: Attributes and Responsibilities" \cite{tdr_2002}. In this, the fundamental concept of \emph{Trusted Digital Repository} (TDR) is defined. A TDR is a \-dig\-it\-al repository which is responsible for ensuring long term access to the digital objects stored in it, so in this case the term \emph{trusted} indicates a controlled relationship between the repository and the user. The principal attributes of this kind of repository are:

\begin{itemize}
\item Compliance with the OAIS model
\item Administrative responsibility
\item Organization viability
\item Financial sustainability
\item Technological and procedural suitability
\item System security
\item Procedural accountability
\end{itemize}

Though authenticity is a fundamental issue in a technological environment, the concepts of \emph{original} and \emph{authentic} are not only related to a technological transformation that a digital object may encounter:

\begin{center}
\emph{``The most fundamental aspect of trust has to do with authenticating the professed identities of human or agent participants. All other trust relations, mechanisms, and system components are created to relate to this fundamental one."}\cite{gladney}
\end{center}

As explained by this statement the first degree of trust must be established among entities made up of humans. The technological aspect should also be considered simply because the authenticity of a digital object can be verified by humans by means of a software. Since it could be very hard to verify the correctness of a software, it is better for a digital object to keep inside the evidences of integrity and authenticity.

In the following we will see how these feature can be essential for an agent to give a repository an appropriate degree of reliability.
\\

In the reminder of this chapter we are going to describe the most common techniques used for extracting information from a digital object. The key concepts of \emph{metadata} and \emph{preservation planning} will be discussed. Moreover some of the most common strategies to contrast digital obsolescence will be pointed out.

\section{An Overview about ``Metadata"}
\label{metadata_extraction}
\begin{center}
\emph{``Preservation metadata is the information necessary to carry out, document, and evaluate the processes that support the long-term retention and accessibility of digital materials."\cite{pronomroar}}
\end{center}

The concept of \emph{metadata} is as important as difficult to define. Several \-de\-fi\-ni\-tions have been proposed, based on different perspectives. A commonly accepted \-de\-fi\-ni\-tion is that \emph{``metadata is data about data"} \cite{intner2006metadata} \cite{wiki:metadata}, which is quite general. Even though the term metadata is usually related to digital contents, catalogue records for library materials may be viewed as metadata too \cite{aus_lib}. These two possible definitions come from a broad view of the concept of metadata. According to the them, metadata describe both digital and nondigital contents.

Other common definitions indicate metadata as \emph{``any data that aids in the identification, description and location of networked electronic resources"} or, even more, \emph{``metadata commonly refers to information available on the Internet"} \cite{intner2006metadata}. In these cases we can see how the concept of metadata is strictly related to digital contents.

A part from the definition, according to \cite{intner2006metadata}, it is possible to identify four principles that metadata have to satisfy:

\begin{enumerate}
\item Modularity: metadata schemas should permit to create new metadata based on parts of existing metadata combined together. 
\item Extensibility: the elements of a metadata schema need to be adaptable to specific fields, for example supporting additional necessary elements for a given application.
\item Refinement: it is important for a metadata schema to allow different levels of detail according to the requirements of the specific application.
\item Multilingualism: this principle is related to multiculturalism issue that involves many aspects, for instance the writing direction (left to right or right to left). Metadata should describe objects taking into account these cultural differences.
\end{enumerate}

As for the definition, there are various points of view concerning the division of metadata into categories. Of course metadata are expected to describe the structure and the behaviour of an object, its functions and how it must be handled by an user. Thus metadata may be classified according to which aspect of the object they describe. An example is provided by the following classification: \emph{structural metadata} about the design and \-spe\-ci\-fi\-ca\-tion of data structures, \emph{descriptive metadata} about the features of the object, for example the resolution of a digital image \cite{wiki:metadata} and \emph{administrative metadata} about the producer and the owner of the digital object \cite{intner2006metadata}. 
Though many categories may be created, the most interesting metadata for this work are preservation metadata. In this type of metadata is stored all the information useful to access the content brought by an object. In the following sections we will see how important this information is to perform a proper preservation action. 

The OAIS reference model \cite{oaisbriefing} plays a key role again since it provides a conceptual archiving model in which the guidelines for the data and metadata structure are established. Two projects for digital preservation named NEDLIB (Networked European Deposit Library) \cite{nedlib} and CEDARS (CURL Exemplars in Digital ARchiveS) \cite{cedars} are based on the OAIS model \cite{intner2006metadata}.

\subsection{Tools for the Extraction of Metadata}
\label{tool_extraction} 
Among the metadata, the format of a digital object is one of the most relevant features needed to understand if a certain file is under risk of obsolescence. Since 2002 the TNA (The National Archives of the UK) \cite{tna} developed a technical \-re\-gi\-stry of file formats named PRONOM \cite{pronomroar}. In this registry the information about the format and the one about the software application environment are joined together. These elements are necessary to know how to render an object and how to guarantee future access to it, for example through a preservation action \cite{pronomroar}. 

An example of software that takes advantage of the information about file formats is AONS (Automated Obsolescence Notification System) \cite{aons}. The AONS I prototype has been developed by the National Library of Australia (NLA) \cite{aus_lib} in collaboration with the Australian National University in the 2006 and can be described as follows:  
\begin{center}
\emph{``is a system [designed] to analyse the digital repositories and determine whether any digital objects contained therein may be in danger of becoming obsolescent. It uses preservation information about file formats and the software which supports these formats to determine if the formats used by the digital objects are in danger"} \cite{aons}
\end{center}
In 2007 another prototype called AONS II \cite{aons2} was developed. This project refined and expanded the functionality of AONS I \cite{aons}.
As its previous version it builds a profile of the formats in a repository making use of format recognition tools such as DROID \cite{droid} and JHOVE \cite{jhove}. 

DROID (Digital Record and Object Identification) \cite{droid} is a software tool \-de\-ve\-lo\-ped by The National Archives to perform automated batch identification of file formats. It uses internal signatures to identify and report the specific file format and version of digital files. These signatures are stored in an XML signature file \cite{droid}. Notice that DROID do not rely on the file extension since it could be shared by \-se\-ve\-ral applications. Signatures are based on key byte sequences that identify uniquely the file format.  

JHOVE (JSTOR/Harvard Object Validation Environment) \cite{jhove} is an extensible software framework for performing format identification, validation and characterization of a digital object \cite{jhove}. All these actions are frequently necessary during digital preservation activities.\\

The format used by a digital object is certainly not the only useful information that can be extracted. As discussed in \ref{metadata_extraction}, each object can be identified by its metadata which can be of several types.

A really easy to use tool for extracting metadata is FITS (File Information Tool Set) \cite{fits}. It identifies, validates and extracts metadata for a wide range of file formats. It is a kind of wrapper since it invokes and manages the output from other open source tools. The output is an XML file containing all the information about the digital object. 

Another metadata extraction tool that must be considered is the NLNZ metadata extractor \cite{nlnz}. Its development began in 2003 supported by the National Library of New Zealand and the software was released in 2007. The purpose of this tool is to extract useful metadata for preservation of digital objects of various formats. Several adapters has been implemented in order to manage several types of objects such as images, Microsoft Office documents, audio and video files, HTML, XML and ARC internet files. The tool opens each file as read-only to guarantee the integrity of the original object \cite{nlnz}.

Like FITS, the NLNZ metadata extractor generates an XML output file containing the metadata necessary for a preservation process.

\section{Preservation Planning}
Once that metadata are extracted and the risk of obsolescence is detected it is necessary to decide how to deal with it. As anticipated in the introduction the main strategies developed to preserve digital object are migration and \-e\-mu\-la\-tion. In both cases it could be very hard to decide which is the best strategy to apply because each situation may have specific requirements and problems to deal with. The decision is usually taken by skilled humans but software can help providing a preservation plan. Such a software has to guarantee full traceability and documentation of all elements influencing the final decision  \citep{planets}. Several aspects are important for a plan, an example is provided by unavoidable losses related to the migration process. The user has to define accurately all the requirements that the planner tool will observe. It is critical to find the right balance between an automated process, of comparison and validation of objects, and the need to inform the planner about the actions to perform.
\\

In 2006 a four-year EU funded project named PLANETS (Preservation and Long-term Access through Networked Services) \cite{planets_web} started with the aim to help preservation planning. As explained in \cite{planets_2008}, Planets involves \-se\-ve\-ral preservation functions such as: preservation planning, characterization, preservation action and an interoperability framework. 

Regarding the \emph{preservation planning} it is made up of a set of services that determine the risk of obsolescence, suggest a collection of plans and choose the best one to be executed. The planner takes advantage of both content characterization and preservation action.

The \emph{content characterization} is responsible for extracting those contents that are \-rel\-e\-vant for the plan. 

The \emph{preservation action} essentially is what the system suggests to perform in order to keep the digital object usable for a certain time. 

\begin{figure}[h!]
\centering
\includegraphics[width=8cm]{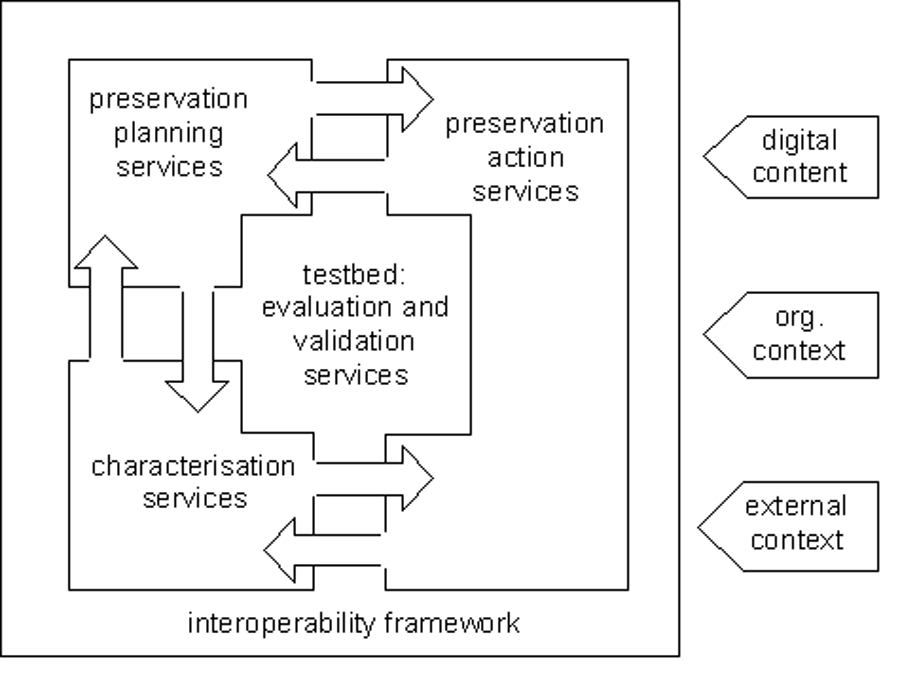}
\caption{Planets architecture \cite{planets_2008}.}
\label{planets_arch}
\end{figure}

These three services are often related to each other and to various sources, for example other open-source projects. This means that an \emph{interoperability framework} becomes necessary to ensure the proper information sharing. The schematic in \ref{planets_arch} shows the architecture of Planets.

If we analyse further the preservation planning service we find that it is carried out by a specific tool called Plato (preservation Planning Tool) \cite{plato}. As explained in \cite{plato} and \cite{planets} it follows three main steps in the preservation planning process. 

\begin{figure}[h!]
\centering
\includegraphics[width=8cm]{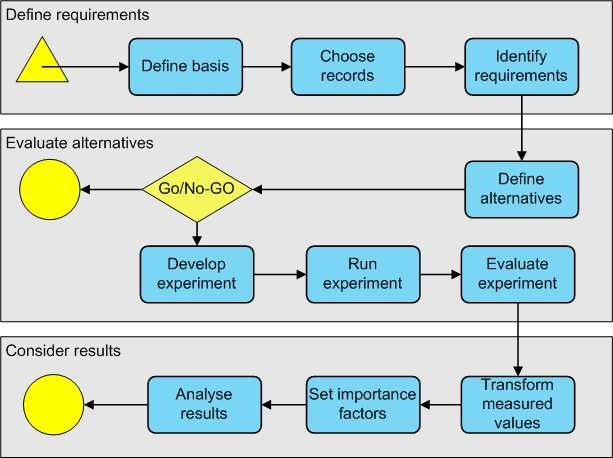}
\caption{Preservation planning workflow \citep{plato}.}
\label{planets_workflow}
\end{figure}

The first and most important one is the requirements definition: the users has to define requirements in a quantifiable way in order to create a so called objective tree which is the basis for a correct evaluation of the possible strategies. This is actually the second step, the evaluation of potential strategies is performed by applying the strategies to a sample object and comparing the results to the requirements. The choice of a most representative sample is fundamental for a correct result. The last step consists of the analysis of the results: it is possible to suggest a certain strategy by giving a proper weight to each requirement. As depicted in Figure \ref{planets_workflow}, each step consists of other processes. In \citep{planets} an accurate description of all processes is provided which will be summed up in the following scheme:

\begin{itemize}
\item \textbf{Requirements definition}
	\begin{enumerate}
	\item \emph{Define basis}: the preservation scenario, the type and number of files, the environment.
	\item \emph{Choose records}: picking of the sample used for the next processes.
	\item \emph{Identify requirements}: the requirements and goals organized in the objective tree for a preservation solution. High-level objectives are categorized as: file characteristic, record \-cha\-rac\-te\-ri\-stic, process \-cha\-rac\-te\-ri\-stic and cost.
	\end{enumerate}
\item \textbf{Alternatives evaluation}
	\begin{enumerate}
	\setcounter{enumi}{3}
	\item \emph{Define alternatives}: description of the possible alternatives and the resources needed to perform them.
	\item \emph{Go/No-Go}: deciding whether proceeding with an alternative among the possible ones and why.
	\item \emph{Develop Experiment}: creation of a specific development plan for each experiment.
	\item \emph{Run Experiment}: applying the chosen alternative to the samples.
	\item \emph{Evaluate Experiments}: estimate how much the results satisfy the requirements.
	\end{enumerate}
\item \textbf{Results evaluation}
	\begin{enumerate}
	\setcounter{enumi}{8}
	\item \emph{Transform Measured Values}: mapping the results into a uniform and comparable scale.
	\item \emph{Set Importance Factors}: giving each objective the appropriate weight.
	\item \emph{Analyse Results}: joining the values of each objective in order to have a single comparable value for each alternative. This process could be a weighted sum, a multiplication, a sum of priority or Austin slight.
	\end{enumerate}
\end{itemize}

\begin{figure}[h!]
\centering
\includegraphics[width=8cm]{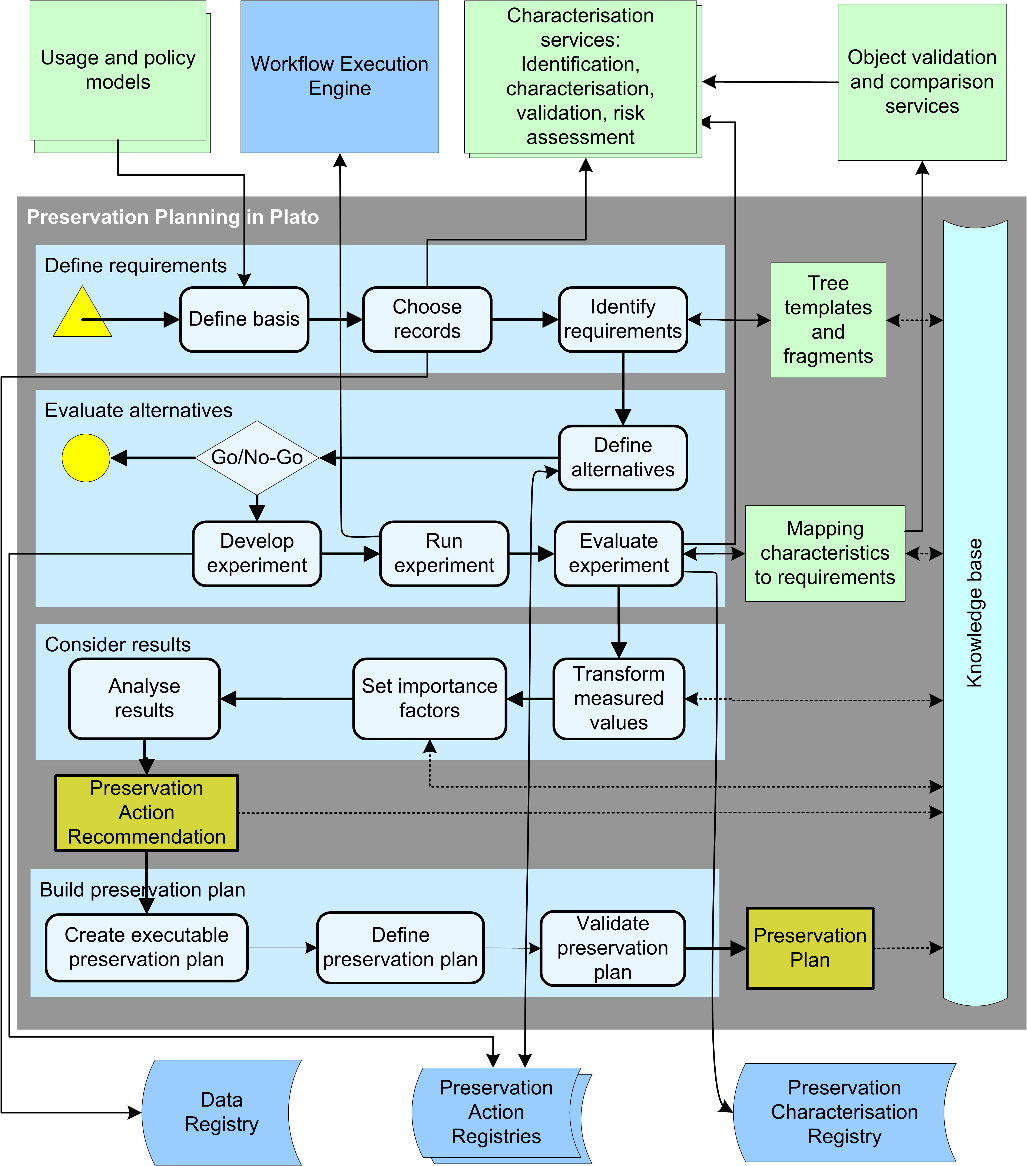}
\caption{Preservation planning environment \citep{planets}.}
\label{pres_env}
\end{figure}

The result of these steps is an objective and well documented ranking of the possible alternatives to perform in the given preservation context.

The planning tool Plato implements this workflow and generates an executable plan in addition. As illustrated in Figure \ref{pres_env} it is necessary to consider the whole environment in which the preservation planning process takes place.

\section{Preservation Strategies}
As asserted in the introduction, migration and emulation are the main techniques in digital preservation. Nevertheless other strategies exist that can be mentioned. 

A general strategy for long-term preservation of digital objects was presented in 2006 by the Online Computer Library Center (OCLC) \cite{oclc}, in which four essential points are underlined. A proper digital preservation strategy has to:

\begin{itemize}
\item Assess the risks for loss of content posed by technology variables such as commonly used proprietary file formats and software applications.
\item Evaluate the digital content objects to determine what type and degree of format conversion or other preservation actions should be applied.
\item Determine the appropriate metadata needed for each object type and how it is associated with the objects.
\item Provide access to the content.
\end{itemize}

From these reference points, there are at least seven different approaches. It is important to underline that the most part of preservation processes takes advantage of migration and emulation which therefore are the most relevant preservation actions.

\begin{enumerate}
\item \textbf{Migration}: this is the most frequently adopted solution and it is regarded as an essential function of digital archives \cite{gladney} \cite{taskforce}. The object is converted into a newer format so that it can be accessed within a more recent software or hardware configuration. The main problem related to migration is the difficulty of ensuring the authenticity of the object so a cryptographic procedure may be necessary. Another aspect to consider is that the migration process could take a long time so, in the worst case, a large collection of objects may be obsolete again when the migration is terminated. 
\item \textbf{Emulation}: this strategy has already been introduced in the introduction. In this case the action is not performed on the obsolete object but on the environment. An older software or operating system is recreated in order to render the object. 
\item \textbf{Refreshing}: this process is related to the deterioration of physical storage media. It is the transfer of a digital object into a newer storage device (of the same kind of the older). The data are supposed not to encounter bitrot or other sort of alteration. 
\item \textbf{Replication}: in this case the digital object is simply duplicated. This simple solution helps to protect data form intentional or accidental damaging but, at the same time, makes it difficult to perform refreshing or migration since multiple copies of the same object are located in different places.
\item \textbf{Encapsulation}: it is a necessary method for those collections of objects that will not be rendered for a long period. It consists of the creation of containers to provide all the useful information about the objects and how to render them in the future.
\item \textbf{Persistent Archive Concept}: it aims to develop a wider infrastructure to guarantee the preservation of a collection of object regardless of the platform. It follows the OAIS reference model.  
\item \textbf{Metadata attachment}: in this alternative approach the focus of preservation are the metadata. As seen in Section \ref{metadata_extraction} a digital object can be described through its metadata but these can become obsolete too. The ASCII format could be a good way to represent metadata since it is backward compatible and easily readable by humans.  
\end{enumerate}

\section{Digital Preservation Platforms}
In this section we introduce some of the most known and commonly adopted \-di\-gi\-tal preservation platforms. All the applications reported in the followings are free and open source, the description of them has been excerpted from \cite{forgetit}. Each of the described solutions are compliant to the OAIS reference model \cite{oaisbriefing} described in Section \ref{OAIS}.

\subsection{Archivematica}
Archivematica \cite{archivematica} is a digital preservation platform with the aim of providing standard-based and long-term access to collections of digital objects. It uses a micro-services design pattern to provide an integrated suite of software tools that allows users to process digital objects from ingest to access in compliance to the OAIS \cite{oaisbriefing} model \cite{archivematica}. The system supports several common metadata standards such as METS \cite{mets} as concerns the ingest and access, PREMIS \cite{premis} and Dublin Core \cite{dublincore} as preservation and descriptive metadata. In addiction it can be integrated with other platform such as DSpace \cite{dspace} providing preservation functionalities.

\subsection{DSpace}
The DSpace \cite{dspace} repository is an out-of-the-box solution for open access repositories which is suitable for both small archives or wider organizations. The platform is worldwide adopted especially by research and non-profit communities. DSpace is also well supported by a large and distributed community of developers and researchers, it is freely available and fully customizable to accomplish the specific requirements of any archiver. The user is allowed to define access and dissemination policies in order to allow only specific users or groups of users to access a certain digital content. In addiction, DSpace \cite{dspace} can be integrated into a cloud storage for long-term preservation of digital objects.

\subsection{Fedora}
The Fedora Repository Project \cite{fedora} is led by the Fedora Project Steering Group and is under the stewardship of the DuraSpace \cite{duraspace} not-for-profit organization providing leadership and innovation for open source technology projects and solutions that focus on durable, persistent access to digital data \cite{fedora}. The system is based on a core repository service which is accessible through a web interface. Images, videos, documents and other types of digital objects are supported and managed by means of several services such as search, messaging and administrative clients. A cloud solution is also provided but not for free.

\subsection{RODA}
KEEP SOLUTIONS \cite{keepsolutions} provides another platform named RODA (Repository of Authentic Digital Objects) \cite{roda}. The platform is based on Fedora \cite{fedora} and takes advantage of several standards such as OAIS \cite{oaisbriefing}, METS \cite{mets} and PREMIS \cite{premis}. RODA \cite{roda} basically supports normalization on ingest for different file formats but can also be extended in order to manage more formats. Each process is performed through several tasks that can also be customized and extended to match the user's requirements. The user can also define policies and strategies for the internal task scheduler to perform preservation actions.

\chapter{Agent Based Models}
\label{capAgents}
In this introductory chapter we will address those basic concepts related to agent based models that are needed in order to better understand the research described in the following chapters. We are going to describe what an agent itself is, the environment in which it operates and finally how it interacts with the other agents as parts of a multi-agent system.

\section{The Agent}
The key issue is what are an agent and its features. First of all it could be useful to underline some important trends in computer science:

\begin{itemize}
\item \textbf{Ubiquity}: the processing capability could be located into different places, so the intelligence needs to be ubiquitous. This trend is related to the cost of perform computing. 
\item \textbf{Interconnection}: nowadays it is almost obvious for a computer to be connected to the Internet. This is an example of how computer system are interconnected; concurrent and distributed systems are now very common in computer science.
\item \textbf{Intelligence}: another fundamental parameter is the growing complexity of the tasks that computers are supposed to solve. 
\item \textbf{Delegation}: this is probably the most relevant trend. The difficult tasks mentioned before are left to computer system that have to deal with them without direct control by the user.
\item \textbf{Human-Orientation}: the ways of programming is constantly  evolving towards a human way of thinking and understanding the world. An example is provided by the evolution from command-line interfaces to high-level programming languages.
\end{itemize}  

These trends act as guidelines for defining and building an agent. Even though there is not any universally accepted definition, it is clear that it must include the concept of autonomy. As mentioned in \cite{wooldridge} 
a possible definition could be:

\begin{center}
\emph{``An agent is a computer system that is situated in some environment,
and that is capable of autonomous actions in this environment in order
to meet its design objectives."}
\end{center}

This computer system exists within an environment and exchanges information with it. The different kind of environments will be discussed further in the next section. 

It is now interesting to look at two common systems, such as a thermostat and a software daemon, as agents. The thermostat interacts with the environment (for example a room) in two different ways. Firstly it acquires information from the environment by detecting the temperature, secondly it modifies the state of the environment by switching the heating on if necessary. A software daemon can be viewed as an agent too since it monitors the environment (in this case an operating system) and, when it is necessary, performs actions to bring the environment into a certain state. In both cases we can see how the computer systems autonomously modify the environment in order to achieve their objectives. 

The examples just discussed are obviously very simple. It is necessary to better focus the concept of environment before introducing more sophisticated kinds of agents. 

\section{The Environment}
\label{env}
Even if basic, the example of the thermostat as agent shows that agents have to interact with the environment. They detect the state in which the environment is and decide which is the best action to perform in order to meet their goals. The process to perform an action that modifies the state of the environment is called a \emph{run}.

\begin{figure}
\centering
\includegraphics[width=8cm]{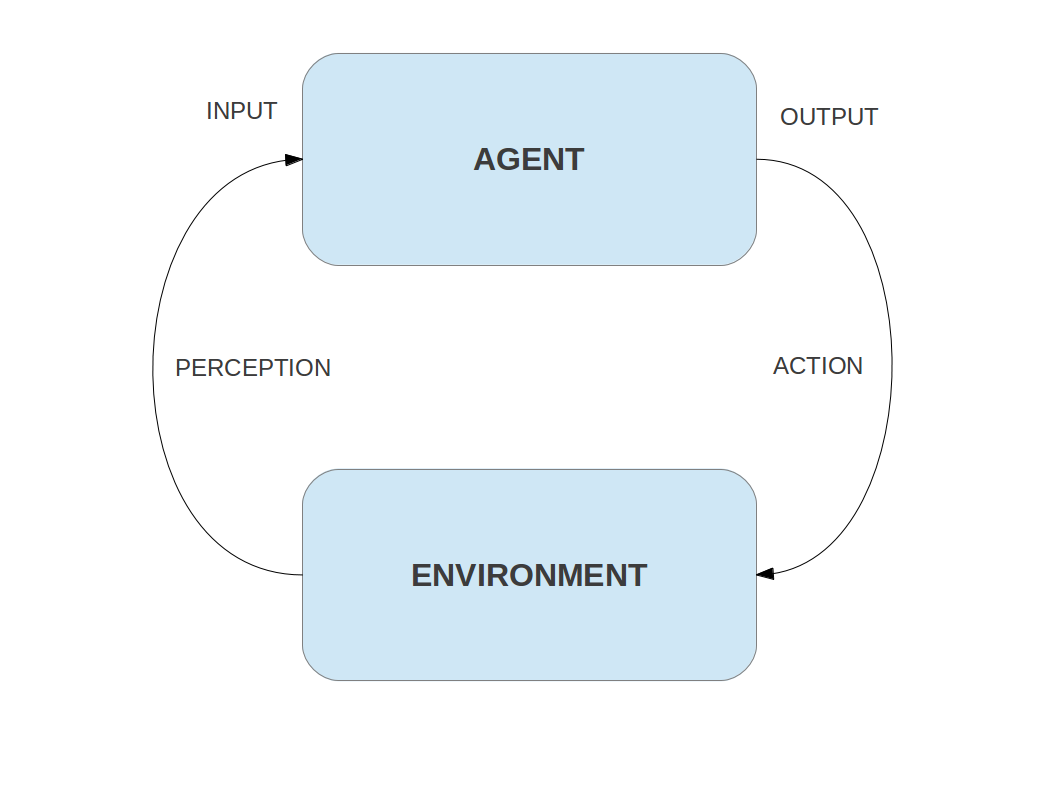}
\caption{Simple interaction between agent and environment.}
\label{int_1}
\end{figure}

This process could be much more difficult than it seems. The decision about which action the agent should better perform will determine the success or the failure in reaching its objectives. This choice is strictly related to the quality of information that the agent can get from the environment. While a correct and detailed information may lead to right actions and thus to success, a poor information could mislead the agent causing failure. The quality of the information available depends on the kind of environment. In \cite{russell}, environments are classified as follows:

\begin{itemize}
\item \textbf{Accessible / inaccessible}: an environment can be defined accessible if the agent can access a complete, accurate and up-to-date information. Most of the real environment are not accessible since a degree of uncertainty is always associated with the information.
\item \textbf{Deterministic / non-deterministic}: this distinction might be regarded as the most relevant. An environment is said to be deterministic if the result of an action is completely defined and predictable. Non-determinism shows two important implications: firstly the agents only have a partial control over the environment, secondly the actions can fail. 
\item \textbf{Static / dynamic}: if seen under the agent's point of view, an environment is static if its state remains unaltered when no agents perform actions on it. Unfortunately most of the environments are dynamic. This means that the state of the environment can change even though no agent tried to modify it. It is easy to see that dynamism leads to non-determinism.
\item \textbf{Discrete / continuous}: a discrete environment can be in a large but finite number of state whereas for a  continuous one the number of state is not countable.
\end{itemize}

By the agent's perspective, the worst kind of environment is the open one: it is inaccessible, non-deterministic, dynamic and continuous.

\begin{figure}
\centering
\includegraphics[width=8cm]{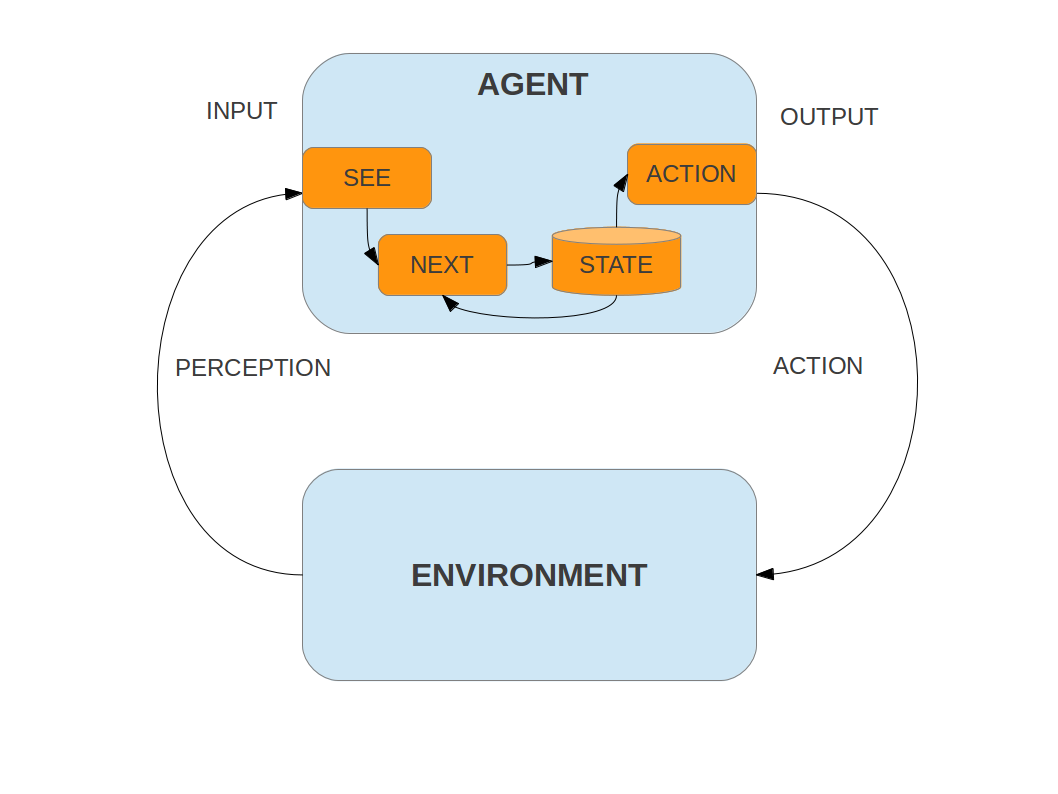}
\caption{Interaction between agent and environment.}
\label{int_2}
\end{figure}

Another fundamental aspect is how agents interact with the environment. As we shall see the environment should be a base for the agent building process. A wrong interaction will make it impossible for the agent to achieve its goals. The schematic in Figure \ref{int_1} shows a basic interaction between an agent and the environment. It is possible to look further inside the agent to better understand how the input is processed and the output is generated.  As depicted in Figure \ref{int_2} the \textit{see} function maps the input into the set of perceptions, the \textit{next} function compares the perceptions with the present intentions (the state) that will be updated if necessary. The \textit{action} function chooses which action to perform according to the intention. Varying these functions different kinds of agent can be built. 

In \citep{wooldridge} the following classification is provided:

\begin{itemize}
\item \textbf{Purely functional agents}: this kind of agent computes a function form the set of perception to set of actions and terminates. If the environment is both static and deterministic the success of the action performed is guaranteed.
\item \textbf{Purely reactive agents}: in this case the agent establishes an ongoing interaction with the environment, it is commonly used when modelling the environment is hard. The previously cited thermostat is an example of reactive agent. 
\item \textbf{Episodic agents}: when the agent keeps in memory a limited number of past events it is called episodic. This agent chooses the action to perform taking into account the previous decisions.
\item \textbf{Real time agents}: repeats actions as frequently as possible and checks the results. The time dependency is underlined.
\end{itemize}

The key point in building an agent is to find the right balance between the pure reactive and the goal-oriented behaviour.

\section{Intelligent Agents}
\label{int_agents}

Once that the concepts of agent and environment have been clarified, we can proceed considering more complex concepts. As underlined in the introduction of this work, a sort of distributed and autonomous intelligence becomes necessary for our purposes. So, as the title of this section suggests, the agents have to be \emph{intelligent agents}.

There are at least three requirements that an agent has to satisfy in order to be classified as intelligent: \emph{reactivity}, \emph{proactivness} and \emph{social ability} as explained in \citep{wooldridge}. It means that an intelligent agent is able to sense environment changes and consequently react to them with respect to its objectives. At the same time it can take initiative without necessarily waiting for the environment to evolve. Finally the intelligent agent will interact with other agents communicating and collaborating always with the aim to reach its goals.

In Figure \ref{env} we saw how the interaction with the environment is determined by the \emph{see}, the \emph{next} and the \emph{action} functions. This structure is matched by the \emph{intentional stance}. In this approach the agent is characterized by mental states such as beliefs, desires and intentions: the so called \emph{BDI paradigm}. With \emph{beliefs} we mean all the knowledge acquired by the agent, the \emph{desires} are all the feasible actions whereas \emph{intentions} are only the actions that the agent will perform. An intentional system can be structured on several level of complexity: as explained in \cite{dennett} a first-order intentional system may have beliefs, desires and intentions whereas a second-order system has beliefs, desires and intentions about other beliefs desires or intentions.

So far we endowed the agent with the possibility to have beliefs, desires and intentions but nothing has been told about how the agent really explicates its own intentions. The key idea here is to tell the agent what to do without explaining how to do it, the proper strategy will be chosen  by the agent itself. Since an agent decides to perform a certain action on the basis of self-interest it is possible to introduce an auxiliary function that maps a state of the environment $E$ into a number. It is called \emph{utility function}:

\begin{equation}
u:E\rightarrow\Re
\end{equation}

In the pessimistic case $u$ is the utility associated with the worst state, in the optimistic case with the best and in the realistic case it is the weighted average of the utilities of all the states at a run. 

Another way to define the utility function is: 

\begin{equation}
u:R\rightarrow\Re
\end{equation}

where $R$ is the collections of the possible runs. This definition may be the better one when a sort of history of the runs performed is required. 

In both the approaches mapping a state of the environment or a run into a real number could be a difficult task. By the way the utility function becomes fundamental since it determines the intention if the agent: it will choose to perform a run that increases its utility. This will be a key concept for the interaction with other agents. 

Thus the aim of the building process is to implement an agent which chooses to execute the run that maximizes its utility function. This is called an optimal agent and in \cite{wooldridge} it is defined as follows:

\begin{equation}
Ag_{opt}=\max_{Ag\in AG}\lbrace \sum\limits_{r\in R(Ag,Env)} u(r)P(r|Ag,Env) \rbrace  
\label{optagent}
\end{equation}

where $AG$ is the set of agents, $R(Ag,Env)$ is the set of all the possible runs for the agent $Ag$ in the environment $Env$ and $P(r|Ag,Env)$ is the probability that the run $r$ occurs given $Ag$ and $Env$. In practice the implementation of some agents may require more memory than the available resources on a certain machine. In this case we must replace $AG$ in equation \ref{optagent} with $AG_{m}$ which is the set of agents that can be implemented on the machine $m$, $Ag_{opt}$ is now called bounded optimal agent.

\section{MAS (Multi-Agent Systems)}
Now that the intelligent agent has been defined, it is possible to place it into a society of intelligent agents: a \emph{multi-agent system}. Though such a system may appear quite hard to manage, computer systems in which many sub-systems have to cooperate together are very common. In this kind of society the environment is shared among the agents, each one manages a part of the environment named sphere of influence. Since spheres of influence that belong to different agents may intersect, or even coincide, these agents need to interact in order to decide how to operate on this common part of the environment. The interaction between agents is the key point of the multi-agent system building problem. \\

As introduced in Section \ref{int_agents}, each agent has got an utility function enclosing its interests and preferences. Thus the decision about which action to perform is lead by the outcome of the utility function. Let us consider the interaction between a finite number on agents that simultaneously choose the action to perform: it is possible to introduce the \emph{state transformer function}

\begin{equation}
\tau : action_{1} \times action_{2} \times ... \times action_{n} \rightarrow \Omega  
\label{optagent_copy}
\end{equation}    

where $\Omega$ is the outcome state of the environment and $action_{i}$ is the action chosen by the i-th agent. This function describes the evolution of the environment due to the operation of all the agents.

Now each agent will retain the outcome state $\Omega$ preferable or not according to its own utility function. That is, a sort of preference ordering is generated. For instance: given two different states $\omega$ and $\omega'$, $\omega$ is weakly preferred by the i-th agent if $u_i(\omega) \geq u_i(\omega')$ is strongly preferred if $u_i(\omega) > u_i(\omega')$. The game theory is an example of situation in which the preferences of agents may be in contrast \cite{wooldridge}, \cite{wiki:game_theory} . 

This interaction may cause a dependence relation between agents. If so this relation could be unilateral, mutual or reciprocal if an agents depends on another one with respect to different objectives. The agents may have no knowledge about the existence of any dependence relation between them.

In order to satisfy their objectives agents then have to communicate and cooperate. Cooperation consists of reaching agreements that satisfy all the agent that are dealing whit the same sphere of influence. This aspect is fundamental for the social intelligence that should emerge from an agent-based model. The agreement reaching process is mainly made up of two other processes named \emph{negotiation} and \emph{argumentation}. In the negotiation process protocols, mechanisms and strategies adopted by the agent are defined. The aim is to ensure the agreement maximizing the sum of the utilities of all the agents that are taking part to the negotiation. In the easiest case, negotiation takes place between two agents (one-to-one negotiation), in situations such as auction the process is a little more complex (many-to-one negotiation) and finally in the extreme case any agent may negotiate with all the others (many-to-many negotiation). 

We can see how close the processes described are to negotiation mechanisms among humans. As happens for humans, agents may have the chance to reconsider their decision during the negotiation process and so the need to justify their choice. This is guaranteed through the argumentation process that mainly consist of convincing others about something. Several ways for convincing others are possible, the most interesting is the logical mode which is adopted in scientific fields (for example in mathematical demonstrations), other possibilities are the emotional mode, the visceral mode and the kisceral mode each one borrowed from human argumentation modes.

\section{Communication among Agents}

The kind of system so far described surely relies on the existence of agents which can exchange information. This fundamental requirement can be satisfied since agents do communicate with each other. In this work the ability to communicate will be of particular interest because the agent will inform others about its internal state and receive instructions As for humans interaction, in order to perform a correct and productive dialogue, the communication needs to adopt a certain protocol to avoid confusion. Agents then has to comply the same protocol and language.

The Knowledge Query and Manipulation Language (KQML) \cite{wooldridge}, \cite{KSE_report} is a message-based language for communication among agents that defines a format for messages. Each message consists of two main parts: a \emph{performative} that specifies the request type and a \emph{parameters part} in which are located further information about the request \cite{wooldridge}. Several versions of this language have been proposed which were characterized by different sets of performatives but a series of issues brought to the implementation of a new language. 

This Agent Communication Language (ACL) \cite{wooldridge} has been developed by the Foundations for Intellgent Physical Agents (FIPA) \cite{wooldridge} \cite{fipa}. As the KQML language it defines the format for messages and a list of performatives but does not adopt a specific language for the content of the message \cite{wooldridge}. This feature \-gua\-ran\-tees more freedom in sharing desired information. The developers focused also on the construction of a proper semantics to this language, in particular a formal language named SL (Semantic Language) \cite{sl} \cite{wooldridge} was chosen. It is based on the theory that regards speech acts as rational actions.

\chapter{Implementation and Modelization Framework Design}
\label{capImple}
The aim of this model is to simulate a distributed environment into which many entities communicate in order to solve their digital preservation issues. \-Dif\-fe\-rent types of entities will be considered such as libraries, universities, personal archives, broadcasters or government entities. Each entity will be able to exchange information with all the others but more relevance will be given to the entities that belong to the same network: even though a personal archive could provide useful information to a library, this should be more interested in what the other libraries have done to deal with a certain problem. Besides this, more attention will be given to information coming from entities that are close by a geographical and cultural point of view. For example a Portuguese university may not take advantage of the new format adopted by an Asiatic university to store its textual documents, while it may be interested in what a Brazilian one has done. 

From this point the key aspect is the level of \emph{trust} that each entity will associate to others. This variable describes the relationship between two entities and tells how strictly related they are. The evolution of the trust level takes place as follows: suppose that an entity encounters a problem such as the impossibility to render a digital object. It will ask other entities among those in its network if any of them has encountered a similar problem. If so, each entity will suggest its own solution and the receiver will decide which one is to be adopted according to its internal state and will perform it. If the solution has been effective and the problem is now solved, the entity increases the level of trust of all the entities that suggested the right solution and decreases the trust of the others. This process allows to identify those entities that provide suitable solutions to the problem. For example a big sound archiver may suggest to duplicate a collection of files in order to preserve them but, since this solution requires a lot of storage memory, it could not be applied by a small personal archive.

\section{Development Environment: GAMA}
\label{gama}

The software employed for the implementation of the agent-based model in this work is GAMA (Gis \& Agent-based Modeling Architecture) version 1.6 \cite{gama}, \cite{gis_abm}. It is a modelling and simulation development environment for building spatially explicit agent-based simulations. It is being developed by several research teams under the umbrella of the IRD/UPMC International Research Unit UMMISCO \cite{gama}. 

GAMA takes advantage of the Gis \& Agent-based Modeling Language (GAML) \cite{gama}. In general terms a GAML model is made up by a certain number of actions which consist of a sequence of statements. As for other programming languages such as Java or C++, each statement uses expressions (groups of keywords, variables and operators) to define the computations to be done. 

\begin{figure}[h!]
\centering
\includegraphics[width=11cm]{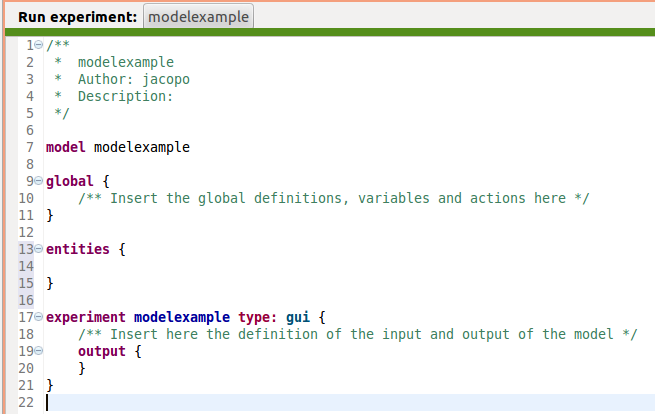}
\caption{Basic structure of a GAML model.}
\label{struct_gaml}
\end{figure}

Figure \ref{struct_gaml} shows that a model is mainly structured in three sections. The first one is the \emph{global} section in which the definitions and behaviours of the environment can be placed, here also the statements for the agents' creation are located. Note that, by the software perspective, the environment is a particular agent too which is named world agent. The second section named \emph{entities} regards the agents' specifications: in this part of the model the various species of agents and their features (facets, behaviours, reflexes, and the like) are defined. It is also possible to describe the aspect of the grid which is regarded as an agent too. In the \emph{experiment} section are given the descriptions of the simulation to run. Here are situated the statements that concern both what will be displayed and which information will be collected during the simulation.

The core of the model is surely made up of the various agent species defined in the entities part. Each species may have a collection of possible actions that the agent can do when asked to or when a particular condition is verified. It is important to underline that species may be related to each other through an inheritance relationship, in this way it is possible to make a species automatically inherit skills, variables, actions and reflexes from its parent. Another useful and interesting structure is the nested structure. The user can declare a new species named micro-species inside another one that thus become the macro-species. 

As we will see later in this work, these two features will be fundamental for the implementation of this model.

\section{Global}
\label{global}
The first part of the model is named \emph{global} and includes all the variables that are common and accessible to every agent. As we shall see, some of these variables will be imported from external source files and other will be updated by the agents during the experiment while other will be simply constants. In addiction, here are instantiated all the variables that are monitored in the experiment or used to draw charts and histograms.

It is possible to locate some statements into the \emph{init} that is performed as soon as the experiment starts. Since the whole environment is regarded as an agent, as described in Section \ref{gama}, the user can declare global action and reflexes to define the global behaviour of the model.

\subsubsection*{Variables}
The model requires a very large number of global variables in order to work \-pro\-per\-ly. The most important of them are described and analysed in the following lines. Before proceeding with the description of important and complex variables, it is necessary look at a couple of simple but fundamental ones. 

\paragraph{operating systems:} this is a simple list of strings representing the name of the operating system adopted by the institutions. In this model we chose the adoptable operating system to be simply one among \emph{Windows}, \emph{Apple} and \emph{Linux}. We decided not to forward us in the various versions and distributions of Microsoft Windows \cite{windows}, Apple \cite{apple} or Linux \cite{linux} since this is not the focus of this work. 

\paragraph{number of pastors:} despite it is a simple integer with its value set to 4, this is a fundamental variable since it states that four types of digital objects are taken into account: audio files, images, text files and videos. As we shall see, many of the following variables have a number of component equal to 4 in order to cover all the possible types. As usual, the choice of limiting the object types to these four has been made to simplify the model and make the simulation easily runnable. By the way this limitation is not a constraint and more articulated models may be developed from this one.  
\\

Let us now begin with the analysis of all the variables related to the formats and the software, two fundamental aspects in this model.

\paragraph{available formats:} this variable is a list of four lists of strings which are the formats available in the model for each type of digital object. By choice we decided to fill these lists up with the first 50 popular formats according to \citep{fileinfo}. Each list of formats is imported from the corresponding text file inserted in the ``includes" folder of the model. In the \emph{init} action each of these four lists is put into the \emph{available\_formats} list at the corresponding index.

\paragraph{mig. time coef.:} as regards the formats, we decided to implement this list of four square matrices that contain migration coefficients. The value of these elements ($M_{ij}$) aims to indicate how long does it take to convert a kB of format $f_i$ into format $f_j$. We would like to underline that these values are set randomly between 10 and 90 when the model runs and they remain constant through all the simulation. As we shall see later, institution agents use them by means of the \emph{migration time} action in order to determine the number of cycles for which the agent is occupied with the migration. The main idea here is to introduce a temporal distinction when migrating from the same format to different destination formats. The user is allowed to replace these elements with realistic values that it may estimate.

\paragraph{available software:} similarly to the previous one, this variable contains the list of the available applications. While formats are the same for every institution, the available applications depend on the operating system adopted by each institution. This variable is thus a bit more complex since it is made up of three lists of lists (one per each operating system), each of these contain four list of strings that, as for the formats, indicate the available application for the given type of digital object in the given operating system. The list of applications are taken from \cite{fileinfo} too. The double step for initializing this variable takes place in the \emph{init} action. As for the format lists, each one has been imported from a corresponding text file.

\paragraph{format vs software:} as happens in real software environments, there is an important relationship between format and application because each application can render a certain number of formats and, on the other hand, each format can be rendered by a list of applications. The aim of the \emph{format\_vs\_software} variable is to establish this relationship in the model so that the software manager of each institution is capable of rendering certain formats with the installed applications. This is a list of four matrices, again, one per type of digital object: columns are the applications while rows are the formats. If the application \emph{j} can render the format \emph{i} the element $m_{ij}$ of this matrix is equal to 1, in the other case 0. As will be explained in the followings, these matrices are read every time that a software manager performs a rendering action. Each matrix has been filled up according to \cite{fileinfo} and saved as a \emph{csv} file in order to be correctly imported from the ``includes" folder.
\\

Other important variables regard the popularity of formats and applications within the model. As we mentioned in the introduction, the risk of obsolescence is not a personal and local parameter, it can be estimated only by considering some global aspects. In this model we decided to consider three global parameters: the total number of digital objects in this format, the number of institutions that have at least one object of the  given format and the number of installed applications can that render it. These parameters, which are used by the institutions to evaluate the risk of a certain format, are embodied by the following variables:

\paragraph{file number for format:} this list is made up of four lists of integers, one for each format of the given object type. When the institutions and their pastors are initialized, they loop over their members (their format collections) and add the corresponding number of file to that lists. The values of these lists are updated every time the number of files of a format collection changes, or when a format collection is created or deleted. In this way it is possible to constantly monitor how many files of each format are there at runtime.  

\paragraph{institution number for format:} this variable is built exactly as the one describe just above. The difference is that, in this case, the value of the integers represents the number of institutions that keep inside a format collection of the format taken into account. It is updated at initialization and when a format collection is created or eliminated.  

\paragraph{software number for format:} the last of this group of variables regards the number of applications, capable of rendering a format, that are installed by the institutions. As usual there is a list of integers for each type of digital objects and these lists are filled by the software manager. This agent evaluates how many of the installed applications are able to perform the rendering of the format and add this number at the proper position in the list. This operation is executed every time an application is installed or removed.  

\paragraph{installed software:} as the previous one, this variable regards the applications. It is a list of three lists, one for each operating system, each one containing other four lists of integers. When the software manager is initialized it loops over the installed applications and increases the corresponding values of these lists. The values are updated every time an application is installed or removed in order to keep updated the number of institutions that have installed a given application. This variable is not related to the formats but it gives important information about which applications are the most adopted by the institutions. It could be useful when an institution has to decide which application to install among a group of them, this parameter should indicate which is the most adopted and thus the most supported one.
\\

The last interesting variable relates the formats and the global size of them that has already been migrated. As we shall see later in Section \ref{decisions_evaluation}, this variable plays an essential role in the evaluation of the agents' decisions.

\paragraph{format migrated sizes:} the name of this variable tries to sum up in a few words the complexity of that variable that relates formats between them with respect to the size, expressed in GB, that has been migrated globally. It is a list of four square matrices, one per object type, the elements of which are the number of GB migrated from one format to another multiplied by the number of cycles required to complete the migration. Each agent updates the proper element of this global variable after every migration. Under the assumption that each format can be migrated to any of the others, the number of the rows and columns equals the number of available formats of the corresponding type. We assume that the higher the number of GB migrated the more relevant the migration is, but also a migration that required an high number of cycles to be performed, even for a smaller size, needs to be considered significant. The latter assumption is due to the fact that institutions can refuse long migrations, as we shall see later, thus the choice to perform a long migration leads to the idea that it was an important action. 

\subsection*{Actions}
The only action available in this part of the model is the \emph{init} action. The variables described above and the graphical environment are set up when this action is executed. Most important, all the agent of the \emph{institution} species are created here. It is also possible to display some general information about the model such as the initial institutions, the available formats and the applications.

\section{Entities}
\label{entities}

In this part of the model are located the declarations of all the agents. This is the section in which there is room for a detailed description and analysis of all the types of agents included in the model. In GAMA the agents are grouped into species, each species can either contain other species (which thus are micro-species) or be included into a macro-species. Each species can be characterized by an inheritance relationship with another species. As mentioned in the beginning of Section \ref{gama}, each agent of any species may contain an arbitrary number of variables of various type but what really describes the agent are its actions and its reflexes. Those two properties characterize the agent's behaviour. In particular, an action is a sequence of statements that the agent will perform when asked to or under certain conditions. Within the declaration of each action other actions may be invoked. The reflexes could be defined as the reactions of the agent to a particular state of the environment. It is not unusual to find the invocation of an action inside the statement of a reflex. The communication among the agents is possible if the species own the skill of communicating. We now go on analysing all the species implemented.

\subsection{Institution}   
\label{institution}
The \emph{institution} species surely is the main species in the model. In order to keep a certain closeness to reality the following classification for the institution agents has been made:

\begin{itemize}
\item \textbf{Public institutions:}
they are characterized by having a large number of digital objects inside and they are keen on a specific type of digital object:
\begin{itemize}
 \item broadcasters: this kind of institution has a very large number of audio-visual digital objects, i.e. audio files, images or videos. One of these three types of objects will be more numerous than the others and will thus characterize the broadcaster's type. A small number of text objects is also contained which can represent the documentation \-re\-gar\-ding the audio-visual objects. 
 \item government institutions: in this work these institutions are seen as public entities having a large number of text objects such as documents, records and the like. There could be a small number of audio-visual objects related to text files.
 \item libraries: as for the government institutions, text objects have a dominant role in this kind of institution that will have a very large number of them. Libraries will also keep inside a large number of audio-visual contents so they certainly are the richest kind of public institution.
 \item universities: at the time this kind of institution is implemented similarly to the government institution.
 \end{itemize} 
\item \textbf{Personal institutions:} they aim to emulate everyone's personal archive, made up of all the digital objects types but in a smaller quantity with respect to public institutions. 

\end{itemize}


The institution species is necessarily endowed with the \emph{communicating} skill in order to exchange information with the other agents. 

\subsubsection{Variables}
Agents of this species have several simple variables inside. They are necessary to identify some features of the institutions, for example its number, the public type and the number of people dedicated to digital preservation (when the institution is public), the operating system  adopted, a list of the pastors and the like. We would like now to focus on two fundamental variables that can be considered the pivots of the entire model.

\paragraph{trust matrix:} as the name suggests, this matrix stores those values that are \-ne\-ces\-sa\-ry for the evaluation of the trust between two institutions. In this work, four parameters have been taken into account to calculate the level of trust: the number of digital objects (of every object type), the geographical distance between the institutions, the ``cultural" distance and the digital preservation staff. This matrix is thus made up of seven rows, one for each parameter, and a number of columns equal to the number of institutions. The matrix is filled when the \emph{set\_trust\_matrix} is invoked. When a new institution joins the model, each institution needs to update its matrix by adding a corresponding column to it. The  elements of this matrix are used by the \emph{trust\_evaluation} action as we shall see later. 

\paragraph{trust weights:} this variable is strictly related to the previous one. Each institution may associate a different level of importance to the trust parameters described above. It is necessary to take account of this by introducing another matrix with the same dimensions of the \emph{trust\_matrix}. In this new matrix each row contains the weights of the corresponding trust parameter. The elements may be updated after every interaction between two institutions. These values are used by the institution when the \emph{trust\_evaluation} action is invoked in order to calculate a weighted mean of the trust components to obtain the level of trust.

\subsubsection{Reflexes}
Institution are, along with pastors, the only agent species having reflexes (as we shall see later pastors only have one simple reflex). It means that these agents are capable of autonomous actions, that could be regarded as reactions to variations of the environment. Some simple reflexes allow the institutions to randomly create or eliminate both a single object or an whole format collection and also install or remove an application. All these reflexes aim to emulate the common actions performed by real archivers. As we did for the variables, in this section we analyse the most important reflexes of these agents.

\paragraph{deal with failures:} as we shall see in Section \ref{pastor}, each time a rendering failure occurs, the involved pastor notifies it to its host institution with a failure message. The institution loops over the incoming warnings and for each one performs the following actions: first of all it evaluates the level of obsolescence risk of the format by means of the \emph{format risk} action described later. If the risk level does not exceed a certain threshold, the format could not considered obsolescent so a local action, for instance the installation of a new application, may be taken into account. Otherwise, the format is regarded as obsolescent thus a migration process may be required. In this case the institution sends a request message to all the other institutions with the aim to receive any suggestion about how to deal with the problem. The content of this request includes a unique tag that allows to identify the issue. 

\begin{figure}[h!]
\centering
\includegraphics[width=13cm]{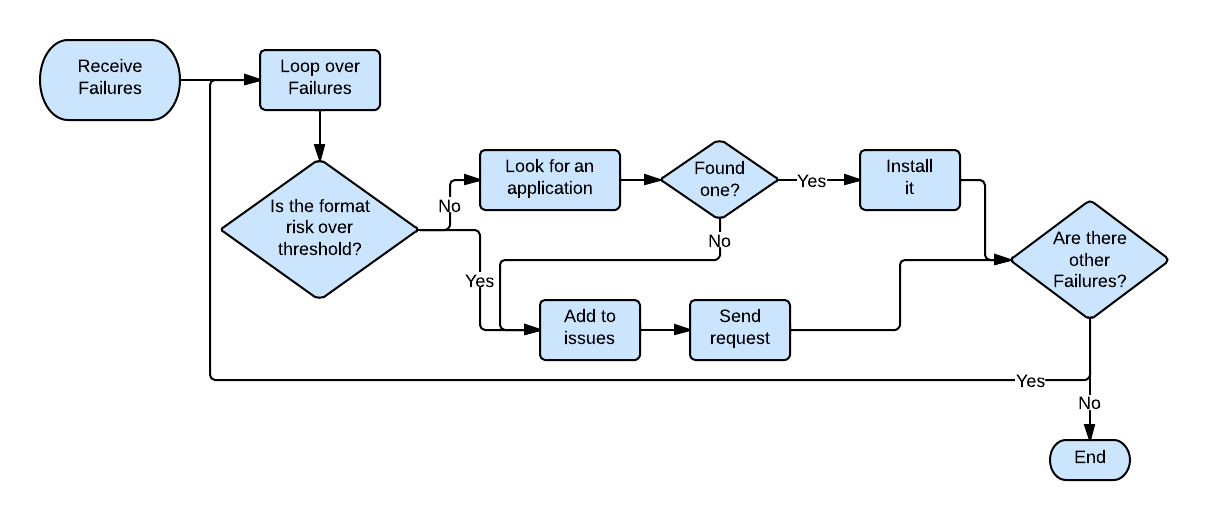}
\caption{Schematic of reflex: deal with failures.}
\label{deal_with_failures}
\end{figure}

\paragraph{suggest solution:} each institution may receive a request from another agent in need, asking for a piece of advice about how to deal with a certain issue. Since, as described in the previous paragraph, each institution sends requests to all the the others the receivers have to deal with a large number of requests. Thus what the agent does first is to loop over the received requests. For each one it looks into its own list of performed migrations and copies the latest performed migration \-star\-ting from the given format into a new list. After that, it sends a propose message to the agent who asked for the suggestion. If no suggestion can be found it sends a propose message with 0 as content.

\begin{figure}[h!]
\centering
\includegraphics[width=13cm]{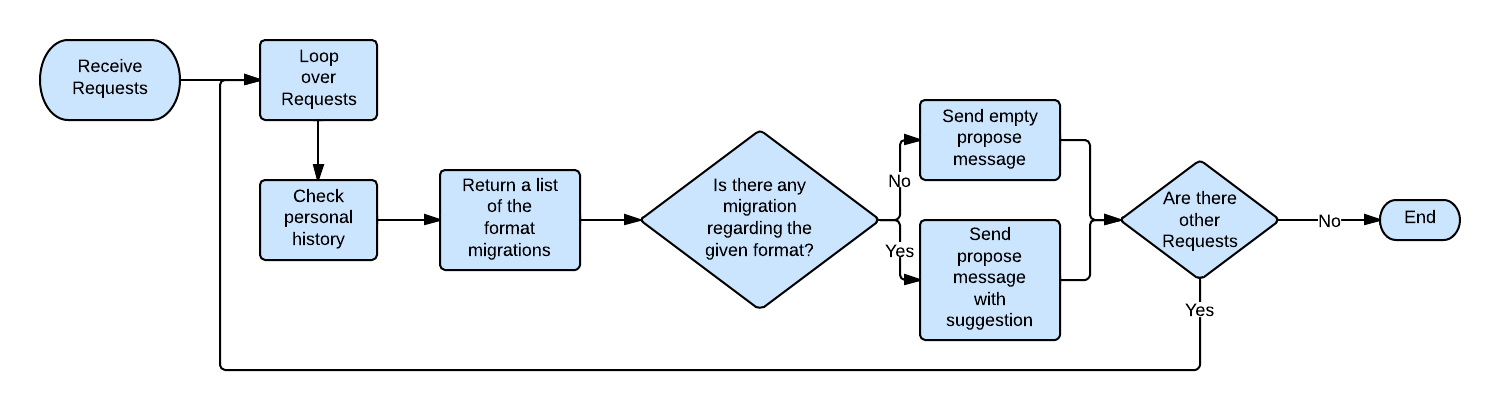}
\caption{Schematic of reflex: suggest solution.}
\label{suggest_solution}
\end{figure}


\paragraph{read suggestions:} the next step is to analyse the received suggestions and choose the most suitable one. When the institution receives any propose message it instantiates a map in order to match the incoming suggestions with the proper issue by means of the unique tag. Once the map is filled, the agent checks the number of pairs to determine whether he received any suggestions or not. Let now discuss what happens in this two cases.

\begin{figure}[h!]
\centering
\includegraphics[width=13cm]{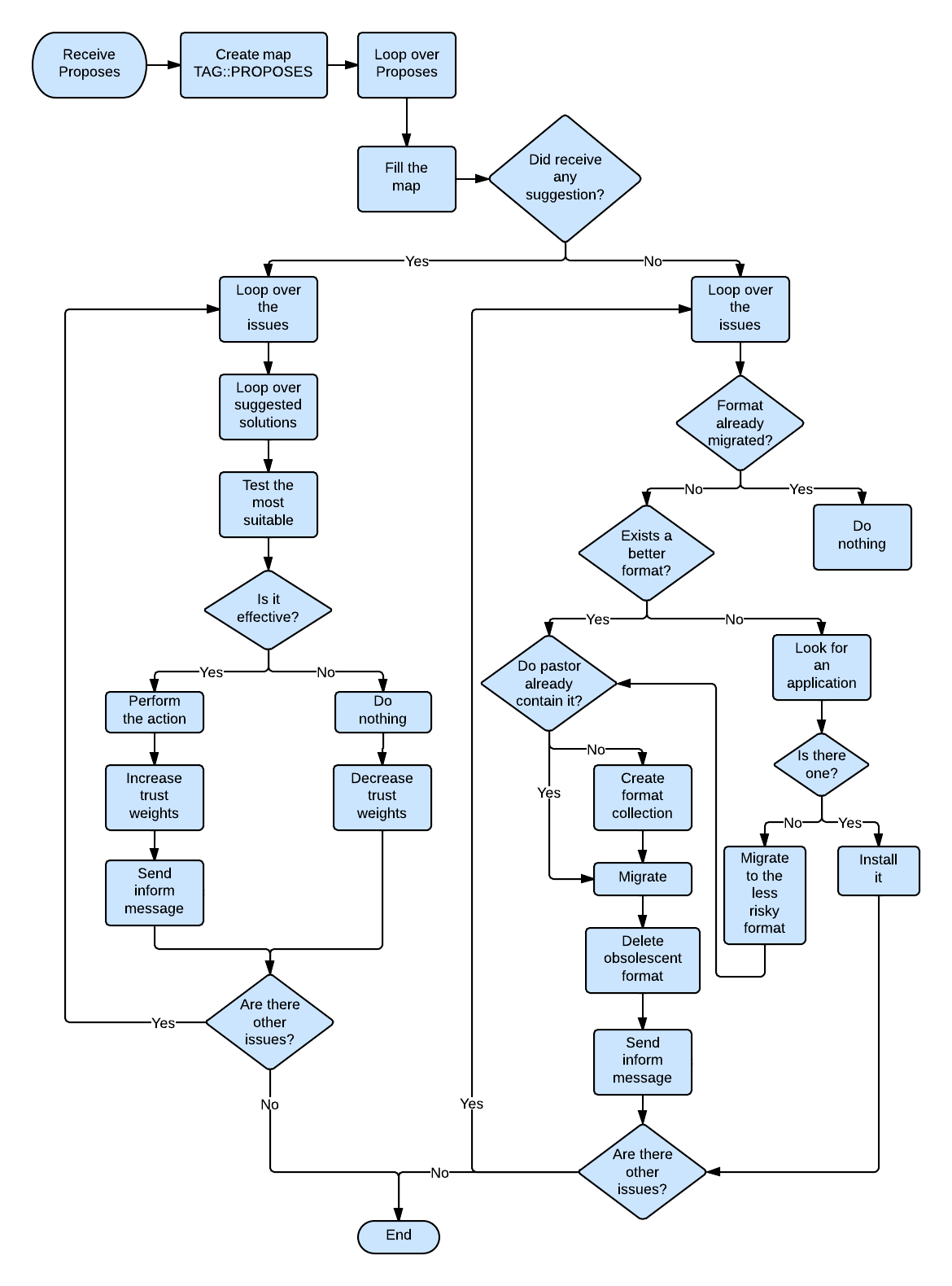}
\caption{Schematic of reflex: read suggestions.}
\label{read_suggestions}
\end{figure}

If the map is empty, of course no suggestions have been sent and so the institution has to deal with its issue without any external help. The agent loops over the issues, and for each one, it checks if the format reported in the given issue has already been migrated or not in the meantime. If so it does nothing, otherwise it asks the software manager to return all the formats of the given type that can be rendered at current state. For each of these formats it evaluates the obsolescence risk level through the \emph{format risk} action and chooses the less at risk one. If the risk level of the chosen format exceeds the one of the format that will be migrated, the agent looks for an application capable of rendering the latter. If it founds at least one, the agent installs it and goes to the next issue, otherwise it decides to migrate to the less at risk format even though its risk is high. In case the risk level of the chosen format does not exceed the one of the format that will be migrated, the agent proceeds with the migration. 


Before performing the migration, the institution checks if the destination format is already contained by the pastor and, if not, it asks it to create a new format collection with the proper name. Once the migration process took place, the institution asks the pastor to remove the obsolescent format collection from its members and the just finished migration is added to the list of the migrations performed by the institution.

In case the map is not empty, the agent analyses all the received suggestions for each issue. It chooses the most suitable one with respect to its current internal state and tests it in order to determine whether it is effective or not.  If so, the agent performs the preservation action and increases the weights of the institution that suggested it. Otherwise nothing is done and the weights are decreased. The schematic in Figure \ref{read_suggestions} depicts the entire reasoning process.

It is important to underline that in both cases, the agent estimates the time required to perform the migration taking into account its computational resources. The main idea is to perform the migration only on the smallest object to be migrated. In this way the agent can evaluate how much time per kB is required and thus which is the time required for the whole action to take place by multiplying for the total size to migrate. Since no real migrations are performed in this model we decided to emulate this process generating a random number that is related to the computational resources (expressed in \emph{HS06} \cite{hs06}) and to the size of the smallest object. The result is then multiplied to the total size of the object to have an idea of the time required for the migration.

\paragraph{listen to informs: } this last reflex is similar to the previous one. While in the previous case the starting point was a request of help that made other institutions suggest something, here the agents autonomously inform all the others about a migration that they decided to perform.

\begin{figure}[h!]
\centering
\includegraphics[width=13cm]{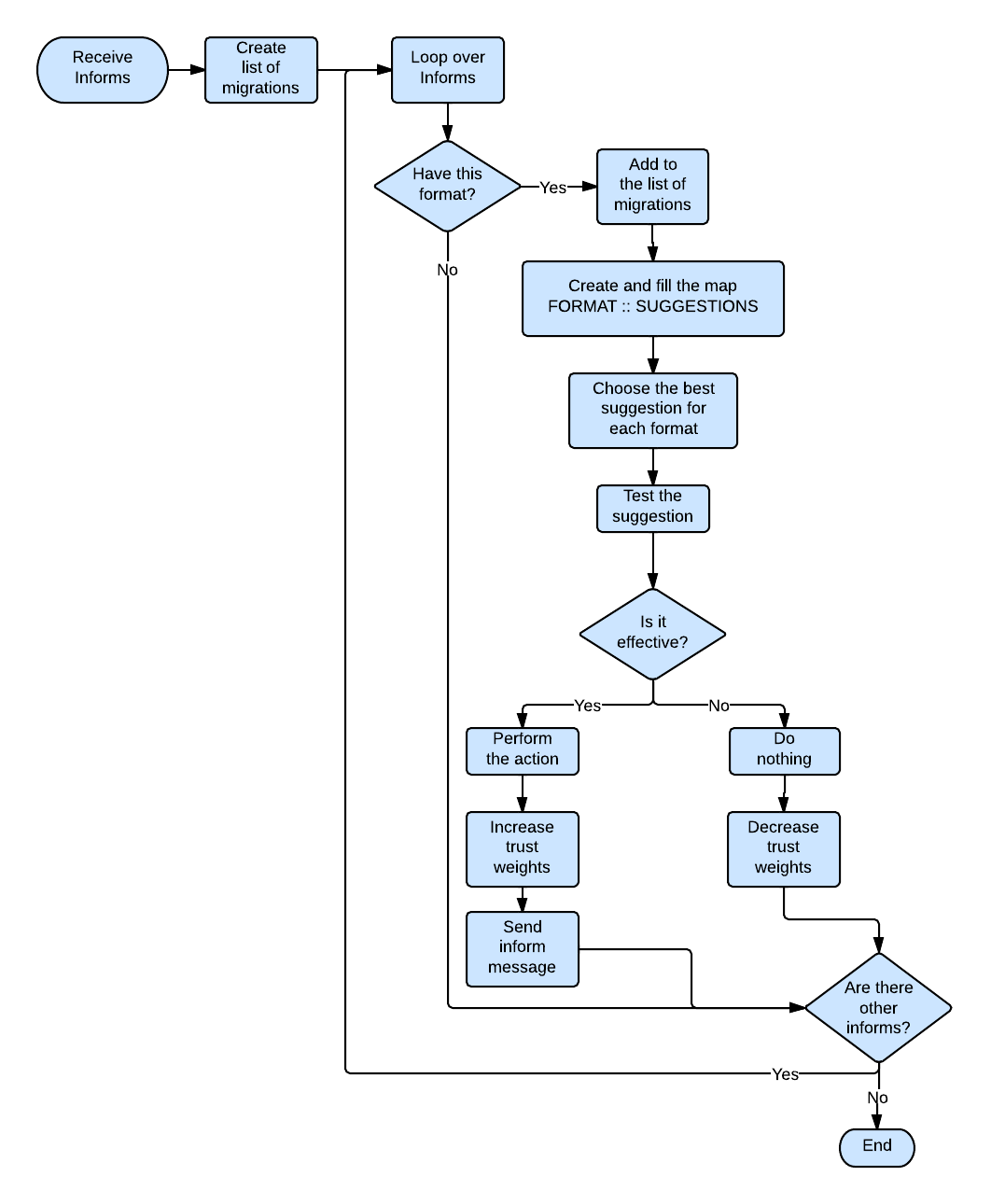}
\caption{Schematic of reflex: listen to informs.}
\label{listen_to_informs}
\end{figure}

\noindent Thus in this case the receiver institution has to loop over a large number of \emph{inform} messages sent by other agents and has the chance to decide if it may be useful to perform one of the suggested actions.

For each object type a list of suggested migrations is created. The agent loops over the informs and checks if the starting format of the suggested migration is also contained by its corresponding pastor. If so this action may be useful to the agent and thus it is added to the list. When the loop is over, the institution creates and fills a map that associates each format with the related suggestions. For each format the most suitable action is chosen and 
tested in case the trust level between the agents is high enough. If the destination format can be rendered by the current software environment and the required time is acceptable the migration is then performed, otherwise nothing is done. The weights of the trust level are consequently updated. The whole process is shown in Figure \ref{listen_to_informs}.

\subsubsection{Actions}
Institutions are endowed with several actions that are fundamental to ensure the communication and interaction mechanism that we aim to implement. Some of the key actions are analysed in the following lines. 

\paragraph{format risk:} institutions take advantage of this action to determine how high is the risk for a digital object to become obsolescent. As discussed in Section \ref{global}, the risk is estimated through global variables that tell how diffused and supported by applications a format is. 

We decided to estimate the global level of obsolescence of a format by means of three components: the first one is the total number of objects of that format that are contained by all the institutions at the time, this indicates how common the format is. The second component is the number of institutions containing at least one object of that format and the third one is the number of installed applications capable of rendering the object. The risk percentage of each component is calculated as follows: 

\begin{equation}
r_i = (\frac{100}{l_i -1}) \cdot [(l_i -1)-p]
\end{equation}

Where $l_i$ is the length of the lists containing respectively the total numbers of objects, the numbers of institutions and the numbers of installed application for each format. The elements of these lists are sorted in ascending order so that the position \emph{p} of each element indicates its degree of obsolescence with respect to the component associated with the list. The risk of obsolescence is, by choice, calculated as follows:

\begin{equation}
R = \frac{1}{N} \sum_{i} r_i
\end{equation}
  
Where $N$ is the number of risk components. The resulting risk percentage is returned so that the institution is allowed to decide how to deal with this format.

\paragraph{destination format:} this action is performed every time the institution decides to migrate some digital objects from a format to another without any external advice. In this case, it is necessary to determine which is the least obsolescent format that can be rendered with the current software environment. To do so, the institution asks the software manager to return a list of formats that it is capable of rendering. The risk level of each of these formats is calculated, and the values are collected into a map in order to be matched with the corresponding format. In the end the pair with the lowest risk level is returned.

\begin{figure}[!h]
\centering
\includegraphics[width=13cm]{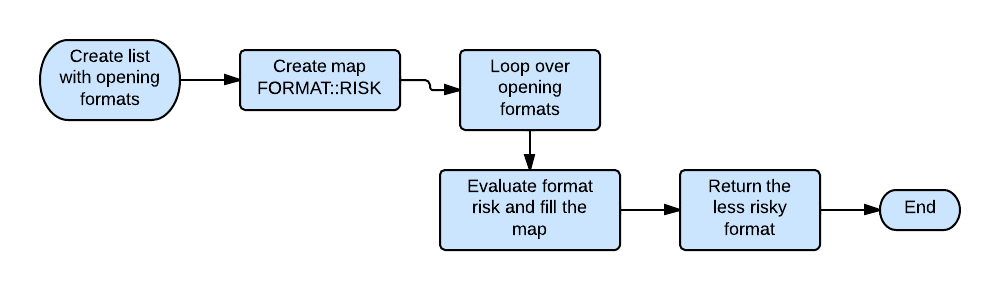}
\caption{Schematic of action: destination format.}
\label{choose_dest_format}
\end{figure}

\paragraph{trust evaluation:} this is no doubt one of the key actions that have been implemented in this model. By means of this action each institution evaluates the level of trust associated with another one with respect to a given type of digital object. This action accepts two integer parameters as arguments: the number of the other institution and an \emph{index} that identifies the type of the object. It returns a floating point number that represents the level of trust. We decided to define the level of trust as made up of four components, one for each parameter taken from the \emph{trust\_matrix}.

The first component ($T^{f}$) is related to the number of files of the given type that the two institutions contain. The main idea is the following. Consider two institution \emph{inst\_i} and \emph{inst\_j}. Suppose we want to evaluate this component of the level of trust between \emph{inst\_i} and \emph{inst\_j}: if \emph{inst\_i} has got less audio files than \emph{inst\_j}, then \emph{inst\_i} should trust \emph{inst\_j} as it should be more experienced in managing this kind of digital objects. The following relation has hence been assumed:

\begin{equation}
T^{f}_{ij} = \frac{1}{2} \cdot (1 + \frac{f_j - f_i}{f_j + f_i})
\end{equation}

Where $f_j$ and$ f_i$ are respectively the number of files of the given index that are owned by \emph{inst\_j} and \emph{inst\_i}. The possible values of this component are limited \-be\-tween 0 and 1.
\\
The second component ($T^{d}$) regards the geographical distance between the institutions. In contrast with the previous component this is obviously symmetric. We assumed that the bigger is the distance between two institutions the less the choices of one should influence the choices of the other and so the less one should trust the other. This component can thus be expressed through the following formula:

\begin{equation}
T^{d}_{ij} = T^{d}_{ji} = 1 - \frac{dist_{ij}}{dist_{MAX}}
\end{equation}

Where $ dist_{ij} $ is the distance between \emph{inst\_i} and \emph{inst\_j}, $ dist_{MAX} $ is the distance between the most distant institutions in the model. It is easy to see how this function is limited between 0 (when the two institutions are the most distant) and 1 (when the institutions are in the same place).
\\
The next component ($T^{c}$) of the trust level involves the ``cultural distance" \-be\-tween the institutions. If we consider the case of text files we immediately notice how important this parameter is. If \emph{inst\_i} needs to decide whether migrating to another format or not, it would trust more those institutions that manage text files that are, for example, written in the same alphabet. The value of this component is symmetric and bounded between 0 and 1 too and it is retrieved from the global matrix variable named \emph{lang\_correlation}.

\begin{equation}
T^{c}_{ij} = T^{c}_{ji} = L_{ij}
\end{equation}

Where $L_{ij}$ is the is the component  $\lbrace  i,j  \rbrace$  of the matrix \emph{lang\_correlation}, here referred as $L$.
\\
The last component ($T^{s}$) taken into account regards each institution's staff \-de\-di\-ca\-ted to the digital preservation. Of course only public institutions will have a certain number of people working in the digital preservation field while privates have none. If we consider again the encounter between \emph{inst\_i} and \emph{inst\_j} some distinctions have to be made: in the simplest case \emph{inst\_j} is private and so this component will be 0 no matter if \emph{inst\_i} is public or private too. On the other hand, if \emph{inst\_j} is public it may have its own staff and we have to distinguish how \emph{inst\_i} should trust it in case \emph{inst\_i} is public or private. This component, which is not symmetric, is calculated as stated in the following expression:

\begin{equation}
T^{s}_{ij} = \begin{cases}
\frac{1}{2} \cdot (1 + \frac{S_j - S_i}{S_j + S_i}) & inst_i\, public \\
\frac{S_j}{S_{MAX(j)}}  & inst_i\, private
\end{cases}
\end{equation}

Where $S_i$ and $S_j$ are the numbers of people dedicated to digital preservation of the two institutions and $S_{MAX(j)}$ is the maximum number of people that an institution of the same type of \emph{inst\_j} can involve in its staff. As happens for the other components, the values of are limited between 0 and 1.
\\
Now that every component has been calculated it is possible to obtain the trust value by a weighted mean of them:
\begin{equation}
T_{ij} = \frac{1}{A} \cdot \sum_{a} w_{a}T^{a}_{ij}
\end{equation} 

In this expression $A$ is the number of trust components which equals 4 in this particular implementation. In addiction, $T^{a}_{ij}$ is each component and $w_{a}$ is the corresponding weight taken from the matrix variable \emph{trust\_weights}. The result of this weighted sum is the floating point number that is returned by this action and that expresses the level of trust between the two institutions taken into account.

\paragraph{analyse migration:} this action is executed after every migration, or refused migration, in order to establish if the decision taken by the agent was the correct one. We would like our agent to perform those migrations that are necessary, but also to avoid those long migrations that may cause the impossibility to perform other important migrations. In order to determine whether the decision to perform or not a migration can be classified as correct, we compare it to what is happening at the global level. Here we see again how a local preservation action is related to global and distributed parameters. The following examples may help to better understand the idea.

Let us suppose that an institution decided to migrate from format $f_i$ to $f_j$. We want to determine if this migration involves a significant amount data at the global level, if so the agent's choice was correct: it did a \emph{good action}. If in contrast, a small amount of data is concerned with this migration and, for instance, the migration from $f_i$ to $f_h$ is prevalent then the agent's choice to migrate was wrong. The institution performed a migration while it was not necessary so we define this event as a \emph{false positive}. 

The institution may also decide to refuse a suggested migration because too long or just because the level of trust associated with the agent who made the suggestion is not high enough. In this case, the institution did a \emph{good action} if the suggested migration was not significant at a global level. The worst case is when the institution discarded a migration that should have been performed because globally relevant. If so, the agent missed the opportunity to perform an effective action and that wrong decision is named \emph{false negative}.

Let us now better define a globally relevant migration and the classification of the actions performed. Starting from the former, the evaluation is made through the values stored in the \emph{format\_migrated\_sizes} matrix described in Section \ref{global}. We estimate the global relevance of the migration with respect to all the other possible migrations beginning from the given initial format as follows:

\begin{equation}
relevance_{ij} (\%) = \frac{100 \cdot A_{ij}}{\sum_k A_{ik}} 
\end{equation} 

Where $A$ is the \emph{format\_migrated\_sizes} matrix, the matrix element $A_{ij}$ indicates that the performed migration was from format $f_i$ to $f_j$ and the sum $\sum_k A_{ik}$ is the sum of all the elements in the row of the starting format $f_i$. We take advantage of the value obtained from this expression to classify the agents' actions as explained in Table \ref{actions_class}.

\begin{table}[!h]	
	\begin{center}
    \begin{tabular}{ |c|c|c|}
    \hline
    Migrated  & Relevance (\%) & Classification\\   
    \hline
	\hline	
	yes & $r_{ij} < 10$ & false positive \\
	\hline
	yes / no & $10 <  r_{ij} < 50$ & indifferent\\   
    \hline
	yes & $r_{ij} > 50$ & good action\\
	\hline      
   	no & $r_{ij} < 10$ & good action \\
	\hline
	no & $r_{ij} > 50$ & false negative\\
	\hline
    \end{tabular}
	\end{center}	
	\caption{Classification of agents' decisions.}
	\label{actions_class}
	\end{table}

We would like to underline that the previous classification values were adopted as a test assumption. The user of our modelling framework can change them at any time in order to match the scenario he wants to simulate.

\paragraph{accept time:} in order to be as close as possible to reality we associated a certain number of cycles to each migration process. In this way the agent which is performing a migration is unable to undertake any other action before it completes the current one. It is possible for the user to define the limit of acceptable required cycles for each of the objects type. We would like to underline the importance of this feature: if an agent decides to perform a long preservation action it may waste the chance to perform any other, maybe more useful, process. In this action the number of cycles required to perform the given migration is compared to the limit of acceptability defined by the user, if the first one exceeds the limit a boolean set to false is returned and the migration will be refused.
	
\paragraph{migration time:} as we just described, we want each migration to last a certain number of cycles. We assume that the duration depends on the amount of bytes to be migrated (total size), the initial format, the destination format and, finally, the computational resources of the institution. We decided to define this relationship as follows:

\begin{equation}
time = \frac{size \cdot M_{ab}}{res} 
\end{equation} 

Where $size$ indicates the total size of the objects to be migrated expressed in bytes, and $M_{ab}$ represents the migration speed expressed in bytes per seconds, this value is read from the corresponding \emph{mig. time coef.} global matrix (it depends on both the starting and the destination format and it is not symmetric). With $res$ we indicate the amount of computational resources of the institution, of course the bigger $res$, the smaller $time$.
The result returned by this function is translated into a number of cycles that the institution has to wait before performing other migrations. Due to the structure of the model, each institution is capable of performing one migration per object type at the same time.

\paragraph{migrate:} by means of this action the institution emulates a real migration process. The action takes as arguments an index that identifies the object type involved, the starting format and the destination format and the required number of cycles which was returned by the \emph{migration time} action. First of all the agent checks that the starting format is still available, after that it determines whether the corresponding pastor contains or not the destination format among its members. If the format is not contained, the pastor is asked to create a new format collection named as the destination format. Now the institution asks the pastor to execute the \emph{convert} action and then to eliminate the format collection corresponding to the starting format. After that the institution updates all the necessary global variables, for instance the number of global migrations or the \emph{format migrated sizes} matrices. Finally the agent informs all the others about the undertaken migration and becomes idle for the number of cycles required to perform the migration.

\begin{figure}[!h]
\centering
\includegraphics[width=13cm]{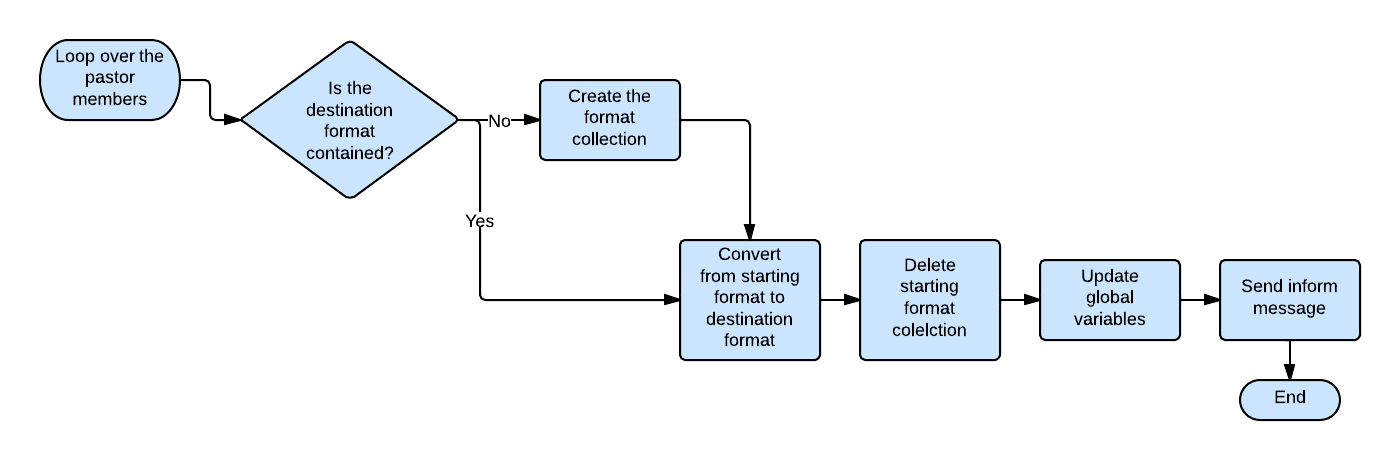}
\caption{Schematic of action: migrate.}
\label{migrate}
\end{figure}

\subsubsection{Nested structure}
In Section \ref{gama} we anticipated that GAMA allows the user to develop agents that contain other agents inside. When a species of agents is defined to keep inside another species is named \emph{macro species} while the ``contained" one is the \emph{micro species}. In this case the \emph{institution} species plays the role of the macro species. The micro species are: 

\begin{itemize}
\item aud\_pastor
\item img\_pastor
\item txt\_pastor
\item vid\_pastor
\item software\_manager
\item format\_collection
\end{itemize}

The first four of this list are children of the \emph{pastor} species, they inherit all the actions and reflexes from their parent species and add other actions. The main tasks for this agents are related to the management of the digital objects of their host institution. They can do so through several actions that will be further described in Section \ref{pastor}. The peculiarity of this species is that they are themselves macro species since they contain agents which are children of the \emph{format\_collection} species. 

By a general point of view it is possible to assert that the \emph{software\_manager} micro species is very similar to the just discussed \emph{pastor} agents. The role of this species is to manage the applications needed to perform the rendering, and maybe the editing, of the digital objects. When the initialization takes place four lists (one per type of digital object) are filled with an arbitrary number of applications, taken from the global lists \emph{available\_software[os]}, that can be used to render objects of the given type. These applications may be considered as installed in the software environment of the institution. In order to simulate what happens for real, some actions have been implemented that allow the software manager to: install or remove an application, open a given format with one of the installed applications (when this is possible), return a list of the formats that can be opened with the applications that are currently installed, display information about the state of the software environment. As happens for the pastors, the software manager cannot perform any of these actions without the consensus of its host institution.

The last micro species of an institution is named \emph{format\_collection}. These agents \-re\-pre\-sent groups of digital objects having the same format. We decided to gather digital objects of the same format rather than implementing each digital object as an agent because this implementation would have required a lot of computational resources and it would have been too hard to run the model. Each format collection is named as the format that it represents, it keeps inside several variables such as: an integer that indicates the number of files of that format, a list of integers that describes how these files are distributed (there could be clusters of files that should be regarded as an unique file), a list that contains the size of each single file or cluster expressed in kB, an integer that indicates the total size and another integer that is updated every time that an object of this format has been rendered. The agent can perform actions to fill the lists described in the previous lines and to display information about its content.

\subsection{Pastor}
\label{pastor}

The agents of species \emph{pastor} are those that are assumed to manage the digital objects of their institution. As described in Section \ref{institution}, each institution has four pastors as members which inherit from the \emph{pastor} species. Each pastor makes use of the \emph{communicating} skill to communicate with its institution and with other pastors when necessary. This agents constantly monitor their collections of digital objects, warn their institution when problems occur (for example a failure in a rendering process) and perform a proper action in agreement with the institution. It is important to underline that even though the collections of digital object are members of the pastors, these agents are unable to perform any action on them without the permission of the institution. The reason for this choice is that the decision about which preservation action perform is always taken by the institution after a reasoning process, where several parameter are taken into account.  

\subsubsection{Variables}
According to our model implementation this species only has got a couple of simple variables. The reason for this choice essentially is that pastors are not strictly autonomous agents since they act when asked to by their institution, so they do not need access many information. These variables are a list of strings containing the names of the included format collections, and an integer that identifies which type of digital objects the pastor do manage. 

\subsubsection{Reflexes}
As anticipated in the previous lines, pastors are not entirely autonomous. This is confirmed by our decision to endow these agents with only one, but fundamental reflex. In fact, this reflex triggers the entire process that leads to migrations. Each pastor periodically loops over its format collections, chooses the one that has not been controlled by longer and tries to render it. After that, it updates the \emph{last checked} variable of this format collection and, if the rendering process fails, it sends a failure message to the institution. 

\subsubsection{Actions}
We implemented some actions that we retain needed for the pastors to manage their format collections. For instance they can create or eliminate both single objects and entire format collections, display all the contained format collections or delete them all.  By the way there are two essential actions that form the core of the model.

\paragraph{check rendering:} this simple but important action is invoked every time that the discussed reflex is performed. The agent asks the software manager to return a list of applications that are able to render the given format. When this list contains only one application, then the current software environment in under risk to become unable to render the format, thus a preservation action should be performed. We decided to let the agent alert the institution when there is one single application and not, for instance, none because without at least one application it would be impossible to perform any migration.

\paragraph{convert:} this action emulates a real migration process by moving a given number of objects from a format to another one. First of all the agent verifies that the destination format is accessible to the institution (it must not be forbidden by legal constraints). If so it proceeds gathering the objects from the initial format collection into a temporary list, then it asks the format collection to remove them and updates the global variables \emph{file number for format} and \emph{institution number for format} described in Section \ref{global}. After that the pastor asks the destination format collection to add the elements, stored into the temporary list, to its internal lists and then updates again the global variables and the migration process terminates.

\begin{figure}[!h]
\centering
\includegraphics[width=13cm]{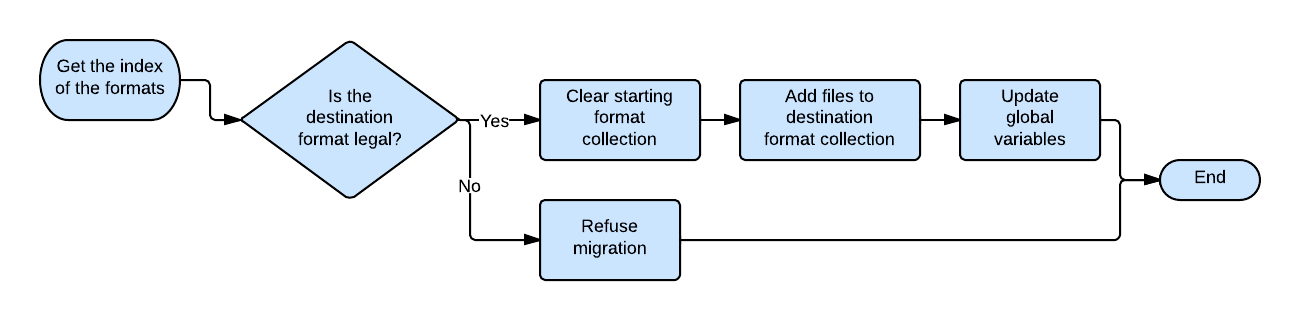}
\caption{Schematic of action: convert.}
\label{convert}
\end{figure}

\subsection{Format Collection}

In this implementation, the \emph{format collection} species is the simplest kind of agent. It represents a group of digital objects of the same format. These agents are characterized by several simple variables such as the name of the format, the number of files, the number of clusters of files and the total size of the collection expressed in kB. There are two basic actions to set up the number of both clusters and single files and to set up their corresponding size. 

At the beginning we decided to implement each digital object as a single agent which seemed the most realistic solution. After a couple of model runs we had to discard this choice because a number of objects greater than a hundred made the simulation very slow and difficult to be performed on a laptop. In contrast, gathering all the objects of the same format as one agent, it is possible to manage a huge number of objects and an acceptable number of formats.

\section{Experiment}

The last part of the model is for all those statements that are related to the experiments. Here the user can set up a real test bed to verify the  behaviour of the model. By default only the windows containing the parameters and the console are shown, but GAMA \cite{gama} allows to display all the required information also into self-updating pie charts, histograms and scatter-plots. Another useful feature is the chance to display monitors in order to observe the behaviour of a \-par\-ti\-cu\-lar agent species or, as happens for this work, the evolution of global variables such as the number of migrations performed by the institutions and the global risk parameters. 

When the model runs all these statements are executed and the declared displays are shown into the proper windows. In the following Figure \ref{exp_exmaple} we can see an example of the display of both a pie chart and three scatter-plots:  

\begin{figure}[h!]
\centering
\includegraphics[width=13cm]{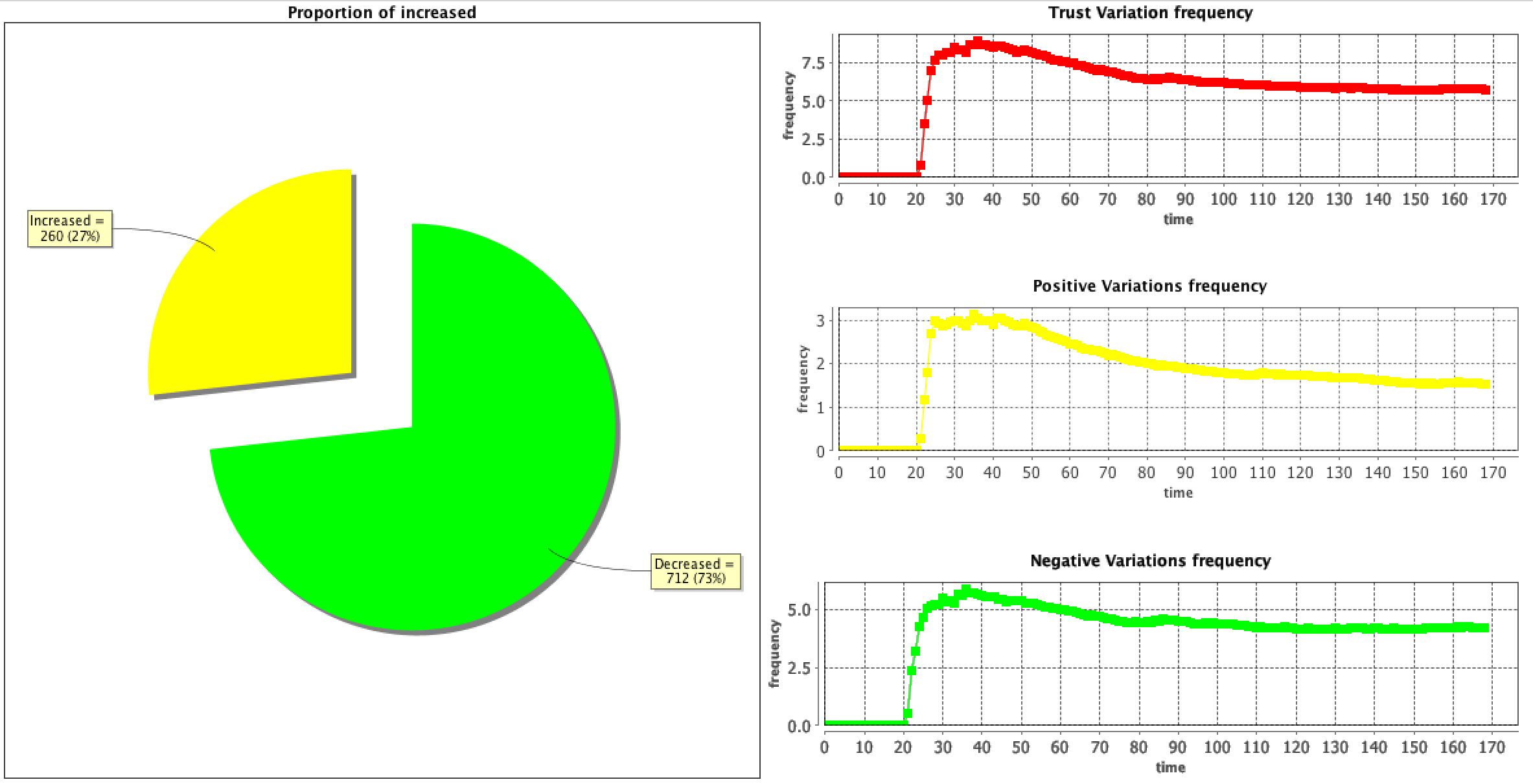}
\caption{Example of an experiment graphics.}
\label{exp_exmaple}
\end{figure}

\begin{figure}[h!]
\centering
\includegraphics[width=13cm]{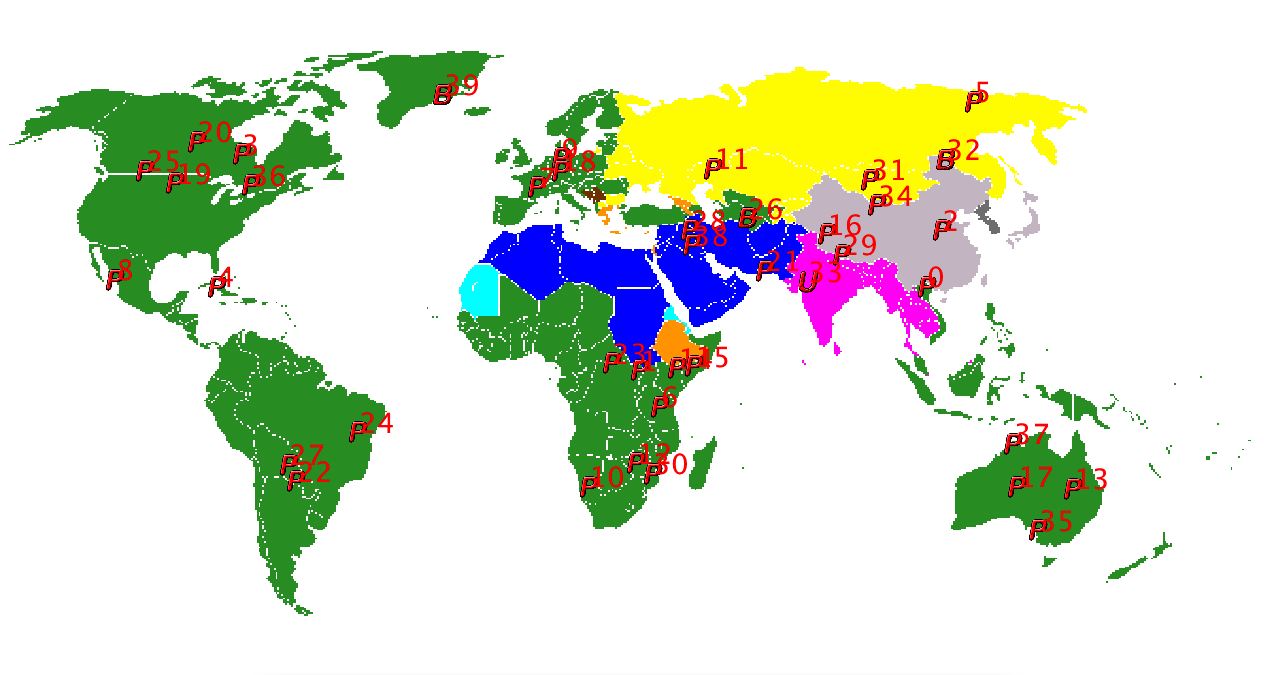}
\caption{Environment and agents within.}
\label{world_map}
\end{figure}

In addiction, it is possible to display the environment into which the agents are located. As we explained in Section \ref{entities}, in this work the agents play the role of archiver institutions distributed all over the world thus the environment may appear as depicted in Figure \ref{world_map}.
As we can see, different colours have been used in order to identify countries where different alphabets are adopted (we discussed the role of alphabets both in the beginning of this chapter and in Section \ref{entities} describing how the cultural component of the trust is evaluated).

The institution agents are identified by red icons made up of a capital letter that indicate the type of the institution (``B" for broadcasters, ``G" for government institutions, ``L" for libraries, ``P" for personal archives and ``U" for universities) and a number corresponding to the agent's number.

The agent's location is set randomly when the institution is created and, obviously, its position is used to evaluate the geographical distance from others expressed in arbitrary units.

In the end, we used the console to display information about the model that is running. For instance we decided to show the number of institutions taking part in the model, the list of available software, the list of available formats and other parameters. In the following chapter we describe further how the experiment part of the model has been used to measure variables of our interest.

\chapter{Testing the Framework}
\label{using_the_model}
Now that the model has been designed and implemented we need to verify the \-con\-sist\-en\-cy of its structure and the stability of its behaviour. The first step is, of course, to run the model and check that no errors occur.

The console played an essential role in this step because we could check the agents' behaviour by asking them to write there what they were doing and the results of their actions. Moreover, by means of the console, we monitored the communication between institution agents to ensure the correct flow of information.

\section{Stability of the Communication Process}
\label{stability}
Once that the implementation has been proved to be correct, we can proceed with the design and implementation of the experimental test bed. Since the communication \-proc\-ess can be considered as the backbone of the model, we want to verify its stability. To do so we made the assumption that each migration process lasts only one time step. In this way we aim to avoid possible instability due to the fact that, while an institution is performing a long migration process, other required migrations cannot be executed. 

We run several experiments modifying some important parameters such as the number of institutions, the threshold that determines whether a format is to be considered under risk of obsolescence and the thresholds over which a suggestion or an inform message is accepted of refused. The values adopted for the experiments are reported in Table \ref{tab1} :

	\begin{table}[h]	
	\begin{center}
    \begin{tabular}{ |c|c|c|c|c|c|}
    \hline
    Parameter  & \multicolumn{5}{|c|}{Values} \\   
    \hline
	\hline	
	Institution nb. & 3 & 5 & 30 & 50 & 100 \\
	\hline
	Risk th. (\%) & 10 & 30 & 50 & 70 & 90 \\    
    \hline
	Suggest th. (\%) & 10 & 30 & 50 & 70 & 90 \\ 
	\hline
	 Inform th. (\%) & 10 & 30 & 50 & 70 & 90 \\   
	\hline      
    \end{tabular}
	\end{center}	
	\caption{Values adopted for experiments on stability.}
	\label{tab1}
	\end{table}

\begin{figure}[h!]
\centering
\includegraphics[width=10cm]{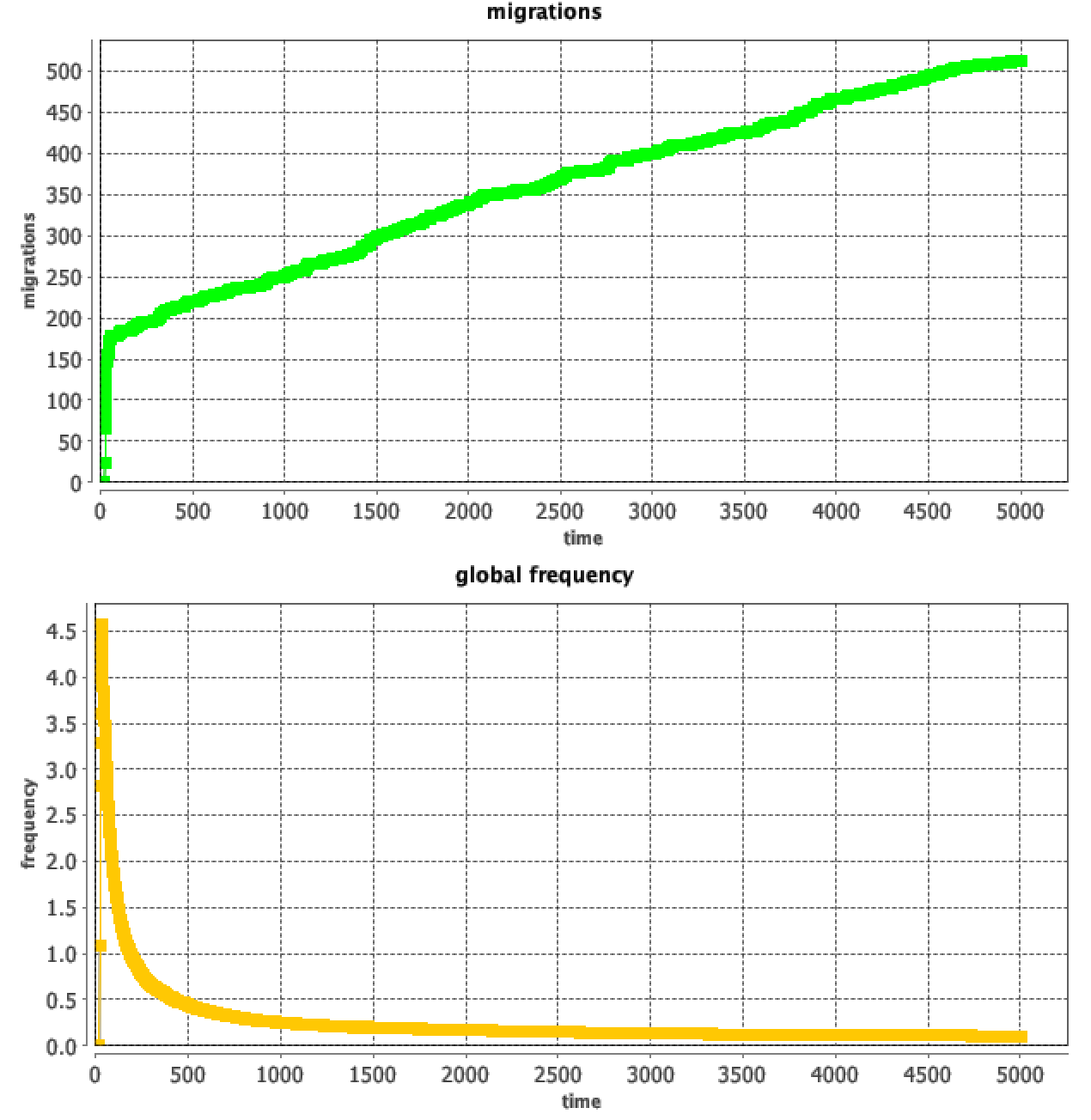}
\caption{Example of scatter-plots of both migrations and frequency of migrations.}
\label{mig&freq}
\end{figure}

For each of these values we run a 5000 cycles simulation and focus on the number of migrations performed by all the institutions and the frequency of these migrations. In the followings we show the results of these experiments. As \-an\-ti\-ci\-pa\-ted in the Section \ref{exp_exmaple} of the previous chapter, GAMA \cite{gama} allows to monitor and draw scatter-plots concerning the variables we want to observe. In this case we are interested in the number of migrations performed and their frequency so we included code into our experiment to show plots such as the one reported in Figure \ref{mig&freq}.

We will address further statistical analyses in the next chapter, where we will focus on the trend of the migrations frequency.

\section{Linear Response Range}
\label{lin_noTime}
In the previous section we saw that the frequency of migrations approaches zero as time goes bye. Actually this is what might appear looking at Figure \ref{mig&freq} but, as we will explain in Section \ref{stab_noTime} it is not possible to affirm that the frequency approaches 0. We may expect the institutions to stop performing migrations because, if the model works fine, all the issues should be solved after a certain number of cycles. This is true in a static environment while, into reality, institutions may decide to cease adopting a format or start adopting another one without an apparent need. In order to better understand this idea we can think about the simple case in which we receive an email attachment in a format that our software environment cannot render. 

The model emulates this situation giving agents the chance to create or delete a format collection at each time step with a certain probability. In this way institutions may encounter new issues to be solved and thus new migrations to be performed. We now want to verify that, without any migration time constraint, the asymptotic value of the frequency of migrations is linear to the probability of creating or deleting a format collection at each cycle. 

We decided to adopt the combination of parameters reported in Table \ref{tab_lin_noTime}

\begin{table}[h]	
	\begin{center}
    \begin{tabular}{ |c|c|c|c|c|c|c|}
    \hline
    Parameter  & \multicolumn{6}{|c|}{Values} \\   
    \hline
	\hline	
	Institution nb. & \multicolumn{6}{|c|}{50} \\
	\hline
	Risk th. (\%) & \multicolumn{6}{|c|}{50} \\   
    \hline
	Suggest th. (\%) & \multicolumn{6}{|c|}{30} \\
	\hline
	 Inform th. (\%) & \multicolumn{6}{|c|}{70} \\
	\hline      
   Probability (\%) & 1 & 2 & 3 & 5 & 7 & 50 \\
	\hline
    \end{tabular}
	\end{center}	
	\caption{Values adopted for experiments on linearity.}
	\label{tab_lin_noTime}
\end{table}

Each simulation lasted 5000 cycles in order to give the model enough time to be into a stable condition. In Section \ref{lin_resp_noTime} we discuss and analyse the results of these simulations.

\section{Feedback Mechanism Stability}
\label{feedback}
Another variable we would like to analyse is the frequency of trust weights variations. Our objective is to determine whether the number of variations of the trust weights matrices approaches a constant value with increasing time. We remember that, as explained in Section \ref{institution}, each institution contains a variable named \emph{trust\_weights} which is updated after every interaction with another agent.

\begin{figure}[!h]
\centering
\includegraphics[width=9cm]{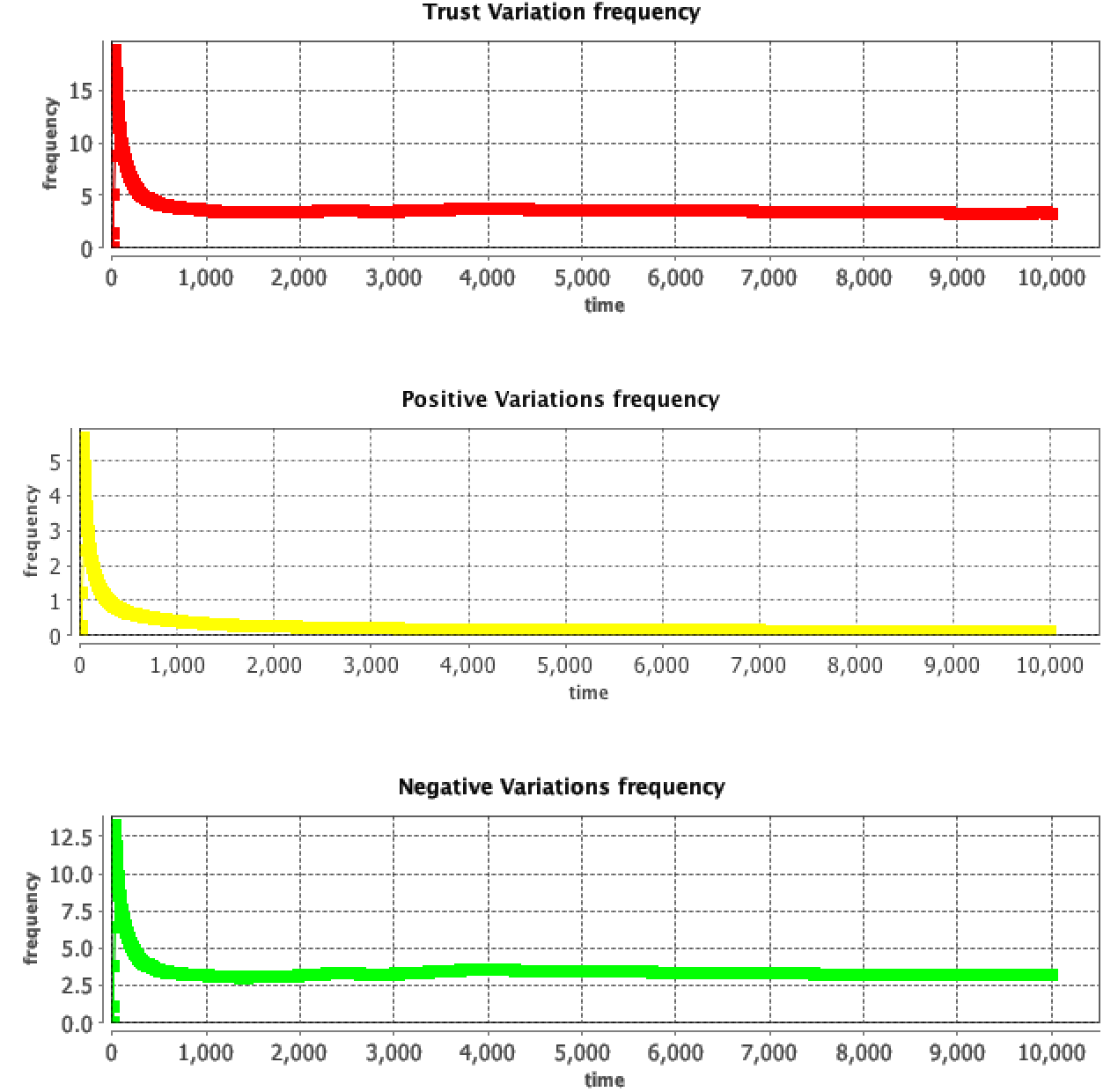}
\caption{Example of plots concerning the trust weights matrices variations.}
\label{trust_var}
\end{figure}

In this particular case institutions increase or decrease by the 10\% the weights related to the other agent in case it suggested an effective or an useless action. Since, at the moment, the time is not involved in the choice of the migration we do not expect any particular behaviour. We run a 10000 cycles simulation and monitor the trend of the frequencies of respectively the total trust variations, the trust positive and negative variations. Figure \ref{trust_var} depicts the results of this experiment. 

The experimental parameters are reported in Table \ref{feedback_params}. The probability value is again the probability for agents to create or delete a format collection at each cycle.

\begin{table}[!h]	
	\begin{center}
    \begin{tabular}{ |c|c|}
    \hline
    Parameter  & Value \\   
    \hline
	\hline	
	Institution nb. & 50 \\
	\hline
	Risk th. (\%) & 50 \\   
    \hline
	Suggest th. (\%) & 30 \\
	\hline
	 Inform th. (\%) & 50 \\
	\hline      
    Probability (\%) & 1 \\
	\hline
    \end{tabular}
	\end{center}	
	\caption{Values adopted for experiments on feedback mechanism stability.}
	\label{feedback_params}
\end{table}

\section{Frequency of ``Time-costing" Migrations}
\label{migFreqTime}
The experiments described in Section \ref{stability}, and the related analysis reported in Section \ref{stab_noTime} proved that communication among agents works efficiently thus information is exchanged properly. We saw how the limit as time approaches infinity of the frequency of migrations equals a finite value that is linear to the probability of creation or elimination of a format collection at each cycle. This behaviour let us assess that the model is stable and that its implementation is solid. However, as we mentioned in Section \ref{stability}, we made the assumption that the migration process is instantly completed which is not true for real migrations. 

To be as close as possible to real life, it is necessary to introduce a temporal dependence so that the migration processes last a certain number of cycles. We discussed how this number of cycles is evaluated in Section \ref{institution} describing the \emph{accept time} and \emph{migration time} actions. By adding this feature we also introduced the \-pos\-si\-bi\-li\-ty for an agent to refuse a suggested migration because too long. This certainly matches with a real situation where an archiver has to consider that a long lasting migration may cause the impossibility to perform other, and maybe more important, migrations.

We decided to monitor the trend of both the number and the frequency of migrations with increasing time. The set of parameters adopted for this simulation is reported in Table \ref{freqTime_params}.

\begin{table}[h]	
	\begin{center}
    \begin{tabular}{ |c|c|}
    \hline
    Parameter  & Value \\   
    \hline
	\hline	
	Institution nb. & 50 \\
	\hline
	Risk th. (\%) & 50 \\   
    \hline
	Suggest th. (\%) & 30 \\
	\hline
	 Inform th. (\%) & 70 \\
	\hline      
    Probability (\%) & 1 \\
	\hline
    \end{tabular}
	\end{center}	
	\caption{Values adopted for experiments on frequency of migrations.}
	\label{freqTime_params}
\end{table}

\begin{figure}[h!]
\centering
\includegraphics[width=11cm]{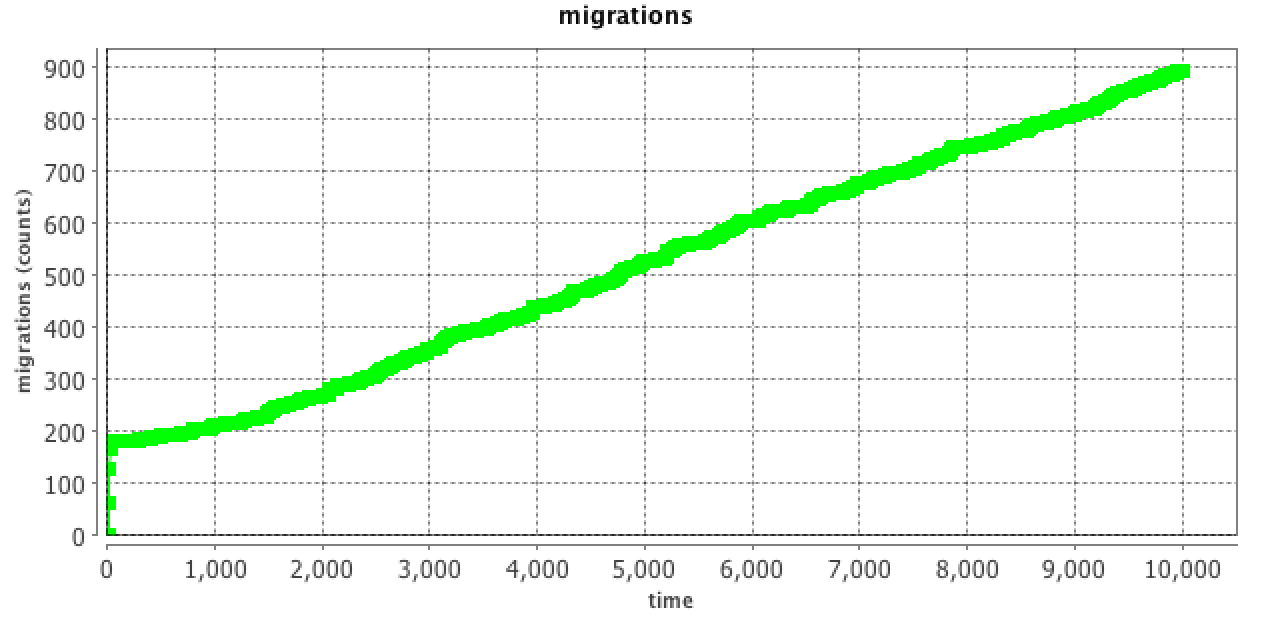}
\caption{Example of plot concerning the global number migrations.}
\label{migTime}
\end{figure}

\begin{figure}[h!]
\centering
\includegraphics[width=11cm]{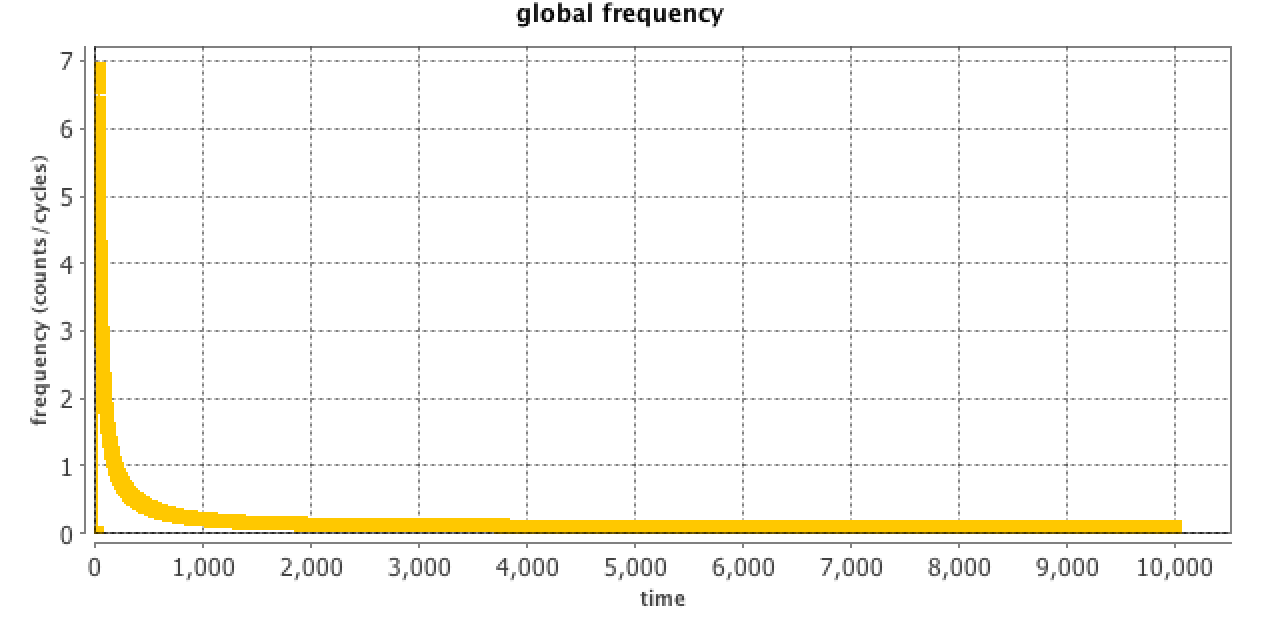}
\caption{Example of plot concerning the global frequency of migrations.}
\label{freqTime}
\end{figure}

We run a 10000 cycles simulation and acquire the data of our interest. As far as the number of migration is concerned, we see in Figure \ref{migTime} an ongoing increase of this variable while, as regards the frequency of migrations, we notice in Figure \ref{freqTime} a significant decay. The statistical analysis of these variables is reported in Section \ref{freqTimeAn} of next chapter.

As we shall see in Section \ref{freqTimeAn}, other twenty simulations were run in order to verify that the behaviour persists under different conditions.

\section{Linear Response Range (``Time-costing" Migrations)}
\label{lin_Time}
As we did in Section \ref{lin_noTime} we would like to determine if the value approached by the frequency of migrations with increasing time is linear to the probability that each institution creates or deletes a format collection from its members. 

Because of the introduction of time costing migrations we need to verify whether the linear behaviour predicted in Section \ref{lin_noTime} is still valid under these conditions. To do so we follow the experimental procedure of Section \ref{lin_noTime} a part from the duration of the simulations that now is 10000 cycles. The combination of parameters is reported in Table \ref{tab_lin_time}.

\begin{table}[h]	
	\begin{center}
    \begin{tabular}{ |c|c|c|c|c|c|c|c|}
    \hline
    Parameter  & \multicolumn{7}{|c|}{Values} \\   
    \hline
	\hline	
	Institution nb. & \multicolumn{7}{|c|}{50} \\
	\hline
	Risk th. (\%) & \multicolumn{7}{|c|}{50} \\   
    \hline
	Suggest th. (\%) & \multicolumn{7}{|c|}{30} \\
	\hline
	 Inform th. (\%) & \multicolumn{7}{|c|}{70} \\
	\hline      
   Probability (\%) & 1 & 2 & 3 & 5 & 7 & 50 & 100\\
	\hline
    \end{tabular}
	\end{center}	
	\caption{Values adopted for experiments on linearity.}
	\label{tab_lin_time}
	\end{table}

Each simulation lasted 10000 cycles in order to give the model enough time to be into a stable condition. In Section \ref{lin_resp_time} we discuss and analyse the results of these simulations. As we shall see in Section \ref{lin_resp_time} we take advantage of other simulations to prove the behaviour is steady.

\section{Migrations ``in Progress"}

\begin{figure}[!h]
\centering
\includegraphics[width=11cm]{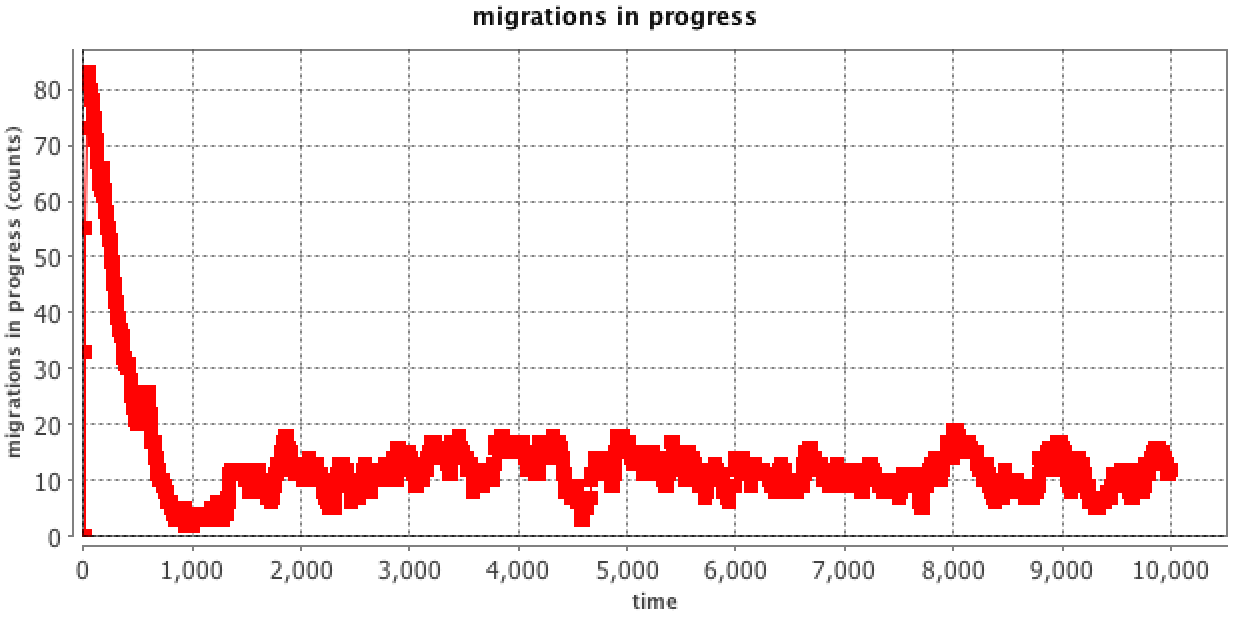}
\caption{Example of plots concerning the number of ongoing migrations.}
\label{in_progress}
\end{figure}

Now that the temporal aspect has been introduced for migration processes we can distinguish between those migrations that are being performed and those that are already completed. Thus it is interesting to monitor the trend of the ongoing migrations. To do so we need to count the number of migrations that the agents are performing and keep that value until they terminate the process. After a certain number of cycles we expect this value to stabilize. The experimental parameters are reported in Table \ref{inProg_params}. Figure \ref{in_progress} shows an example of the trend that we observed.

\begin{table}[h]	
	\begin{center}
    \begin{tabular}{ |c|c|}
    \hline
    Parameter  & Value \\   
    \hline
	\hline	
	Institution nb. & 50 \\
	\hline
	Risk th. (\%) & 50 \\   
    \hline
	Suggest th. (\%) & 30 \\
	\hline
	 Inform th. (\%) & 70 \\
	\hline      
    Probability (\%) & 1 \\
	\hline
    \end{tabular}
	\end{center}	
	\caption{Values adopted for experiments on ongoing migrations.}
	\label{inProg_params}
\end{table}

In Section \ref{ongoing_analysis} we analyse the results of the experiments concerning this quantity.

\section{Evaluation of Agents' Decisions}
\label{decisions_evaluation}
So far we described the experiments performed in order to get information about the trend of migrations, their frequency and the stability of the whole system. Though everything seems working fine, we must not forget that the aim of this work is to provide a model into which agents communicate and cooperate to find effective solutions to digital preservation issues. It means that, not only our system needs to be stable, but also the decisions taken by the agents have to be correct. 

\begin{figure}[h]
\centering
\includegraphics[width=13cm]{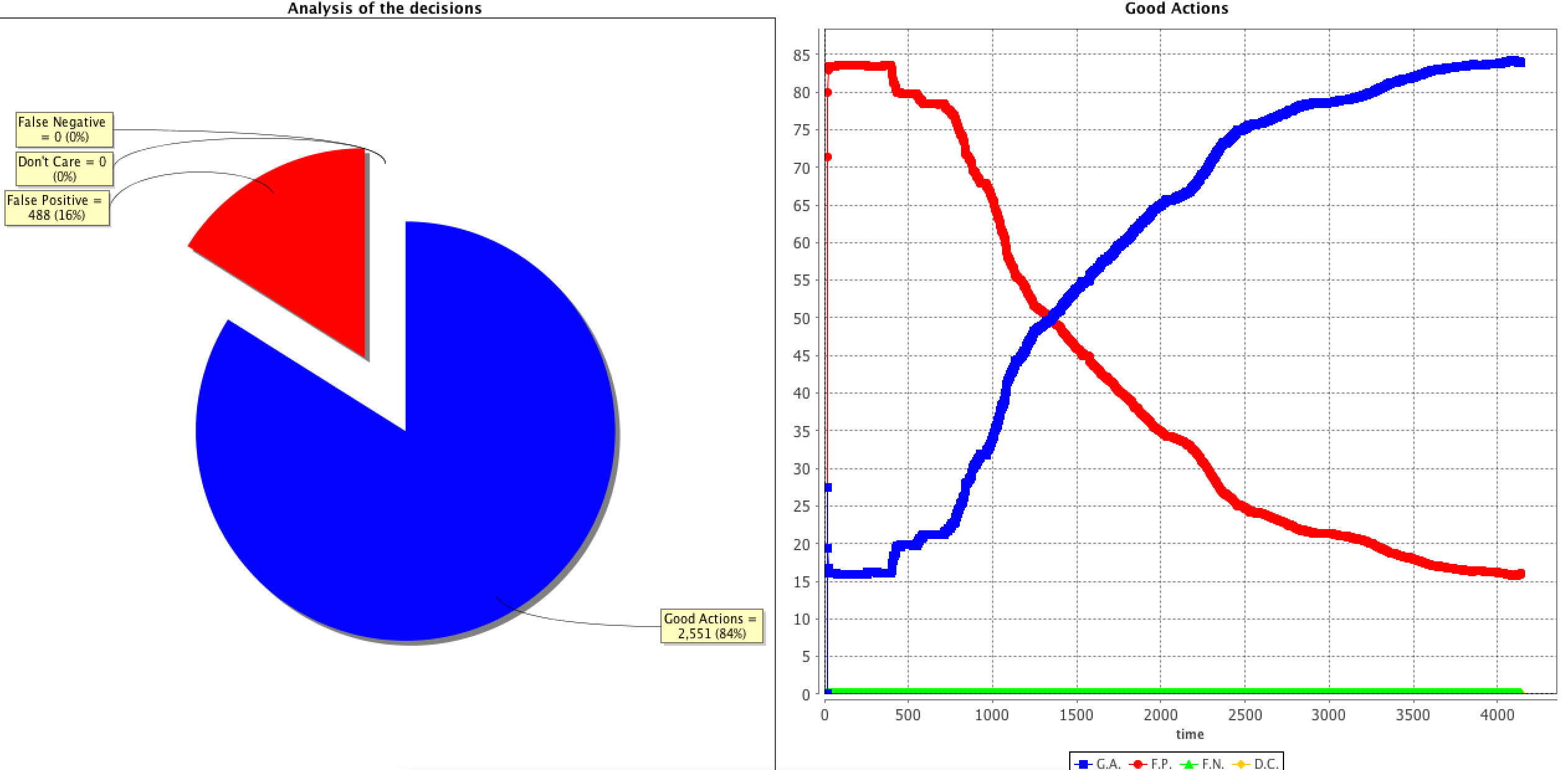}
\caption{Example of plot concerning the actions evaluation.}
\label{class_plot}
\end{figure}

In this section we describe the experiment implemented with the aim to evaluate the goodness of the agents' decisions. As we discussed in Section \ref{institution}, institution agents can essentially decide to take into account a suggestion, and thus perform a migration, or refuse it and maybe install an application or simply do nothing. In particular, we are interested in evaluating the decision to accept or refuse a migration.  With the parameters reported in Table \ref{class_params} we run a 10000 cycles simulation to monitor the number of \emph{good actions}, \emph{false positive}, \emph{false \-ne\-ga\-ti\-ve} and \emph{indifferent} decisions (see the \emph{evaluate migration} action in Section \ref{institution}). Figure \ref{class_plot} shows an example of what we could observe.

\begin{table}[h]	
	\begin{center}
    \begin{tabular}{ |c|c|}
    \hline
    Parameter  & Value \\   
    \hline
	\hline	
	Institution nb. & 50 \\
	\hline
	Risk th. (\%) & 50 \\   
    \hline
	Suggest th. (\%) & 30 \\
	\hline
	 Inform th. (\%) & 70 \\
	\hline      
    Probability (\%) & 1 \\
	\hline
    \end{tabular}
	\end{center}	
	\caption{Values adopted for experiments on actions evaluation.}
	\label{class_params}
\end{table}

We remind that the probability value in Table \ref{class_params} identify the probability of creation or deletion of a format as described in Sections \ref{lin_noTime} and \ref{lin_Time}. In addition, Figure \ref{class_plot} shows a pie chart concerning the percentage of \emph{good actions} in blue, \emph{false positive} in red, \emph{false negative} in green and \emph{indifferent} actions in orange. The scatter plot depicts the temporal evolution of the pie chart. The Section \ref{decisions} of the following chapter is dedicated to the analysis of these results.

\chapter{Statistical Analysis}
\label{statAnalysis}
In this chapter we analyse and discuss the results of the experiments performed as described in Chapter \ref{using_the_model}. The statistical analysis of our data has been performed making use of OriginLab 8.5 \cite{origin} software. The structure of this chapter matches the one of Chapter \ref{using_the_model} because into each of the following sections we analyse the corresponding experiment described before.

\section{Stability of the Communication Process}
\label{stab_noTime}

The graphs reported in Section \ref{stability} show that the frequency of migration decreases with increasing time. We decided to fit our observed values with the following exponential function:

\begin{equation}
f(t) = a \cdot \exp (-b \cdot \sqrt{t}) + c 
\label{exp}
\end{equation} 

Where the time $t$ appears under square root because of the interaction of the agents. We now report our data fitted with this function. Each fit involves data that belong to a certain time range, in particular from 200 to 5000 cycles. In this way we aim to avoid the effects of the initial transitory phase. For each fit a reduced chi square value is calculated. The error associated with the number of migration is its square root according to the Poisson's distribution. Since no error is associated with the number of cycles, the expression of the frequency of migrations with the corresponding error is the following: 

\begin{equation}
frequency = \frac{migrations}{time} \pm \frac{\sqrt{migrations}}{time}
\label{error}
\end{equation}

\subsection{Variation of the Number of Institutions}
\label{instNbVar}
\begin{figure}[!h]
\centering
\includegraphics[width=12cm]{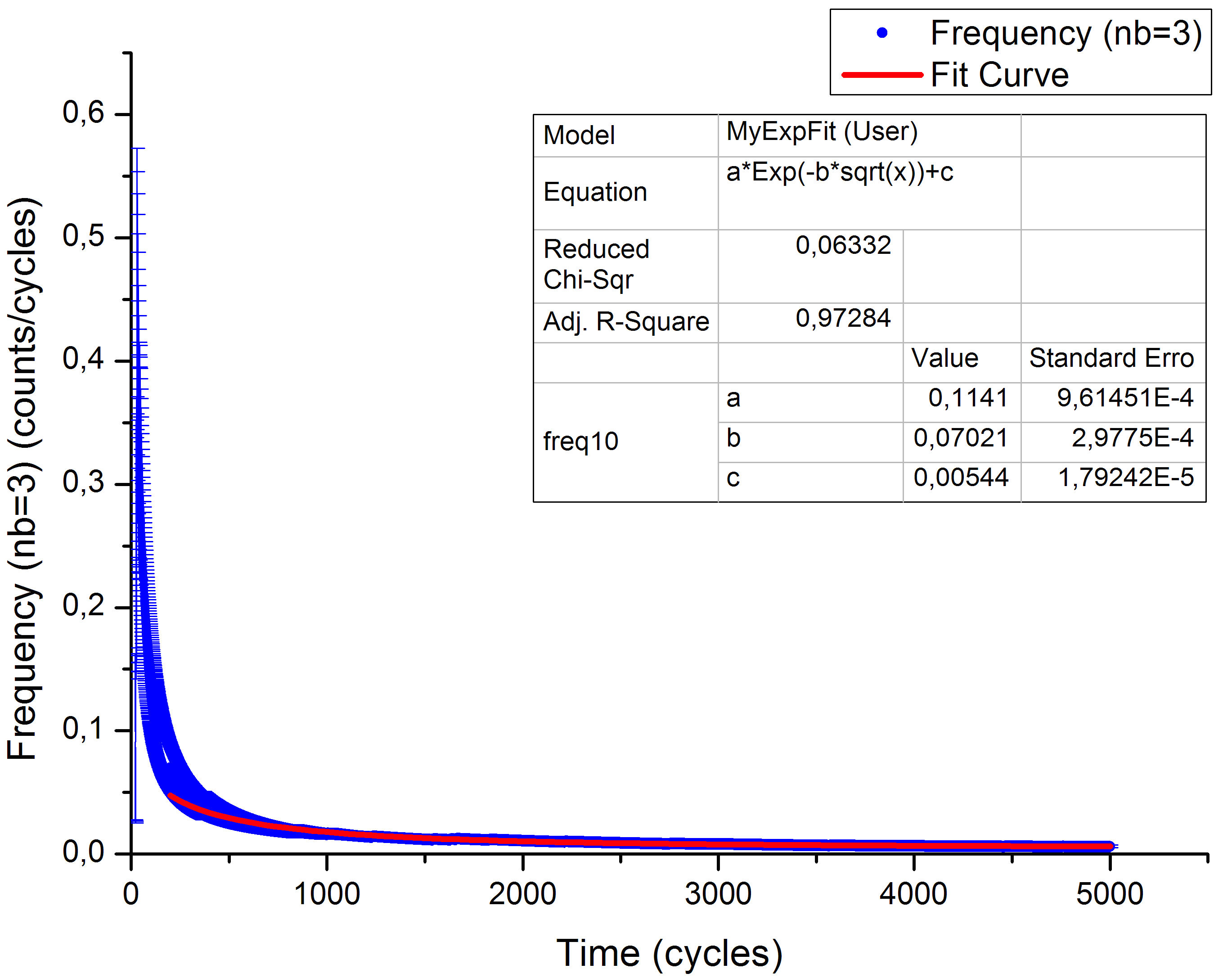}
\caption{Plot of migrations frequency with institution number = 3.}
\label{fig:inst3}

\centering
\includegraphics[width=12cm]{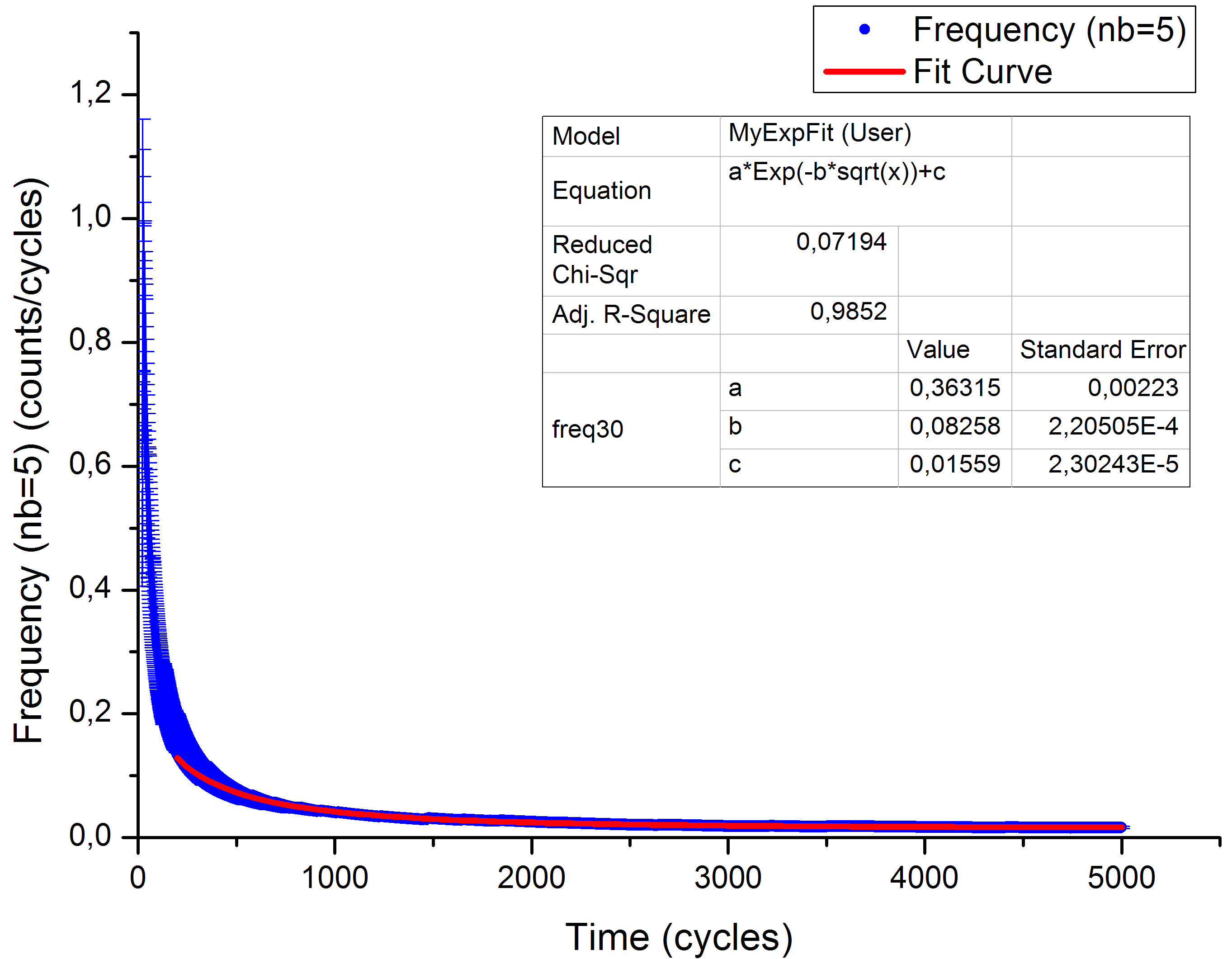}
\caption{Plot of migrations frequency with institution number = 5.}
\label{fig:inst5}
\end{figure}

\begin{figure}[!b]
\centering
\includegraphics[width=12cm]{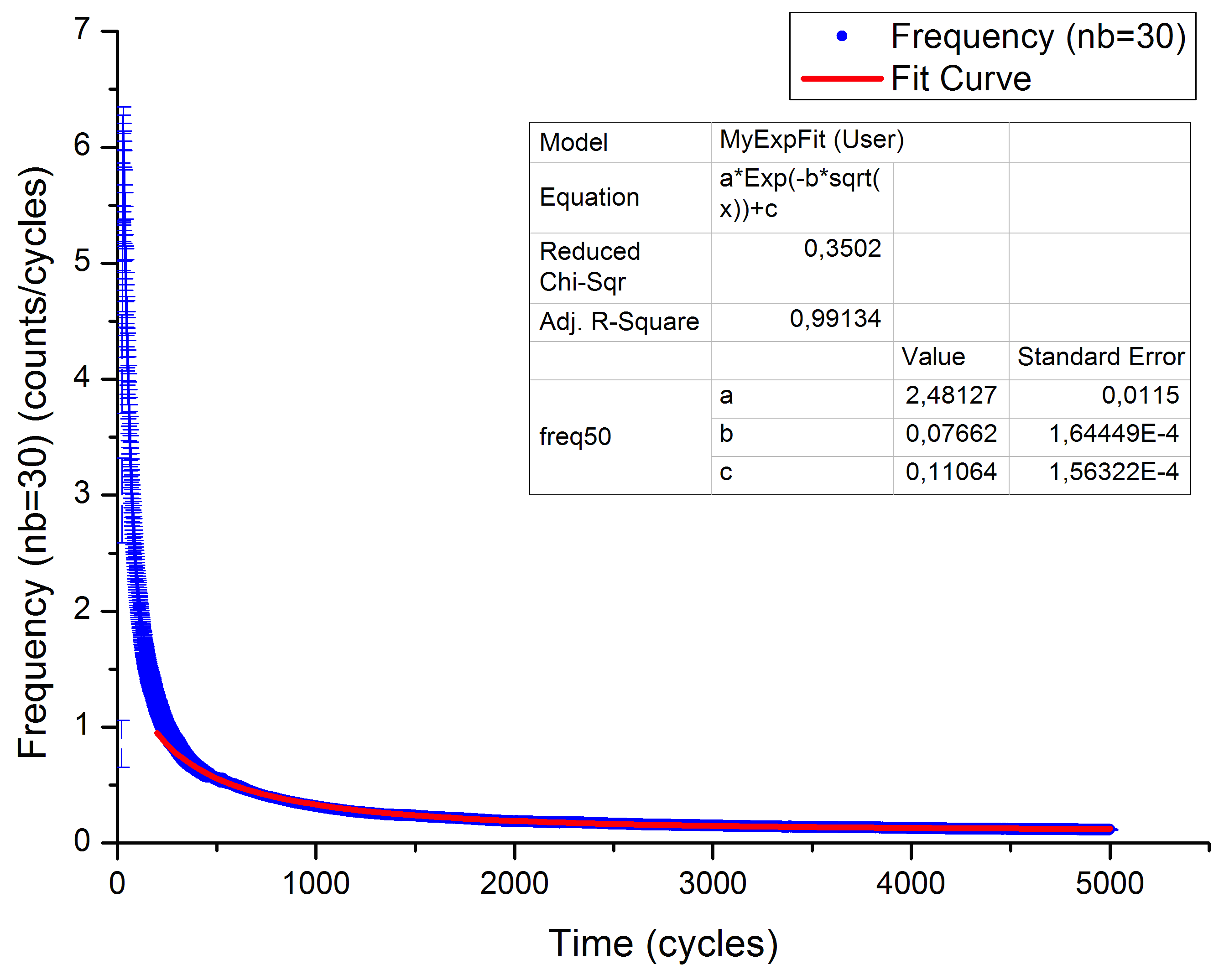}
\caption{Plot of migrations frequency with institution number = 30.}
\label{fig:inst30}

\centering
\includegraphics[width=12cm]{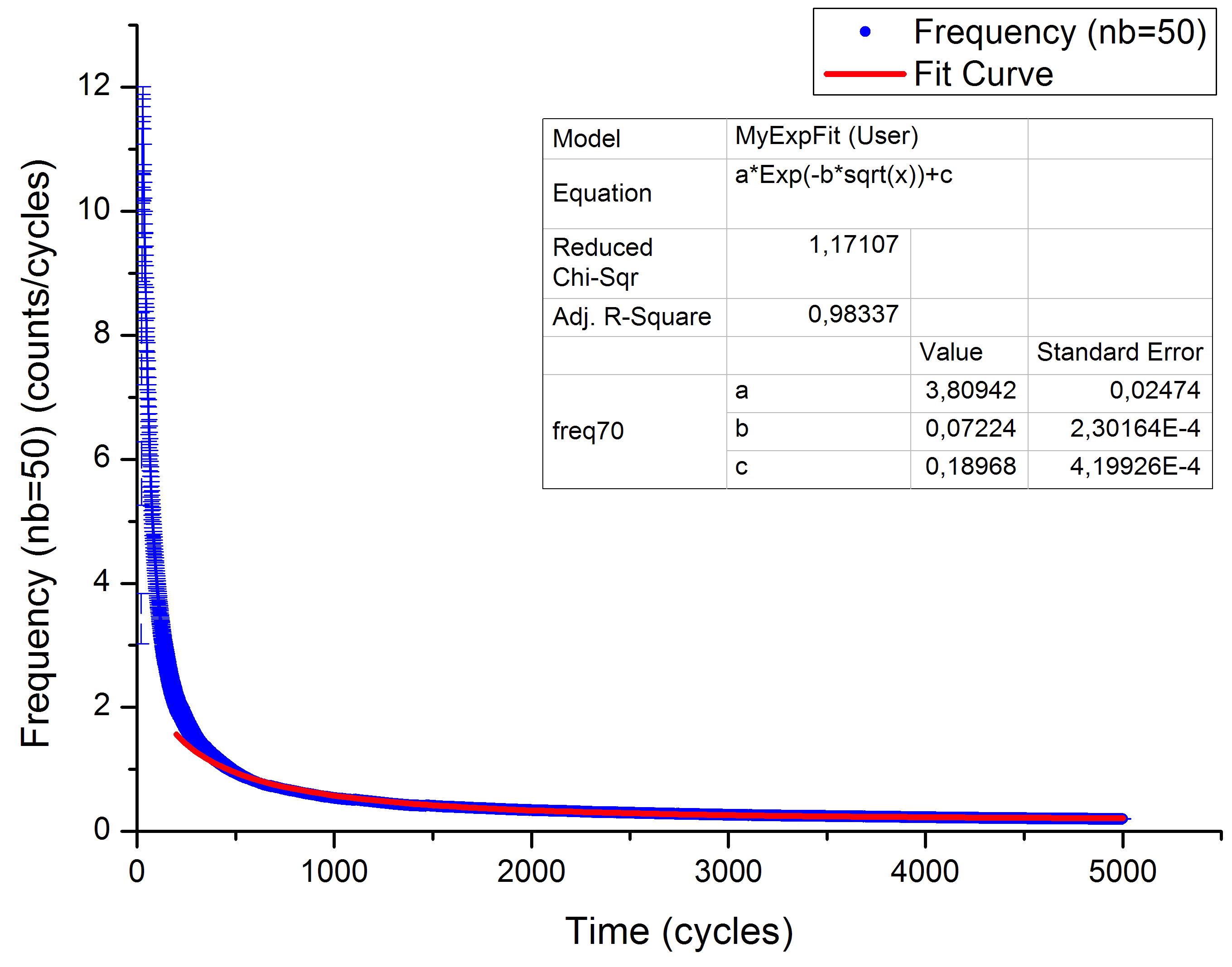}
\caption{Plot of migrations frequency with institution number = 50.}
\label{fig:inst50}
\end{figure}

\begin{figure}[!h]
\centering
\includegraphics[width=12cm]{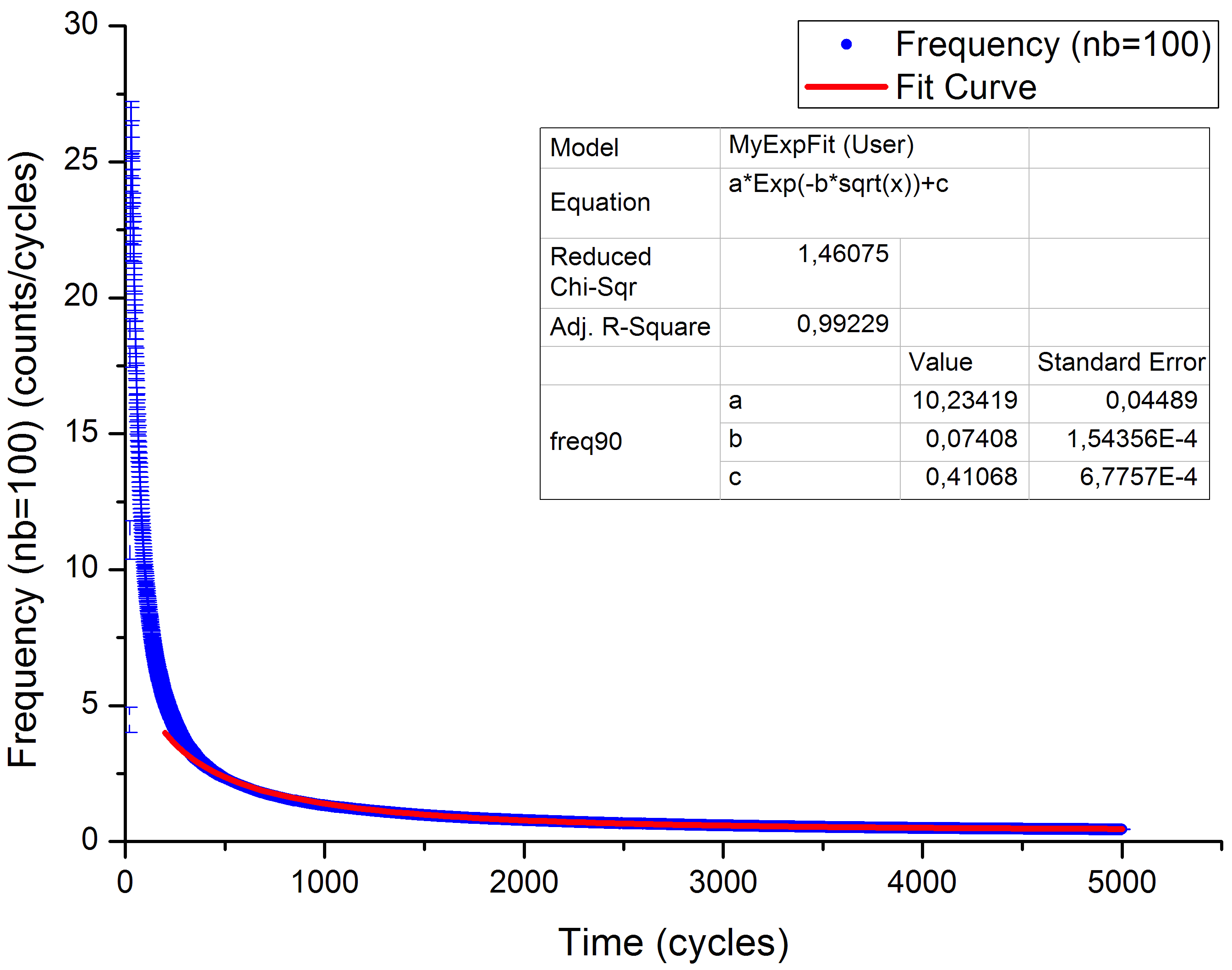}
\caption{Plot of migrations frequency with institution number = 100.}
\label{fig:inst100}

\end{figure}

As the plots in Figure \ref{fig:inst3}, \ref{fig:inst5}, \ref{fig:inst30}, \ref{fig:inst50}, \ref{fig:inst100} depict, the function fits our data properly with very low reduced chi square values reported in Table \ref{chi2inst}. We notice that reduced chi square increases with the number of institutions.

\begin{table}[!h]	
	\begin{center}
    \begin{tabular}{ |c|c|c|c|c|c|}
    \hline
    Institution Number values   &  3 & 5 & 30 & 50 & 100 \\   
    \hline
	\hline	
	Reduced $\chi^2$ values & 0.063 & 0.072 & 0.35 & 1.17 & 1.46 \\
	\hline    
    \end{tabular}
	\end{center}	
	\caption{Reduced $\chi^2$ values, variable number of institutions.}
	\label{chi2inst}
	\end{table}
	
It is also possible to determine the asymptotic frequency value with the related error which is the $c$ parameter of the fit function \ref{exp}. The values are reported in Table \ref{asymptNB}.

\begin{table}[!h]	
	\begin{center}
    \begin{tabular}{ |c|c|c|c|c|c|}
    \hline
    Inst. Nb.   &  3 & 5 & 30 & 50 & 100 \\   
    \hline
	\hline	
	Freq. ($\frac{counts}{cycles}$) & $5.440 \cdot 10^{-3}$ & $1.5590 \cdot 10^{-2}$ & $1.1064 \cdot 10^{-1}$ & $1.8968 \cdot 10^{-1}$ & $4.1068 \cdot 10^{-1} $\\
	\hline
	Err. ($\pm \frac{counts}{cycles}$) & $1.8 \cdot 10^{-5}$ &$ 2.3 \cdot 10^{-5}$ & $1.6 \cdot 10^{-4}$ & $4.2 \cdot 10^{-4}$ & $6.8 \cdot 10^{-4}$ \\
	\hline        
    \end{tabular}
	\end{center}	
	\caption{Asymptotic frequency values with errors, variable number of institutions.}
	\label{asymptNB}
	\end{table}
	
\clearpage

\clearpage

\subsection{Variation of the Risk Threshold}
\label{RiVar}

\begin{figure}[!h]
\centering
\includegraphics[width=12cm]{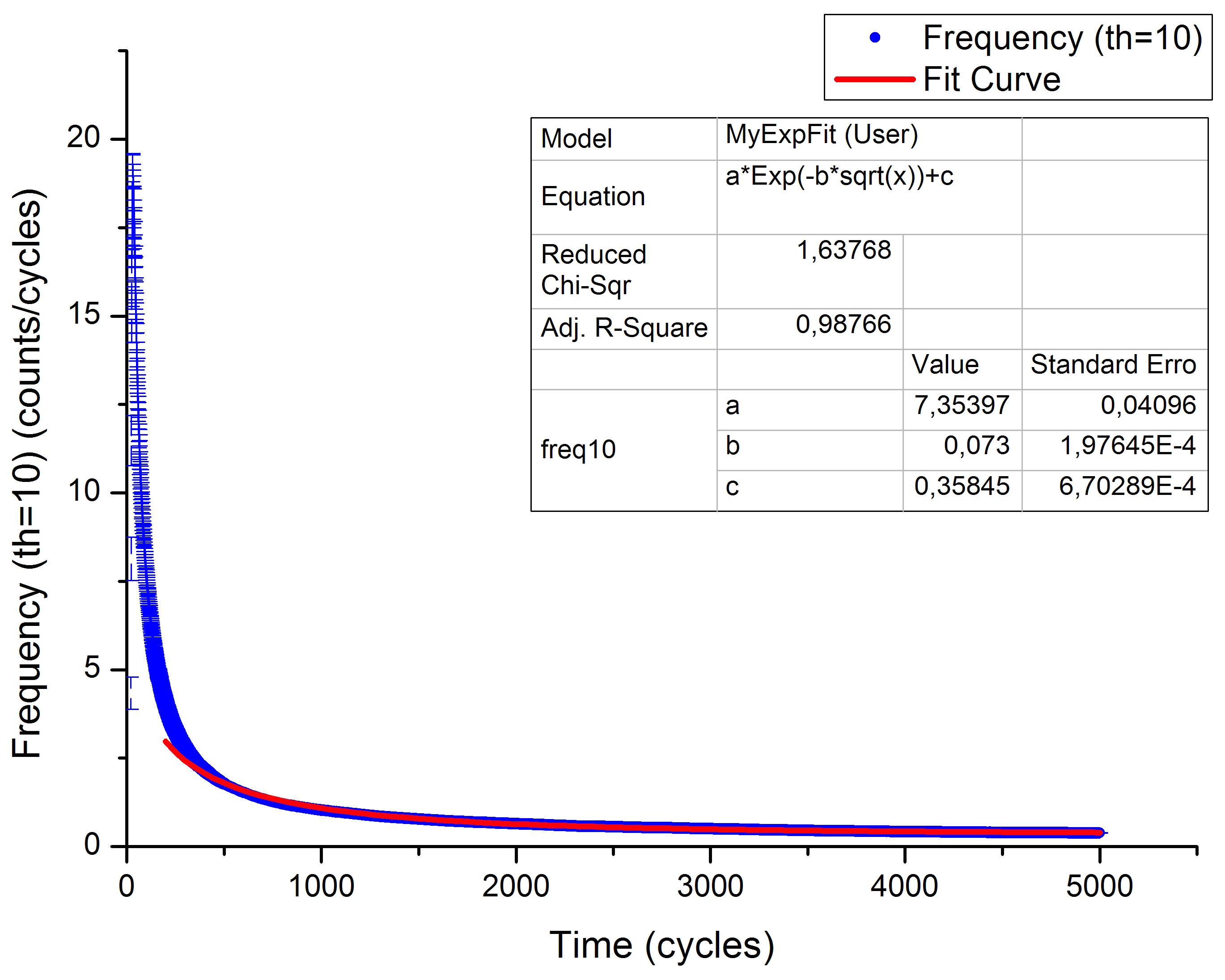}
\caption{Plot of migrations frequency with risk threshold = 10.}
\label{fig:risk10}
%
\centering
\includegraphics[width=12cm]{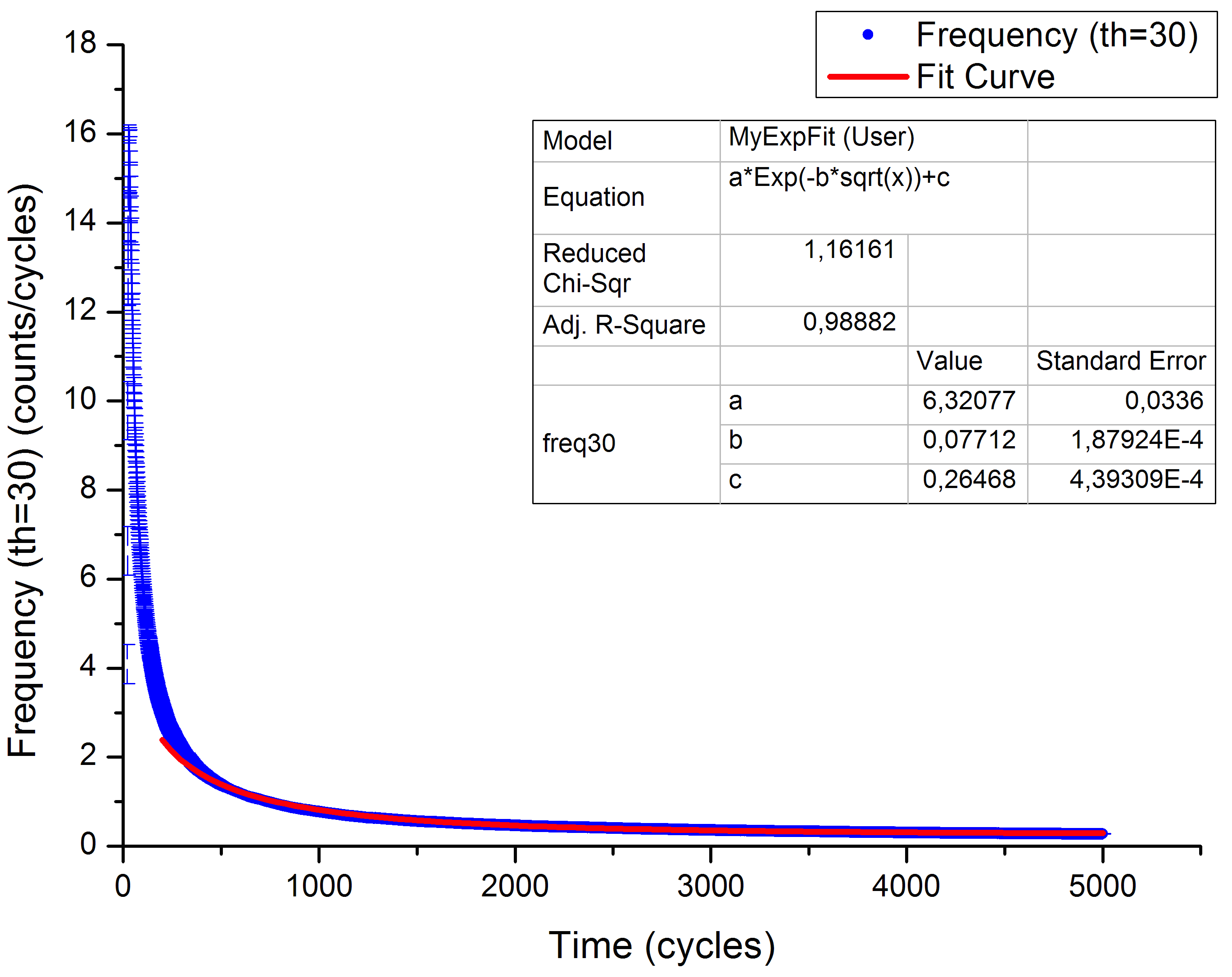}
\caption{Plot of migrations frequency with risk threshold = 30.}
\label{fig:risk30}
\end{figure}

\begin{figure}[!h]
\centering
\includegraphics[width=12cm]{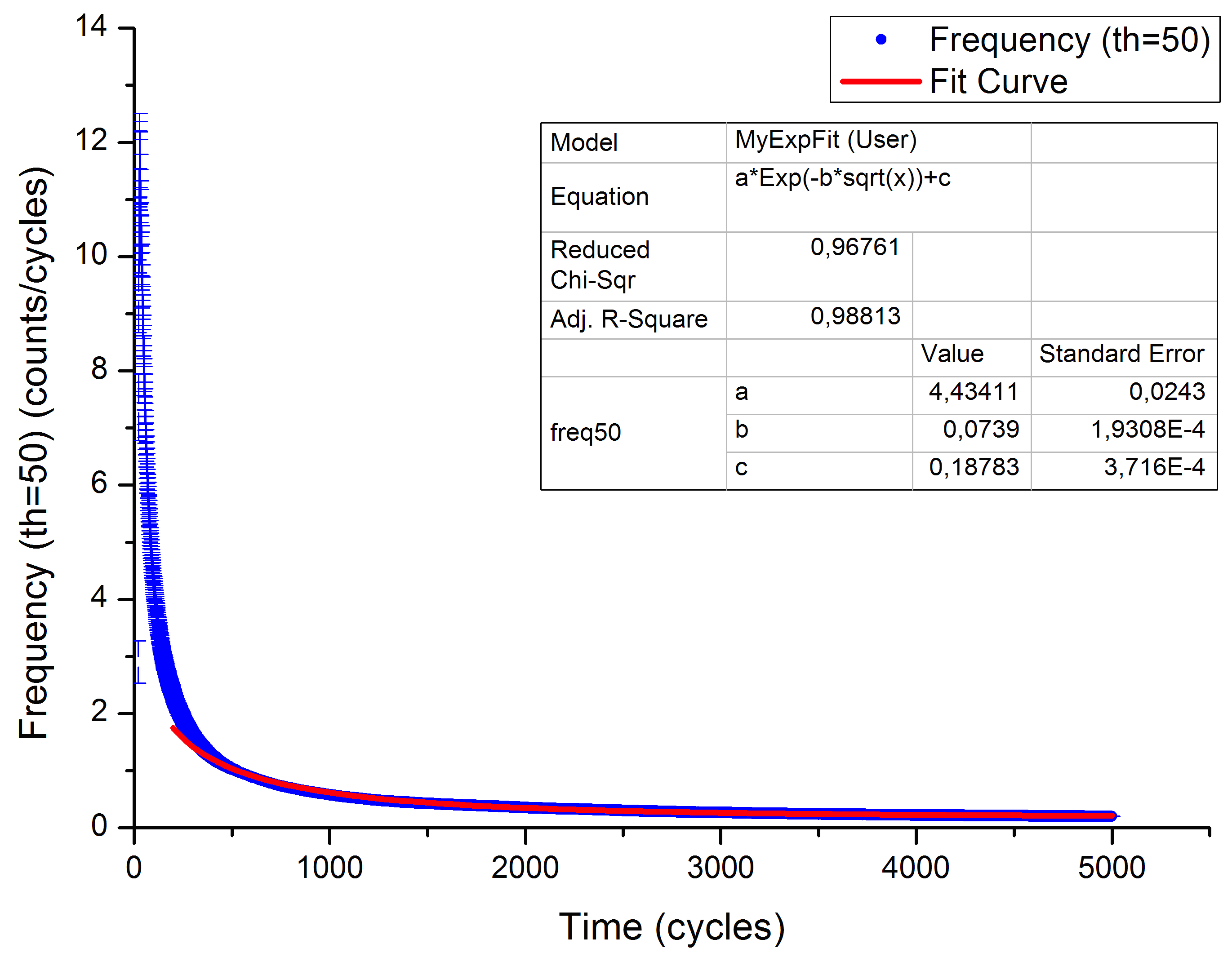}
\caption{Plot of migrations frequency with risk threshold = 50.}
\label{fig:risk50}
%
\centering
\includegraphics[width=12cm]{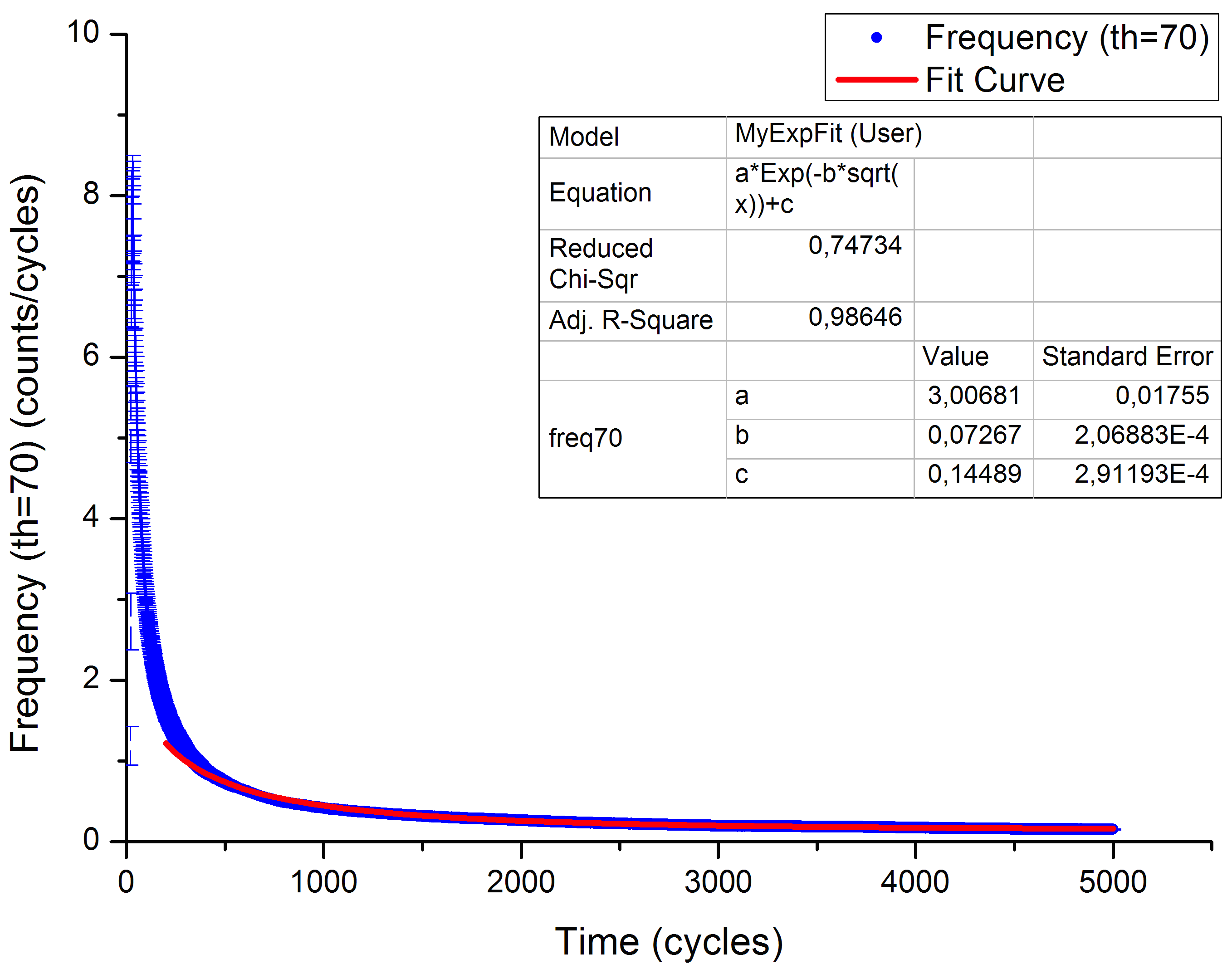}
\caption{Plot of migrations frequency with risk threshold = 70.}
\label{fig:risk70}
\end{figure}

\begin{figure}[!h]
\centering
\includegraphics[width=12cm]{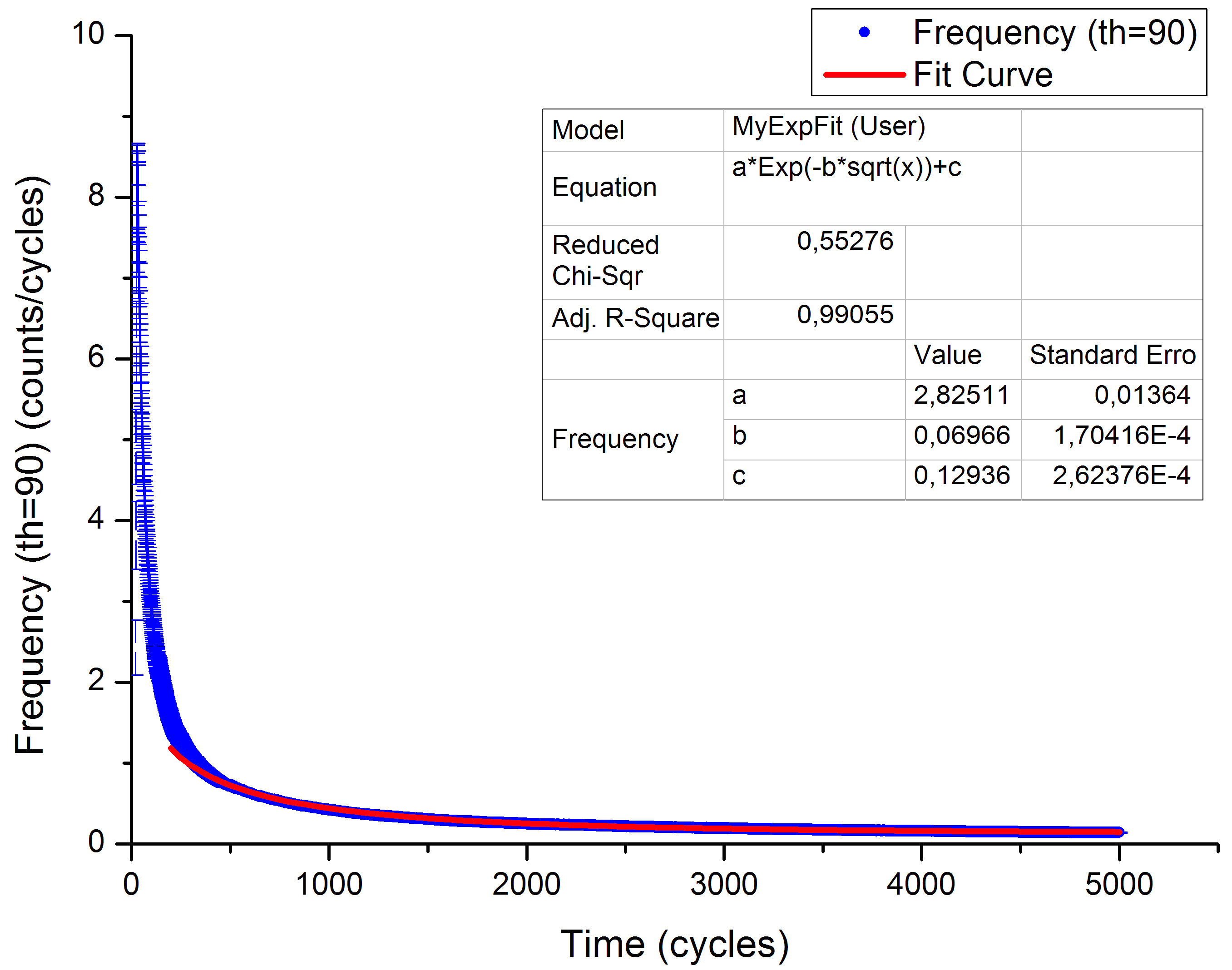}
\caption{Plot of migrations frequency with risk threshold = 90.}
\label{fig:risk90}

\end{figure}

As far as the risk threshold is concerned, the plots in Figure \ref{fig:risk10}, \ref{fig:risk30}, \ref{fig:risk50}, \ref{fig:risk70}, \ref{fig:risk90} show that the function fits our data properly again with very low reduced chi square values reported in Table \ref{chi2rt}. In this case we notice that the reduced chi square decreases with increasing the threshold.

\begin{table}[!h]	
	\begin{center}
    \begin{tabular}{ |c|c|c|c|c|c|}
    \hline
    Risk Threshold values (\%)  &  10 & 30 & 50 & 70 & 90 \\   
    \hline
	\hline	
	Reduced $\chi^2$ values & 1.64 & 1.16 & 0.97 & 0.75 & 0.55 \\
	\hline    
    \end{tabular}
	\end{center}	
	\caption{Reduced $\chi^2$ values, variable risk threshold.}
	\label{chi2rt}
	\end{table}

As in Section \ref{instNbVar} it is possible to obtain the asymptotic values of the frequency of migrations. They are reported in Table \ref{asymptRISK}.

\begin{table}[!h]	
	\begin{center}
    \begin{tabular}{ |c|c|c|c|c|c|}
    \hline
     Risk Th. (\%)  &  10 & 30 & 50 & 70 & 90 \\ 
    \hline
	\hline	
	Freq. ($\frac{counts}{cycles}$) & $3.5845 \cdot 10^{-1}$ & $2.6468 \cdot 10^{-1}$ & $1.8783 \cdot 10^{-1}$ & $1.4489 \cdot 10^{-1}$ & $1.2936 \cdot 10^{-1} $\\
	\hline
	Err. ($\pm \frac{counts}{cycles}$) & $6.7 \cdot 10^{-4}$ &$ 4.4 \cdot 10^{-4}$ & $3.7 \cdot 10^{-4}$ & $2.9 \cdot 10^{-4}$ & $2.6 \cdot 10^{-4}$ \\
	\hline        
    \end{tabular}
	\end{center}	
	\caption{Asymptotic frequency values with errors, variable risk threshold.}
	\label{asymptRISK}
	\end{table}

\clearpage

\clearpage

\subsection{Variation of the \emph{Suggest} Threshold}
\label{SuVar}
\begin{figure}[!h]
\centering
\includegraphics[width=12cm]{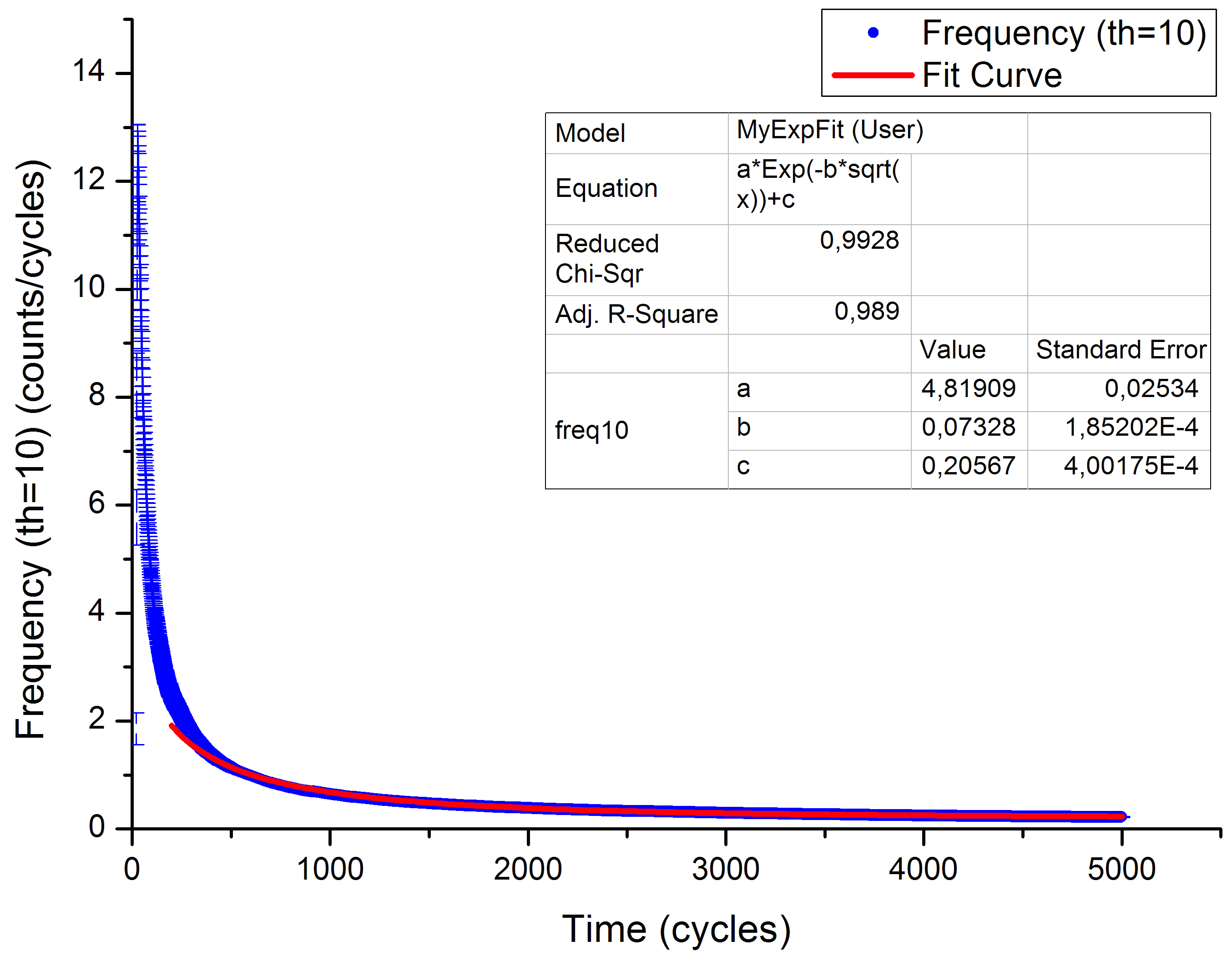}
\caption{Plot of migrations frequency with \emph{suggest} threshold = 10.}
\label{fig:st10}
%
\centering
\includegraphics[width=12cm]{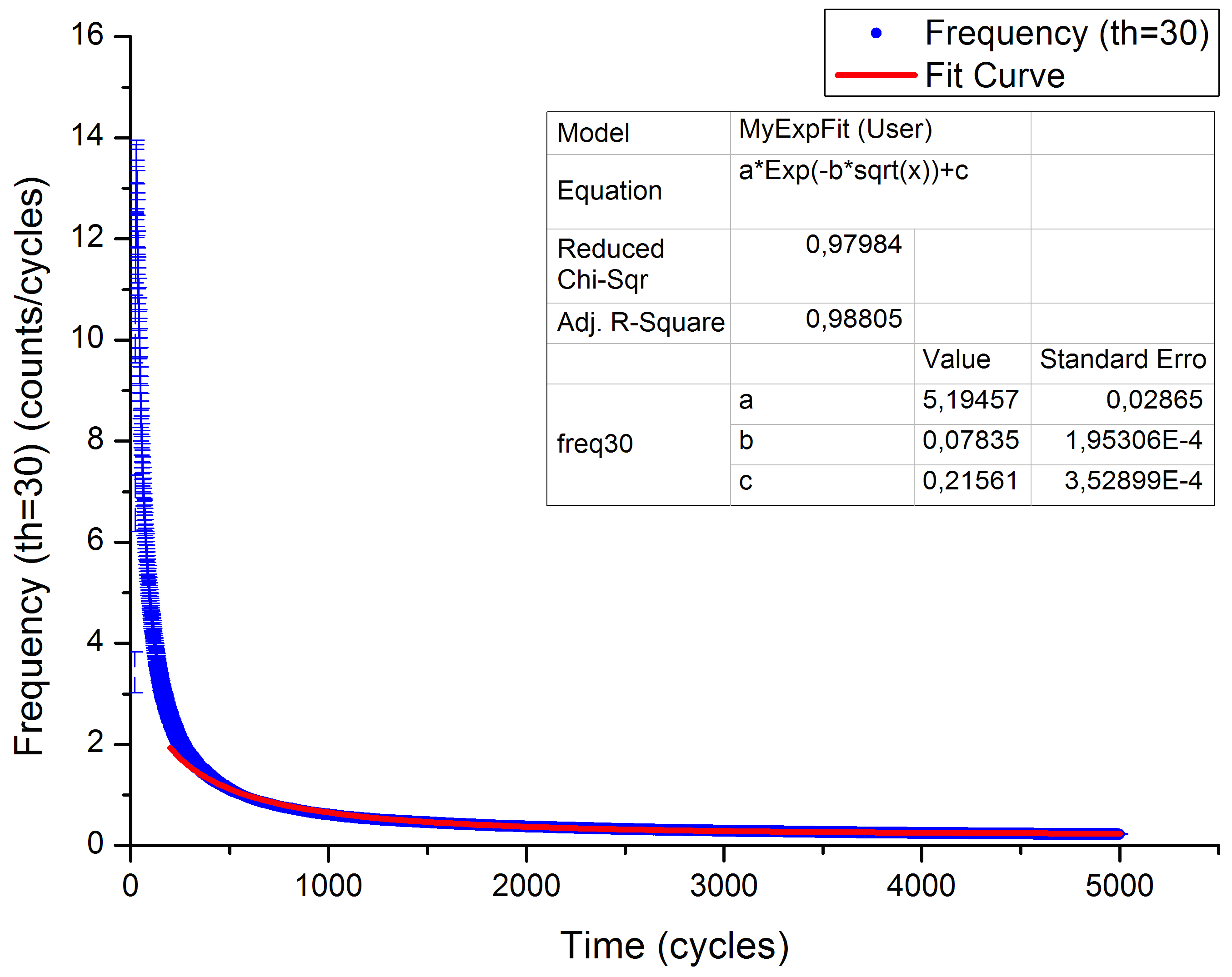}
\caption{Plot of migrations frequency with \emph{suggest} threshold = 30.}
\label{fig:st30}
\end{figure}

\begin{figure}[!h]
\centering
\includegraphics[width=12cm]{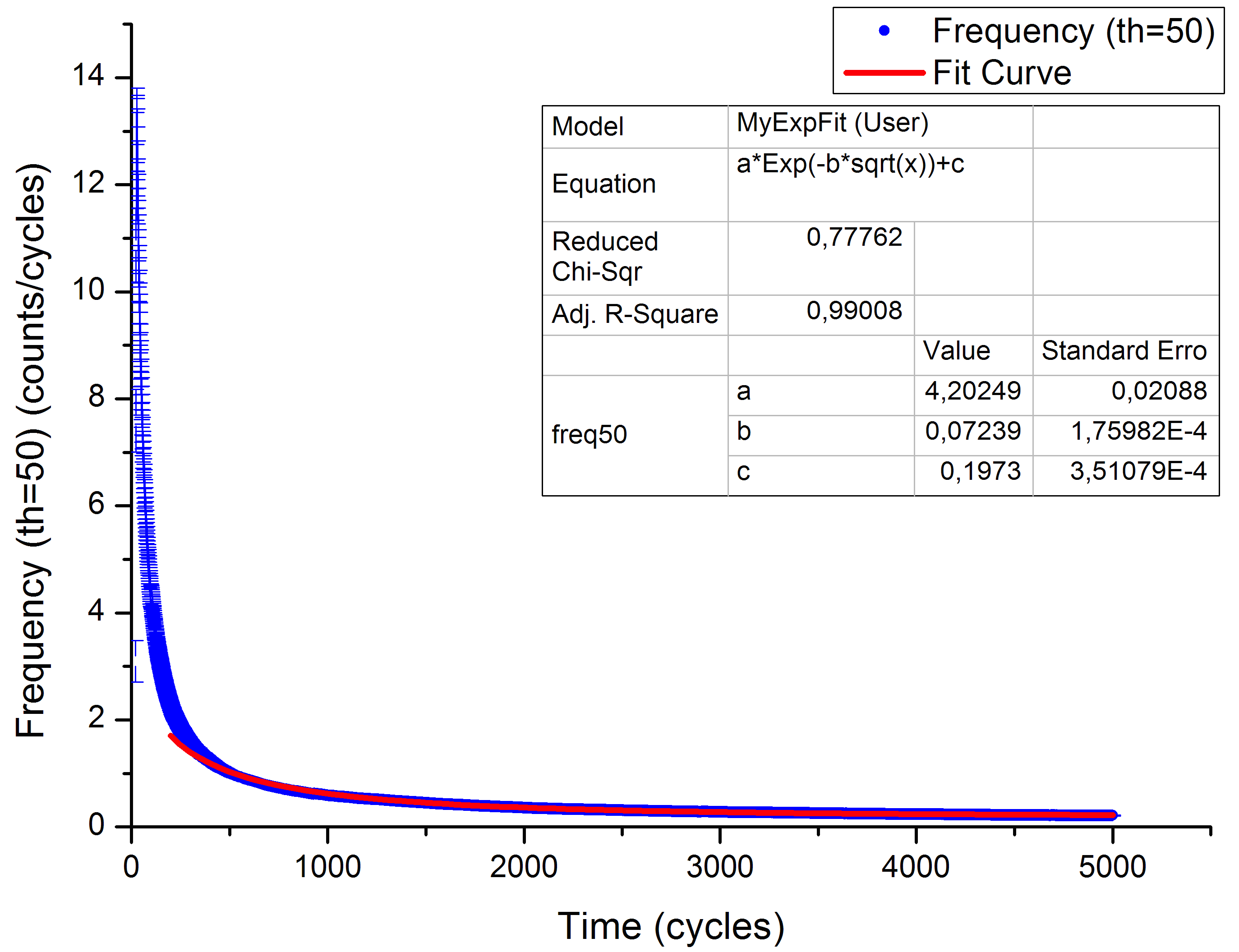}
\caption{Plot of migrations frequency with \emph{suggest} threshold = 50.}
\label{fig:st50}
%
\centering
\includegraphics[width=12cm]{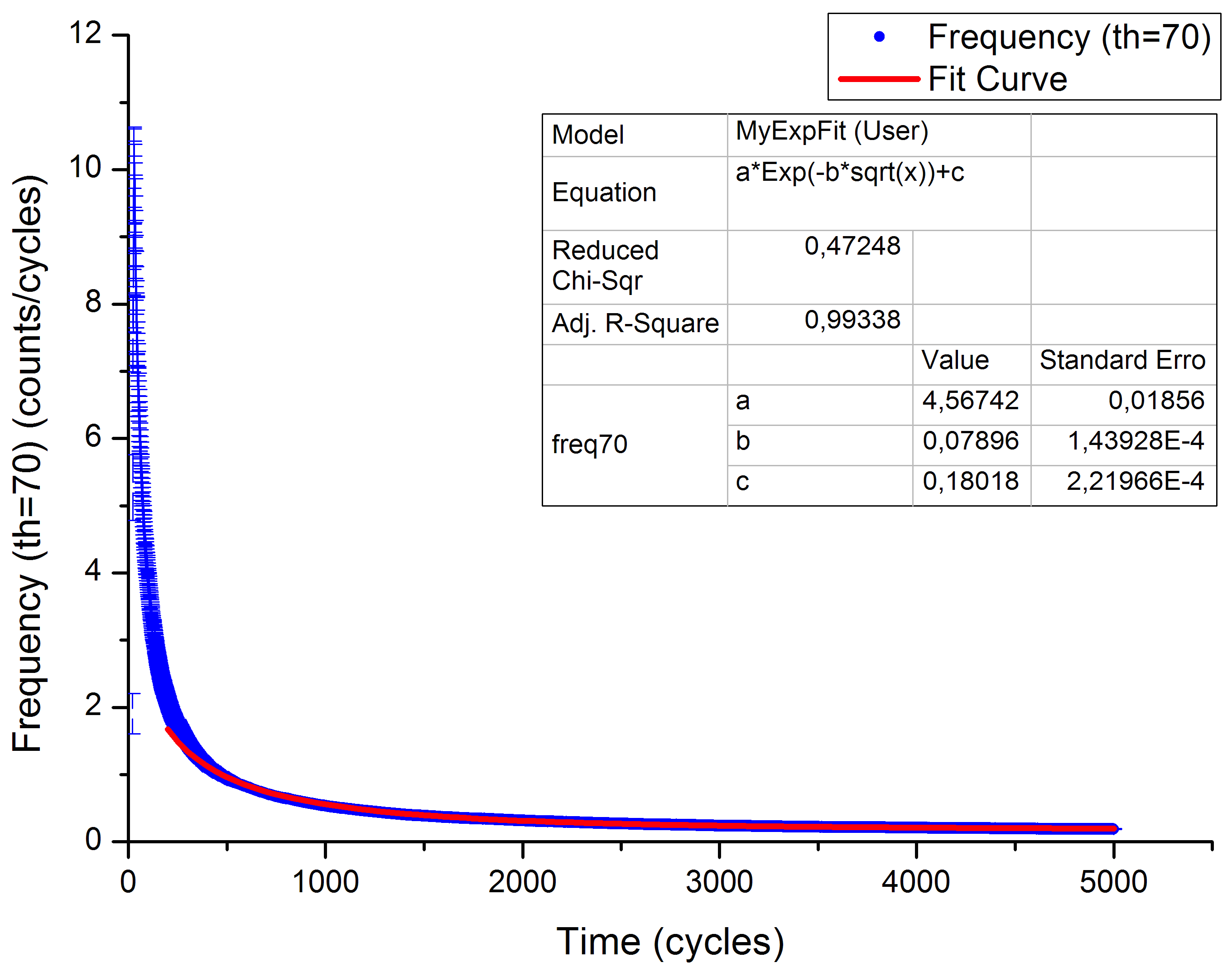}
\caption{Plot of migrations frequency \emph{suggest} threshold = 70.}
\label{fig:st70}
\end{figure}

\begin{figure}[!h]
\centering
\includegraphics[width=12cm]{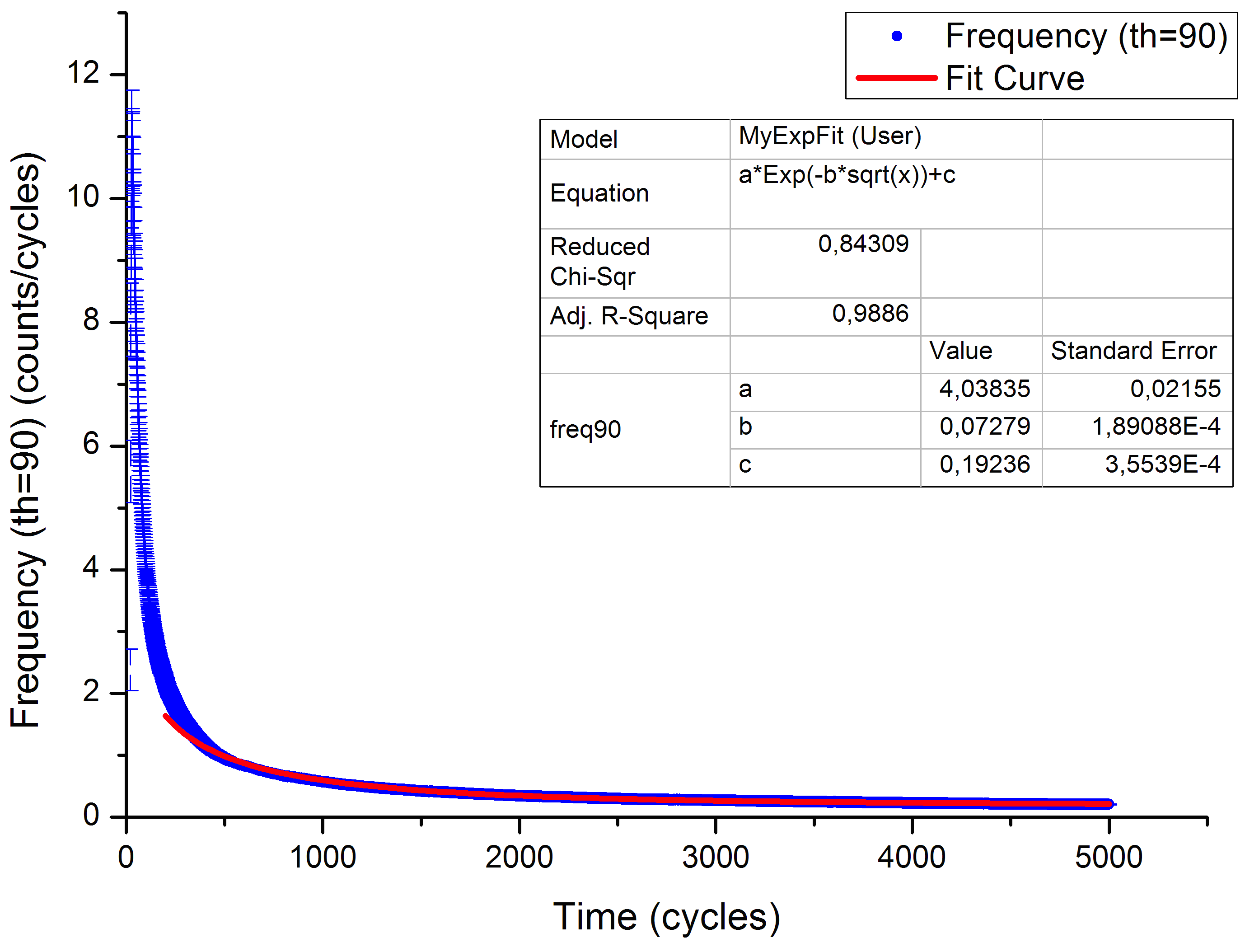}
\caption{Plot of migrations frequency with \emph{suggest} threshold = 90.}
\label{fig:st90}

\end{figure}

The plots in Figure \ref{fig:st10}, \ref{fig:st30}, \ref{fig:st50}, \ref{fig:st70}, \ref{fig:st90} show that the function fits our experimental data about the threshold that determines if a suggested migration is accepted or not. All the reduced chi square values in Table \ref{chi2st} remain below 1, which means that function fits well our data in each of the conditions taken into account.

\begin{table}[h!]	
	\begin{center}
    \begin{tabular}{ |c|c|c|c|c|c|}
    \hline
    Suggest Threshold values (\%)  &  10 & 30 & 50 & 70 & 90 \\   
    \hline
	\hline	
	Reduced $\chi^2$ values & 0.99 & 0.98 & 0.78 & 0.47 & 0.84 \\
	\hline    
    \end{tabular}
	\end{center}	
	\caption{Reduced $\chi^2$ values, variable \emph{suggest} threshold.}
	\label{chi2st}
	\end{table}
	
In Table \ref{asymptSUG} we report again the values of the asymptotic migrations frequency for each of the simulations performed.

\begin{table}[!h]	
	\begin{center}
    \begin{tabular}{ |c|c|c|c|c|c|}
    \hline
     Sugg. Th. (\%) & 10 & 30 & 50 & 70 & 90 \\  
    \hline
	\hline	
	Freq. ($\frac{counts}{cycles}$) & $2.0567 \cdot 10^{-1}$ & $2.1561 \cdot 10^{-1}$ & $1.9730 \cdot 10^{-1}$ & $1.8018 \cdot 10^{-1}$ & $1.9236 \cdot 10^{-1} $\\
	\hline
	Err. ($\pm \frac{counts}{cycles}$) & $4.0 \cdot 10^{-4}$ &$ 3.5 \cdot 10^{-4}$ & $3.5 \cdot 10^{-4}$ & $2.2 \cdot 10^{-4}$ & $3.6 \cdot 10^{-4}$ \\
	\hline        
    \end{tabular}
	\end{center}	
	\caption{Asymptotic frequency values with errors, variable \emph{suggest} threshold.}
	\label{asymptSUG}
	\end{table}

\clearpage

\clearpage

\subsection{Variation of the \emph{Inform} Threshold}
\label{InVar}

\begin{figure}[!h]
\centering
\includegraphics[width=12cm]{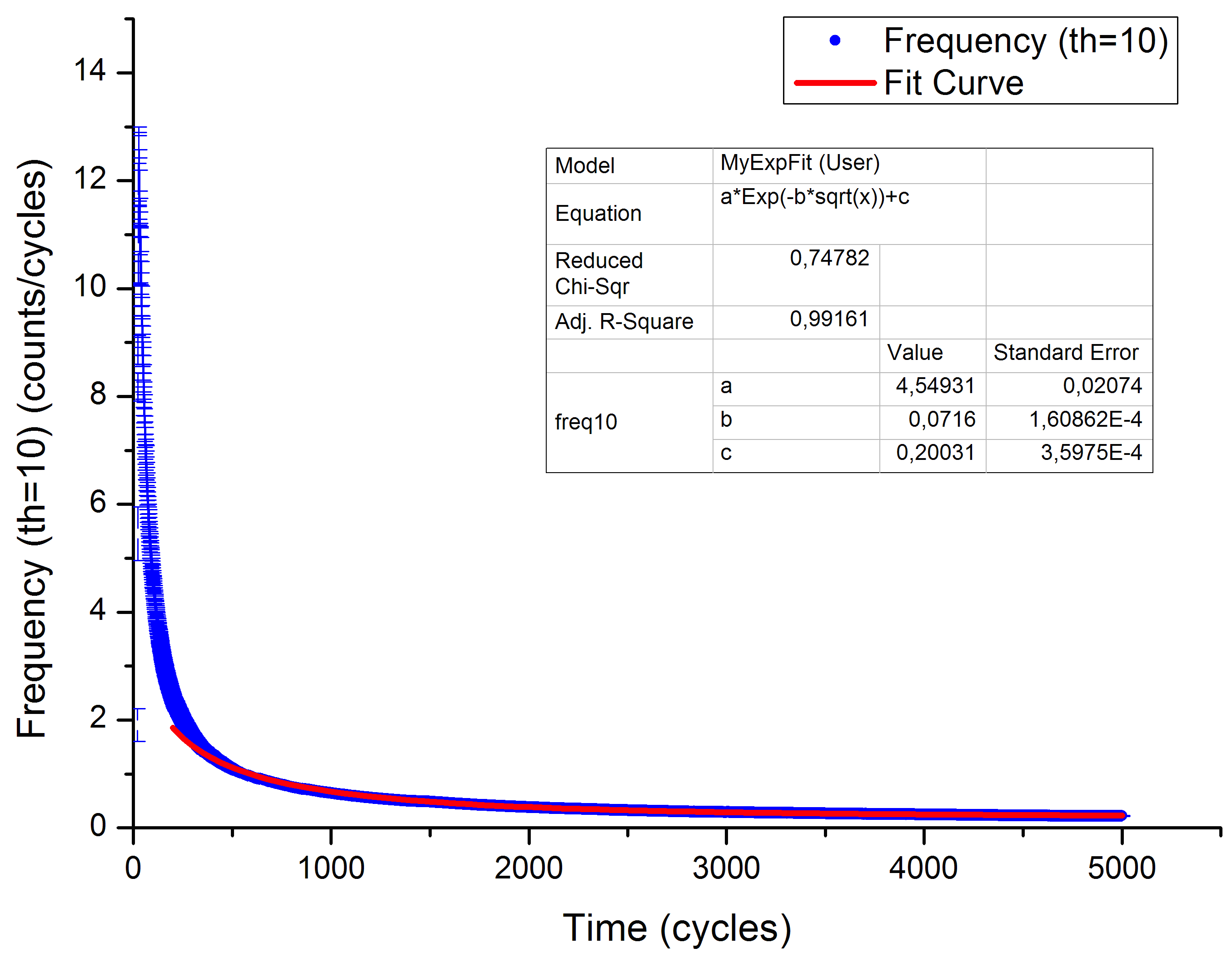}
\caption{Plot of migrations frequency with \emph{inform} threshold = 10.}
\label{fig:it10}
%
\centering
\includegraphics[width=12cm]{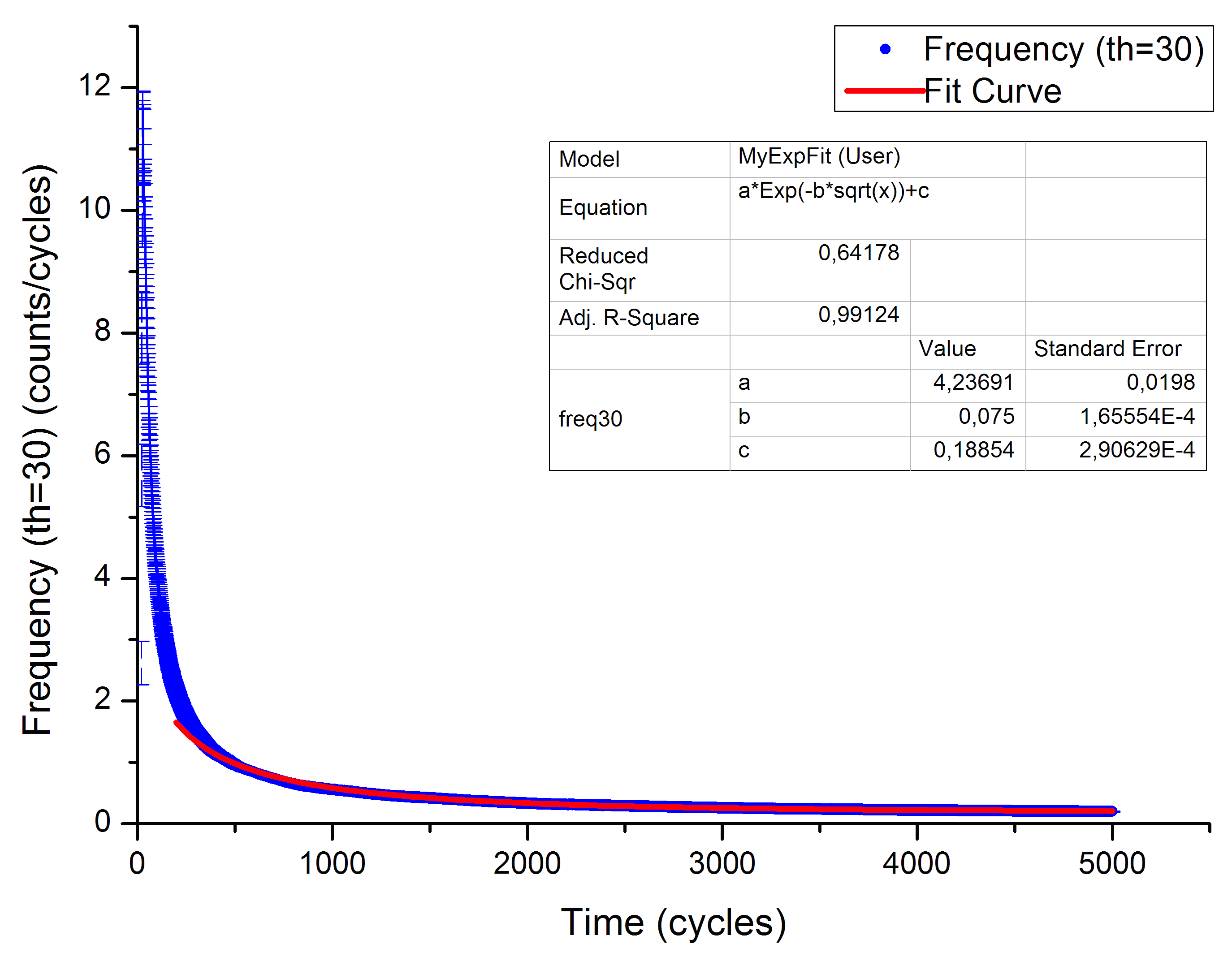}
\caption{Plot of migrations frequency with \emph{inform} threshold = 30.}
\label{fig:it30}
\end{figure}

\begin{figure}[!h]
\centering
\includegraphics[width=12cm]{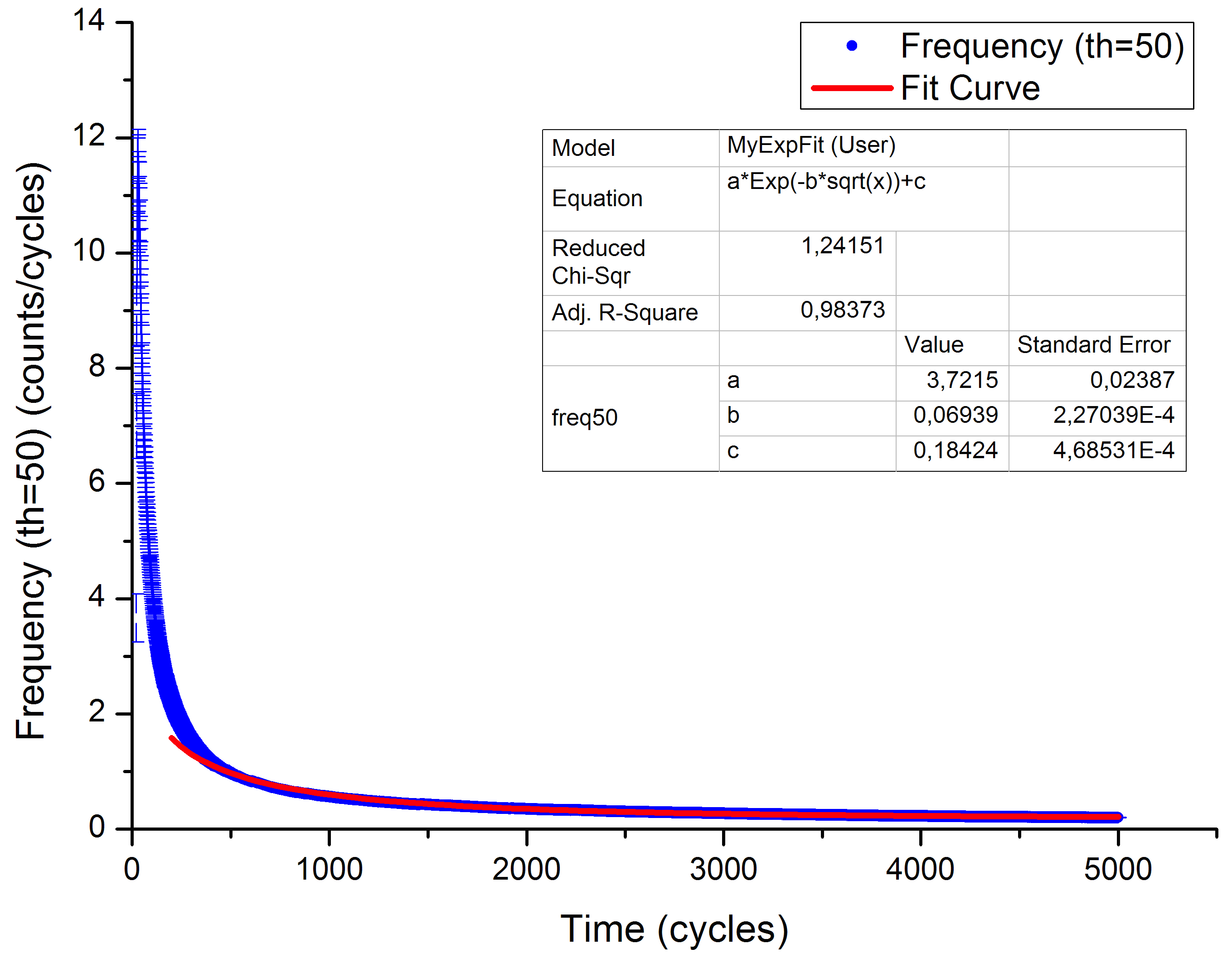}
\caption{Plot of migrations frequency with \emph{inform} threshold = 50.}
\label{fig:it50}
%
\centering
\includegraphics[width=12cm]{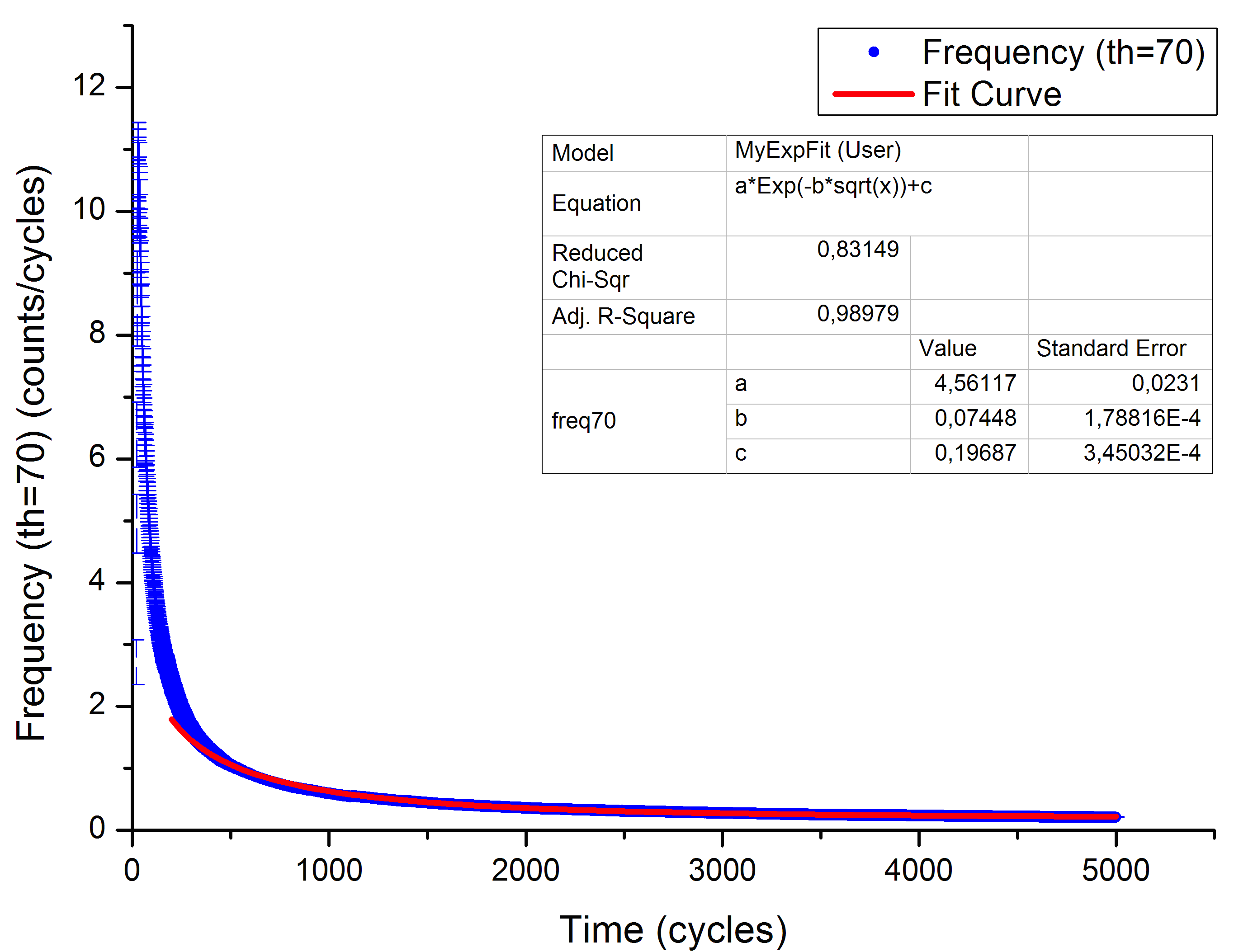}
\caption{Plot of migrations frequency \emph{inform} threshold = 70.}
\label{fig:it70}
\end{figure}

\begin{figure}[!h]
\centering
\includegraphics[width=12cm]{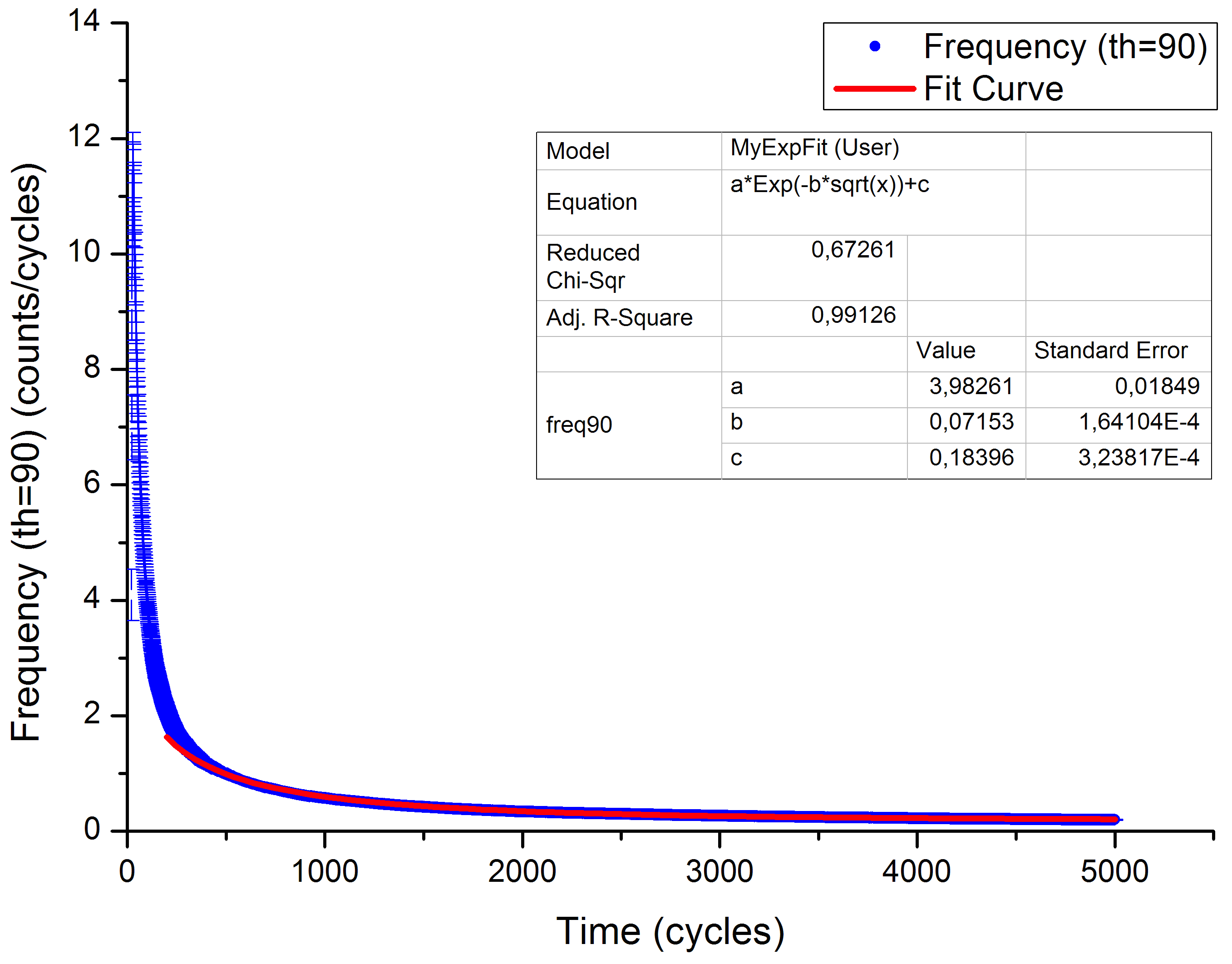}
\caption{Plot of migrations frequency with \emph{inform} threshold = 90.}
\label{fig:it90}

\end{figure}

The last experimental parameter is the threshold that agents use to decide how to deal with a migration suggestion coming from an inform message. As the plots in Figure \ref{fig:it10}, \ref{fig:it30}, \ref{fig:it50}, \ref{fig:it70}, \ref{fig:it90} depict, the function fits well our data with low reduced chi square values reported in Table \ref{chi2it}. We notice here that only when the threshold equals the 50 \% the reduced chi square exceeds 1. 

\begin{table}[h!]	
	\begin{center}
    \begin{tabular}{ |c|c|c|c|c|c|}
    \hline
    Inform Threshold values (\%)  &  10 & 30 & 50 & 70 & 90 \\   
    \hline
	\hline	
	Reduced $\chi^2$ values & 0.75 & 0.64 & 1.24 & 0.83 & 0.67 \\
	\hline    
    \end{tabular}
	\end{center}	
	\caption{Reduced $\chi^2$ values, variable \emph{inform} threshold.}
	\label{chi2it}
	\end{table}
	
Even for this set of simulations we report the values of the asymptotic migrations frequency for each of the simulations performed. Table \ref{asymptINF} shows the values obtained.

\begin{table}[!h]	
	\begin{center}
    \begin{tabular}{ |c|c|c|c|c|c|}
    \hline
     Info. Th.(\%) &  10 & 30 & 50 & 70 & 90 \\  
    \hline
	\hline	
	Freq. ($\frac{counts}{cycles}$) & $2.0031 \cdot 10^{-1}$ & $1.8854 \cdot 10^{-1}$ & $1.8424 \cdot 10^{-1}$ & $1.9687 \cdot 10^{-1}$ & $1.8396 \cdot 10^{-1} $\\
	\hline
	Err. ($\pm \frac{counts}{cycles}$) & $3.6 \cdot 10^{-4}$ &$ 2.9 \cdot 10^{-4}$ & $4.7 \cdot 10^{-4}$ & $3.5 \cdot 10^{-4}$ & $3.2 \cdot 10^{-4}$ \\
	\hline        
    \end{tabular}
	\end{center}	
	\caption{Asymptotic frequency values with errors, variable \emph{inform} threshold.}
	\label{asymptINF}
\end{table}

\clearpage

\subsection{Fit Result Summary}
We report the result of each fit in Table \ref{stabFitTable}. All the fit described in Sections \ref{instNbVar}, \ref{RiVar}, \ref{SuVar} and \ref{InVar} returned the three parameters $a, b$ and $c$ of the fitting function \ref{exp} with the corresponding error.

\begin{table}[!h]	
	\begin{sideways}	
    \begin{tabular}{ |c|c|c|c|c|}
    \hline
     Institution Nb. & $a \pm \sigma_a (\frac{counts}{cycles})$ & $b \pm \sigma_b (cycles^{-\frac{1}{2}})$ & $c \pm \sigma_c (\frac{counts}{cycles})$ & $red .\chi^2$\\  
    \hline
	3 & $(1.1410 \pm 0.0096 )\cdot 10^{-1}$ & $(7.021 \pm 0.030)\cdot 10^{-2}$ & $(5.440 \pm 0.018) \cdot 10^{-3}$ & 0.063 \\
	\hline
	5 & $(3.632 \pm 0.022) \cdot 10^{-1} $ & $(8.258 \pm 0.022) \cdot 10^{-2} $ & $(1.5590 \pm 0.0023) \cdot 10^{-2} $ & 0.072\\
	\hline
	30 & $2.481 \pm 0.012 $ & $(7.622 \pm 0.0016) \cdot 10^{-2} $ & $(1.1064 \pm 0.0016) \cdot 10^{-1} $ & 0.35 \\
	\hline
	50 & $3.809 \pm 0.025 $ & $(7.224 \pm 0.023) \cdot 10^{-2} $ & $(1.8968 \pm 0.0042) \cdot 10^{-1} $ & 1.17 \\
	\hline
	100 & $10.234 \pm 0.045 $ & $(7.408 \pm 0.015) \cdot 10^{-2} $ & $(4.1068 \pm 0.0068) \cdot 10^{-1} $ & 1.46\\
	\hline
	\hline
	Risk Th. & $a \pm \sigma_a (\frac{counts}{cycles})$ & $b \pm \sigma_b (cycles^{-\frac{1}{2}})$ & $c \pm \sigma_c (\frac{counts}{cycles})$ & $red .\chi^2$\\
	\hline
	10 & $7.354 \pm 0.041$ & $(7.300 \pm 0.020) \cdot 10^{-2} $ & $(3.5845 \pm 0.0067) \cdot 10^{-1} $ & 1.64 \\
	\hline
	30 & $6.321 \pm 0.034$ & $(7.712 \pm 0.019) \cdot 10^{-2} $ & $(2.6468 \pm 0.0044) \cdot 10^{-1} $ & 1.16 \\
	\hline
	50 & $4.434 \pm 0.024 $ & $(7.390 \pm 0.019) \cdot 10^{-2} $ & $(1.8783 \pm 0.0037) \cdot 10^{-1} $ & 0.97 \\
	\hline
	70 & $3.007 \pm 0.018 $ & $(7.267 \pm 0.021) \cdot 10^{-2} $ & $(1.4489 \pm 0.0029) \cdot 10^{-1} $ & 0.75 \\
	\hline
	90 & $2.825 \pm 0.014 $ & $(6.966 \pm 0.017) \cdot 10^{-2} $ & $(1.2936 \pm 0.0026) \cdot 10^{-1} $ & 0.55 \\
	\hline
	\hline
	Suggest Th. & $a \pm \sigma_a (\frac{counts}{cycles})$ & $b \pm \sigma_b (cycles^{-\frac{1}{2}})$ & $c \pm \sigma_c (\frac{counts}{cycles})$ & $red .\chi^2$\\	
	\hline
	10 & $4.819 \pm 0.025$ & $(7.328 \pm 0.019) \cdot 10^{-2} $ & $(2.0567 \pm 0.0040) \cdot 10^{-1} $ & 0.99 \\
	\hline
	30 & $5.195 \pm 0.029$ & $(7.835 \pm 0.020) \cdot 10^{-2} $ & $(2.1561 \pm 0.0035) \cdot 10^{-1} $ & 0.98 \\
	\hline
	50 & $4.203 \pm 0.021$ & $(7.239 \pm 0.018) \cdot 10^{-2} $ & $(1.9730 \pm 0.0035) \cdot 10^{-1} $ & 0.78 \\
	\hline
	70 & $4.567 \pm 0.019$ & $(7.896 \pm 0.014) \cdot 10^{-2} $ & $(1.8018 \pm 0.0022) \cdot 10^{-1} $ & 0.47\\
	\hline
	90 & $4.038 \pm 0.022$ & $(7.279 \pm 0.019) \cdot 10^{-2} $ & $(1.9236 \pm 0.0036) \cdot 10^{-1} $ & 0.84\\
	\hline
	\hline
	Inform Th. & $a \pm \sigma_a (\frac{counts}{cycles})$ & $b \pm \sigma_b (cycles^{-\frac{1}{2}})$ & $c \pm \sigma_c (\frac{counts}{cycles})$ & $red .\chi^2$\\
	\hline
	10 & $4.549 \pm 0.021$ & $(7.160 \pm 0.016) \cdot 10^{-2} $ & $(2.0031 \pm 0.0036) \cdot 10^{-1} $ & 0.75\\
	\hline
	30 & $4.237 \pm 0.020$ & $(7.500 \pm 0.017) \cdot 10^{-2} $ & $(1.8854 \pm 0.0029) \cdot 10^{-1} $ & 0.64 \\
	\hline
	50 & $3.722 \pm 0.024$ & $(6.939 \pm 0.023) \cdot 10^{-2} $ & $(1.8424 \pm 0.0047) \cdot 10^{-1} $ & 1.24\\
	\hline
	70 & $4.561 \pm 0.023$ & $(7.448 \pm 0.018) \cdot 10^{-2} $ & $(1.9687 \pm 0.0035) \cdot 10^{-1} $ & 0.83\\
	\hline
	90 & $3.983 \pm 0.018$ & $(7.153 \pm 0.016) \cdot 10^{-2} $ & $(1.8396 \pm 0.0032) \cdot 10^{-1} $ & 0.67\\
	\hline
	
    \end{tabular}
	\end{sideways}	
	\caption{Fit result: parameters, errors, reduced chi square values.}
	\label{stabFitTable}
\end{table}

\clearpage

\include{LinNoTimeAnalysis}

\include{FeedbackAnalysis}

\include{FreqTimeAnalysis}

\include{LinTimeAnalysis}

\include{InProgressAnalysis}

\include{EvaluationAnalysis}

\chapter{Conclusions}
\label{capConcl}
The research activities reported in this thesis concern an agent-based model which provides a framework where users can simulate various digital preservation scenarios. The user can test as well the effect of many initial conditions and the behaviour of the system and \-si\-mu\-la\-te those processes in which a digital object faces the risk of obsolescence, a migration process has to be performed and the most appropriate format has to be adopted. 

Agents monitor the system and communicate with each other in order to share information about their internal state and find out the most suitable preservation action to be performed. Information is evaluated and propagated according to a weighting based on the level of trust assigned to both the agents who raised the issue and suggested the solution. 

The definition of the trust level is based on a number of trust parameters such as the cultural and geographical distances, the expertise of the involved agents and the file format numerosity. This choice is not a constraint to the model that can easily be extended to match the users' requirements. The trust network is defined according to educated guesses; different rules may be applied and tested since the deliverable of the present research activity is the modelling framework itself and not the specific models implemented on it.

The objective was both to demonstrate that the framework implementation is stable and to prove how a multi-agent system can either perform an autonomous preservation action or suggest a list of best candidate solutions to the user. It benefits the management of several kinds of digital archive, especially those with limited resources dedicated to preservation such as small personal collections and many public institutions. Moreover the proposed solution can be applied to the ``access copies" of digital contents preserved in digital libraries, such as the copies \-pu\-blished on web sites and electronic shelves, in order to perform automatic updates.

\section{Statistical Analysis Results}
In this section we quickly go through the results of the experiments described in Chapter \ref{using_the_model} and we discuss the results obtained in Chapter \ref{statAnalysis} in order to point out what they proved.

First of all we would like to remind that the aim of this work is to provide a framework that users can adopt to run simulations in order to test specific \-di\-gi\-tal preservation scenarios. Therefore, as explained in Sections \ref{stability} and \ref{stab_noTime}, an essential requirement for such a framework is to have a stable behaviour of the communication system. That was proved by the analysis performed in Section \ref{stab_noTime}, where we demonstrated with twenty different user cases that the frequency of migrations, without any time constraints, faces an exponential decay until an asymptotic value is approached.

Moreover, in Sections \ref{lin_noTime} and \ref{lin_resp_noTime}, we highlighted that not only the framework is stable, but also that a linear response to the initial condition exists. For instance, the asymptotic value approached by the frequency of migrations, mentioned in the previous lines, depends on the probability that at each time step an institution encounters a new format or ceases to adopt one. This probability is one of the many conditions that users can set in order to customize the framework and match their experimental requirements. We proved that, as far as migration time is not an issue, the trend of the asymptotic frequency is linear to the probability for a certain range of values.

The feedback mechanism has been also proved to be stable. In Sections \ref{feedback} and \ref{feed_mech_an}, we analysed the frequency of variations of the \emph{trust weights} matrix and observed that the trend of the frequency of both positive and negative variations faces a slow linear decrease when the system is into a stable condition after a certain number of time steps. The significant difference between the value of positive and negative variation frequency indicates that, even tough they were educated guesses, our assumptions were effective enough to make the system capable of identifying the sources of wrong suggestions and discard them.

A temporal dependence, due to the time needed to perform the migrations, has hence been introduced. We decided to test the behaviour of the framework when a certain number of cycles was associated with each migration. We also endowed the institution with the capability of refusing migrations in case the number of cycles required was too high. In Sections \ref{migFreqTime} and \ref{freqTimeAn} we analysed again the frequency of migrations and observed that its trend faces an exponential decay with increasing time as happened in the stability experiments. This is a significant result because it means that users are allowed to adopt specific temporal units and migration durations without affecting the framework behaviour.

Keeping the temporal dependence, we run other simulations to test whether the linear response was still verified. The analysis performed in Sections \ref{lin_Time} and \ref{lin_resp_time} is similar to the one done without the introduction of the temporal dependence. As mentioned in Section \ref{lin_resp_time}, the duration of migrations affected the almost perfect linear response observed in Section \ref{lin_resp_noTime} with a sort of distortion. On the other hand we observed the same saturation effect when high probability values were adopted.

With the same user case considered in Section \ref{migFreqTime}, where the focus was the trend of migration frequency, we were interested in the migration processes also in Section \ref{in_progress} and \ref{ongoing_analysis}. In this case we monitored the number of those migrations that were being performed by the institution at time. In particular we observed that this value oscillates due to the duration of the migrations. These oscillations, by the way, occur around a constant value which is further confirmation of the stable behaviour of the framework.

The last but most promising results concern the analysis of the agents' decisions. By means of this experiment we noticed how the institution agents seem to learn how to distinguish and choose to perform only useful migrations among the suggested candidates. Moreover, the percentage of good actions and unnecessary migrations (false positives) is complementary at each time step. That means that either indifferent actions or missed migrations (false negatives) never happened. We can thus conclude that the institutions perform an extra action in the worst case, but they never miss a migration that was necessary to solve one of their preservation issues.



\section{Summary and Future Works}
The work presented in this thesis provided a novel approach to the decision processes concerning common digital preservation issues. In particular, we focused on the migration process endowing the agents with the capability of communicating, cooperating and propagating information about the performed actions in order to help each other in finding the most suitable solution to a given preservation issue.

The framework has been designed and implemented with the aim to provide a \-fle\-xi\-ble test bed in which the user is able to simulate several dynamic and distributed digital preservation scenarios and to probe different approaches in defining the trust rules for the network. The stability of the framework has been proved under various user cases. In addiction a significant dependency from the initial conditions has been observed which allows the users to evaluate the effect of different initial rules on the environment. Every set of initial conditions can be introduced by the user in order to verify how they affect the stability of the evolution of such a complex system.

Several models with specific user cases have been tested demonstrating how institutions could benefit from an interaction as the one presented in this work.

The next step could be the design and the implementation of a software application capable of either performing autonomous preservation actions or helping the user in taking decisions about the best preservation strategy to select. A small network may be set up in order to verify that the communication processes take place efficiently so that information can be propagated through the network. The nodes of the network would play the role of the institution agents, keeping inside all the logic for the communication, the decisional processes and also the preservation actions that would no longer be simulated, but executed on actual digital objects for real.



\bibliographystyle{unsrt}

\bibliography{biblio3}

\end{document}

%% file: LinNoTimeAnalysis.tex
\section{Linear Response Range}
\label{lin_resp_noTime}
\begin{figure}[!h]
\centering
\includegraphics[width=11cm]{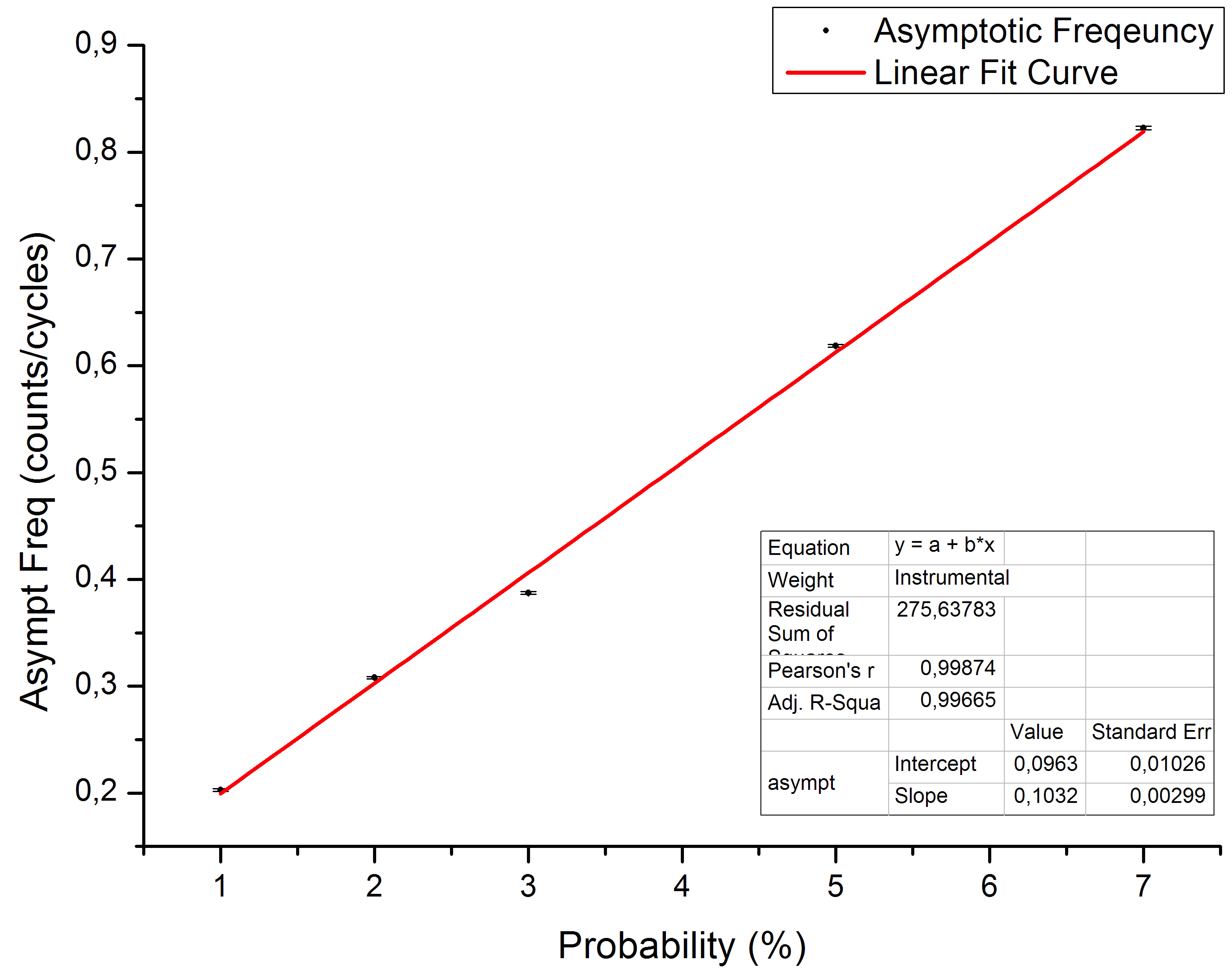}
\caption{Plot of asymptotic values of migrations frequency.}
\label{linNoTime1-7}
\end{figure}

In this section we analyse the data collected in the experiment described in Section \ref{lin_noTime}. First of all we focus on small probability values, which means between 1\% and 7\% that are, in our opinion, the more realistic condition. Figure \ref{linNoTime1-7} shows the fit result.

\begin{table}[!h]	
	\begin{center}
    \begin{tabular}{ |c|c|c|c|}
    \hline
	 $a \pm \sigma_a (\frac{counts}{cycles})$ & $b \pm \sigma_b (\frac{counts}{cycles \cdot probability}) $ & $\chi ^2$ & $\rho$\\
	\hline
	\hline
	$0.0963 \pm 0.0103 $ & $ 0.1032 \pm 0.0030$ & 275.64 & 0.999 \\
  	\hline 
    \end{tabular}
	\end{center}	
	\caption{Linear fit result.}
	\label{LinNoTimeFitTable}
\end{table}

Looking at the fit result in Table \ref{LinNoTimeFitTable}, the following considerations could be done: though the linear fit curve seems to fit well our data, the chi square value is very high, in particular it equals 275.64 with 5 data. This is due to the fact that the errors are three orders of magnitude smaller than the data. These errors are \-e\-va\-lu\-ated by the analysis software when the frequency of migration is fitted with the exponential function \ref{exp} described in Section \ref{stab_noTime}. On the other hand we notice that the value of Pearson's coefficient is higher than 0.99 which means we are allowed to assert that the linearity of our data is verified.

\begin{figure}[!h]
\centering
\includegraphics[width=11cm]{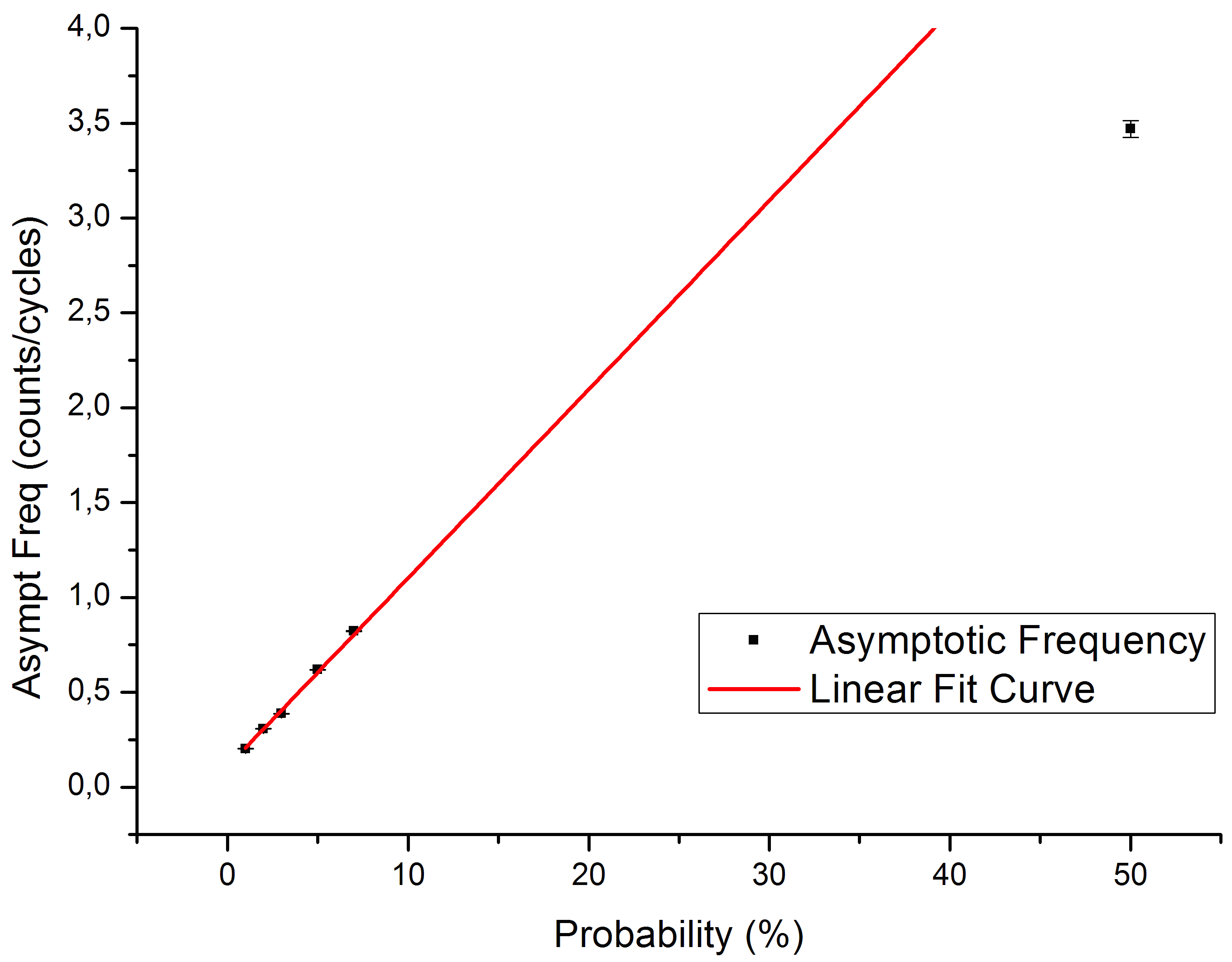}
\caption{Plot of asymptotic values of migrations frequency with 50\% probability.}
\label{linNoTime1-50}
\end{figure}

As indicated in Table \ref{tab_lin_noTime}, we also run a simulation with a 50\% probability value. With such a high probability we expect a slower exponential decay and also a higher asymptotic frequency value due to the increased rate of creation and deletion of \emph{format collections}. In Figure \ref{linNoTime1-50} we see how the 50\% probability point is below the linear fit curve. 

This behaviour allows us to assume that a sort of saturation occurs when high probability values are adopted. We remind that all these experiments are performed without any time constraint on migration processes. As we shall see later the temporal aspect may cause some distortions.

%% file: FeedbackAnalysis.tex
\section{Feedback Mechanism}
\label{feed_mech_an}

The objective of the analysis of the simulations described in Section \ref{feedback} is to check if the feedback system presents a stable behaviour or not. As already explained we run a 10000 cycles simulation and consider the frequency of the total variations, positive and negative variations of the \emph{trust\_weights} matrix. As Figure \ref{trust_var} depicts these quantities stabilize with increasing time. First of all we fit our data in the region between 1000 and 5000 cycles. After that we verify the stable trend in the region from 5000 to 10000 cycles with a linear fit. In both cases the expression of the fit function is the following:

\begin{equation}
y(t) = a + b \cdot t
\end{equation} 

As happens for the frequency of migrations, also the number of variations is a count therefore the quantities and their related errors have the following expression:

\begin{equation}
frequency = \frac{vatiations}{time} \pm \frac{\sqrt{variations}}{time}
\label{error}
\end{equation}  

\begin{figure}[!h]
\centering
\includegraphics[width=12cm]{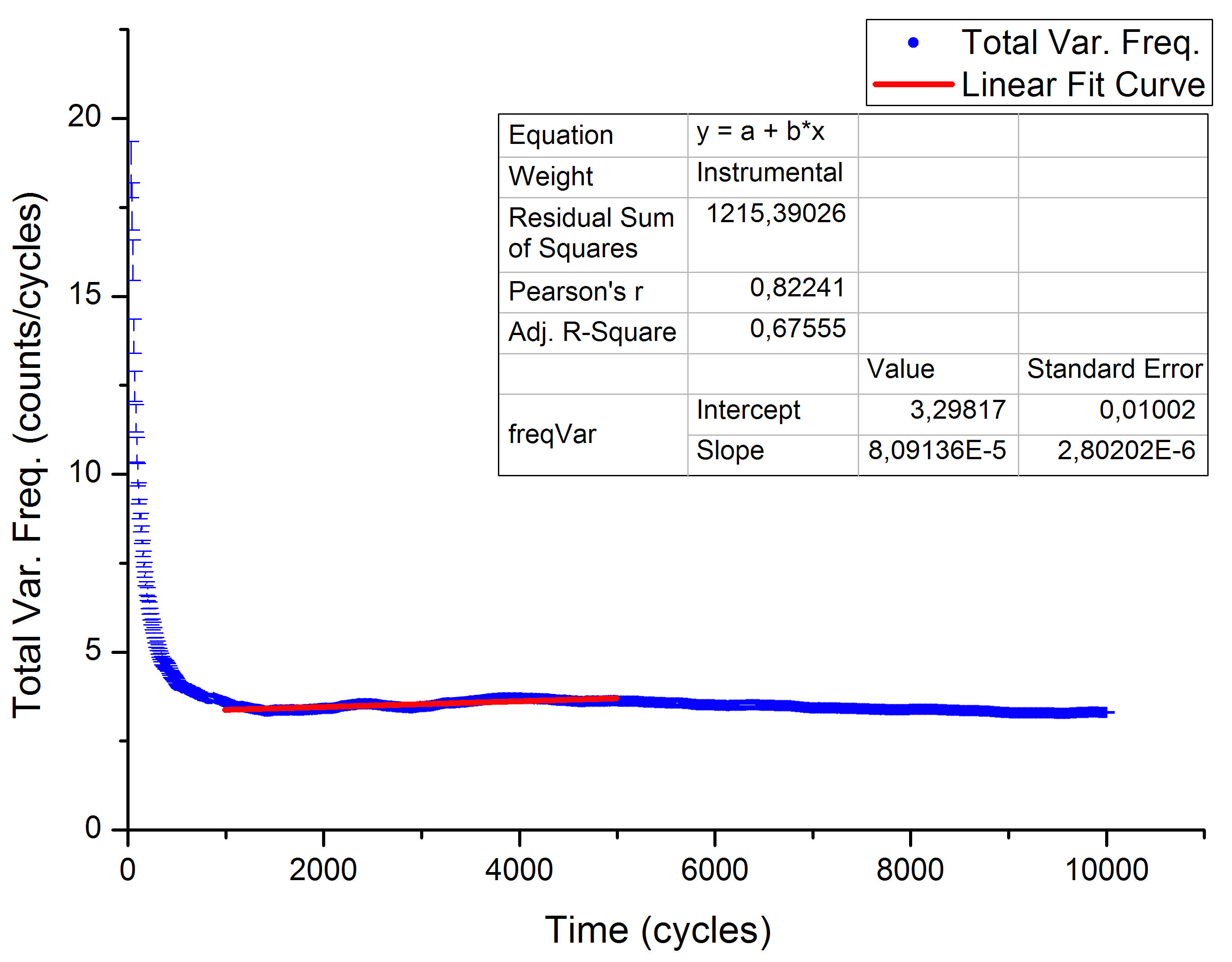}
\caption{Plot of total variations frequency.}
\label{totVarFreq2}
\end{figure}

\begin{figure}[!h]
\centering
\includegraphics[width=12cm]{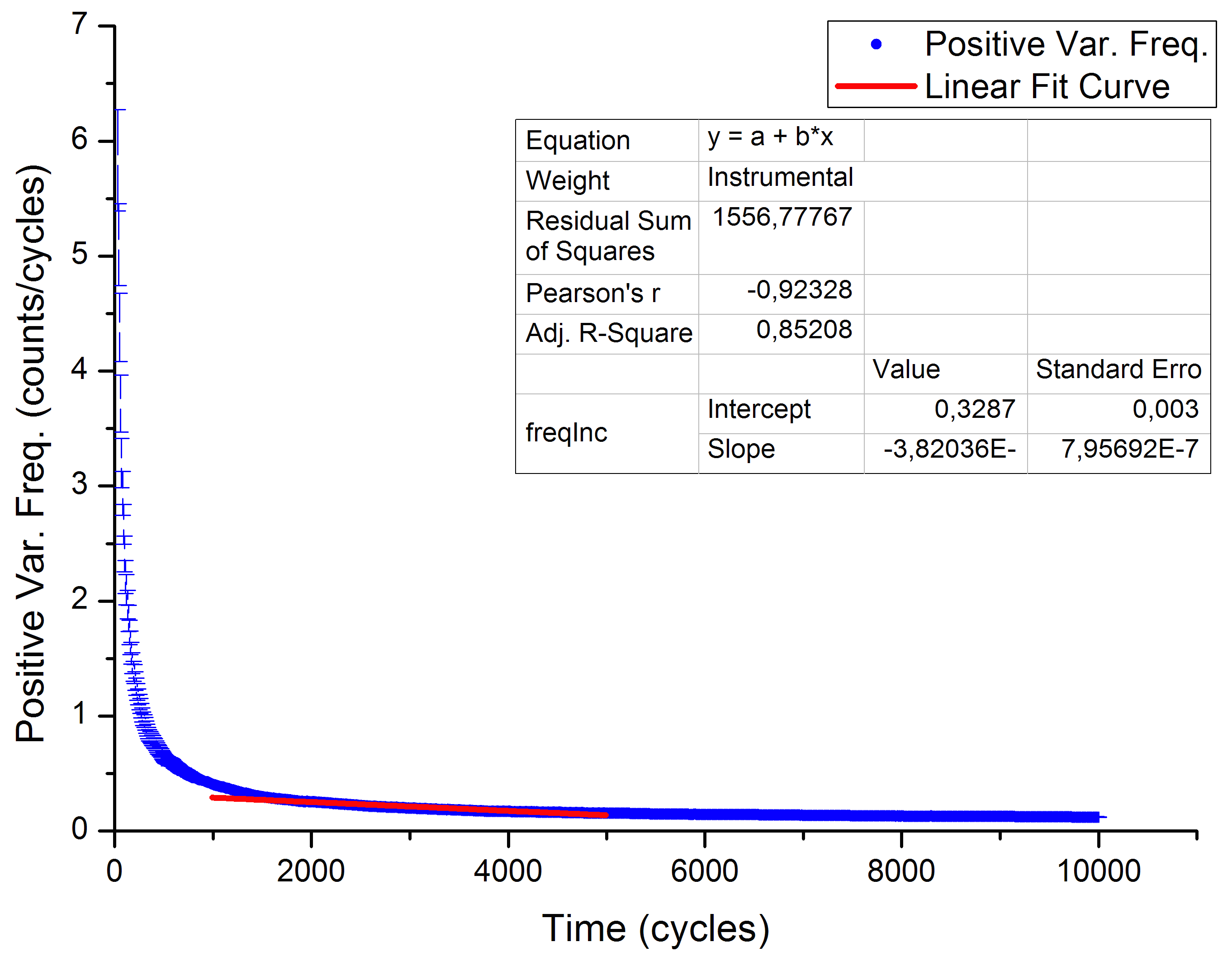}
\caption{Plot of positive variations frequency.}
\label{posVarFreq2}
%
\centering
\includegraphics[width=12cm]{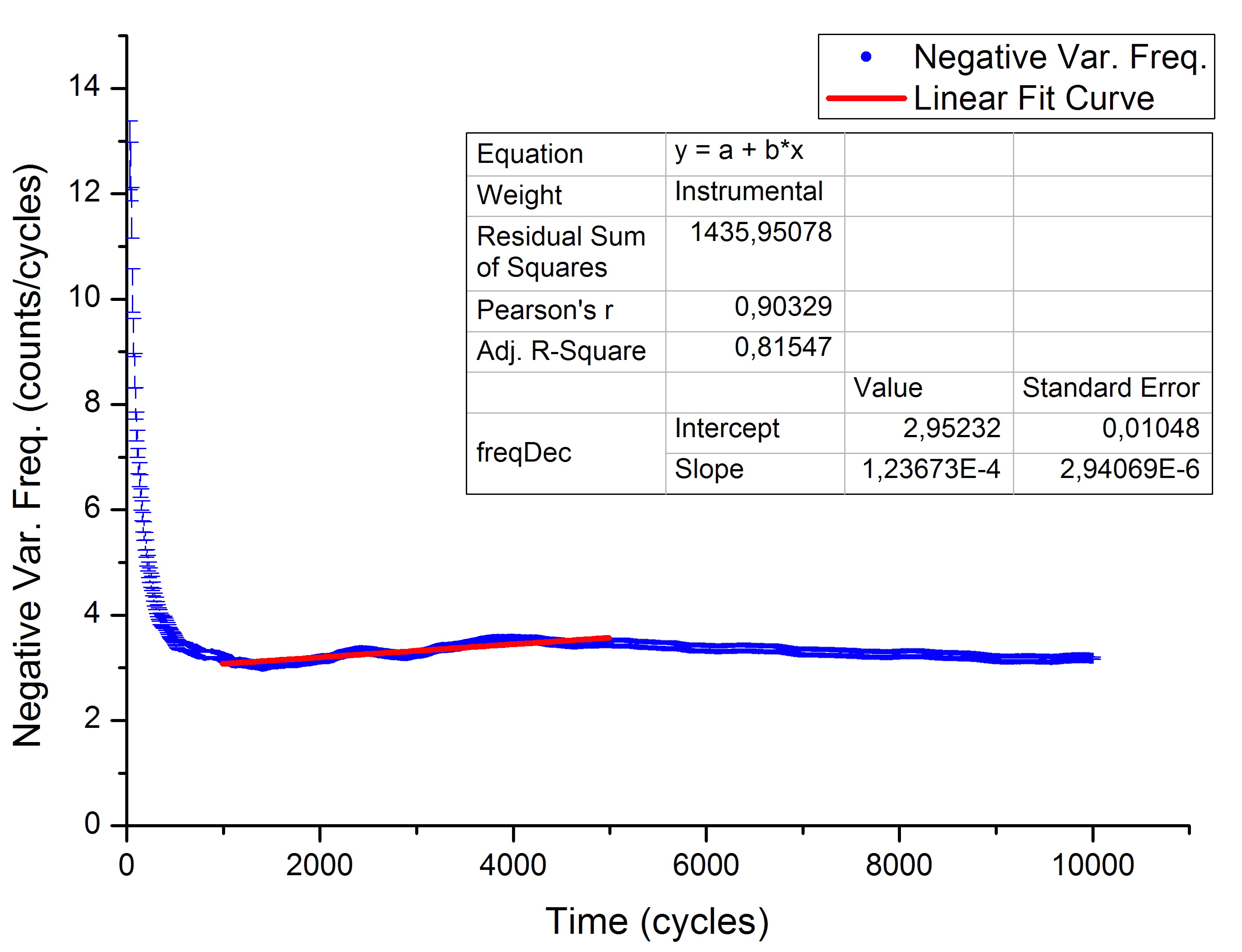}
\caption{Plot of negative variations frequency.}
\label{negVarFreq2}
\end{figure}

The plot reported in Figure \ref{totVarFreq2} shows that the linear fit of the total variations frequency is not very meaningful in this first region. Even tough the Pearson's coefficient equals 0.82, which could be acceptable, the reduced chi square value is very high, in particular it equals 3.04. Similar results come from the analysis of both positive and negative variation frequency. As far as positive variations are concerned the fit reported in Figure \ref{posVarFreq2} returns a negative Pearson's coefficient, in particular -0.92, and again a high reduced chi square value that equals 3.89. The last fit in Figure \ref{negVarFreq2} is about the the negative variations frequency returns a Pearson's coefficient equals to 0.903 and a reduced chi square value 3.59. All this results are reported in Table \ref{varFitTab2}. Of course the results of this analysis are not satisfying and thus we can only affirm that the trend is neither constant nor linear. 

\begin{table}[!h]	
	\begin{center}
    \begin{tabular}{ |c|c|c|c|}
    \hline
    Fit Results  & Total & Positive & Negative  \\   
    \hline
	\hline	
	a ($\frac{counts}{cycles}$) & 3.30 & $3.287 \cdot 10^{-1}$ & 2.9532  \\
	\hline
	$\sigma_{a}$ ($\frac{counts}{cycles}$) & $0.01$ & $3.0 \cdot 10^{-3}$ &  $1.05\cdot 10^{-2}$ \\
	\hline
	b ($\frac{counts}{cycles^2}$) & $8.09\cdot 10^{-5}$ & $3.82 \cdot 10^{-6}$ & $1.237 \cdot 10^{-4}$  \\
	\hline
	$\sigma_{b}$ ($\frac{counts}{cycles^2}$) & $2.8 \cdot 10^{-6}$ & $8.0 \cdot 10^{-7}$ & $2.9 \cdot 10^{-6}$  \\
	\hline  
	$\rho$ & 0.82 & -0.92 & 0.90  \\
	\hline
	red. $\chi^2$ & 3.04 & 3.89 & 3.59  \\
	\hline            
    \end{tabular}
	\end{center}	
	\caption{Summary of fit results.}
	\label{varFitTab2}
\end{table}

\clearpage

\begin{figure}[!h]
\centering
\includegraphics[width=12cm]{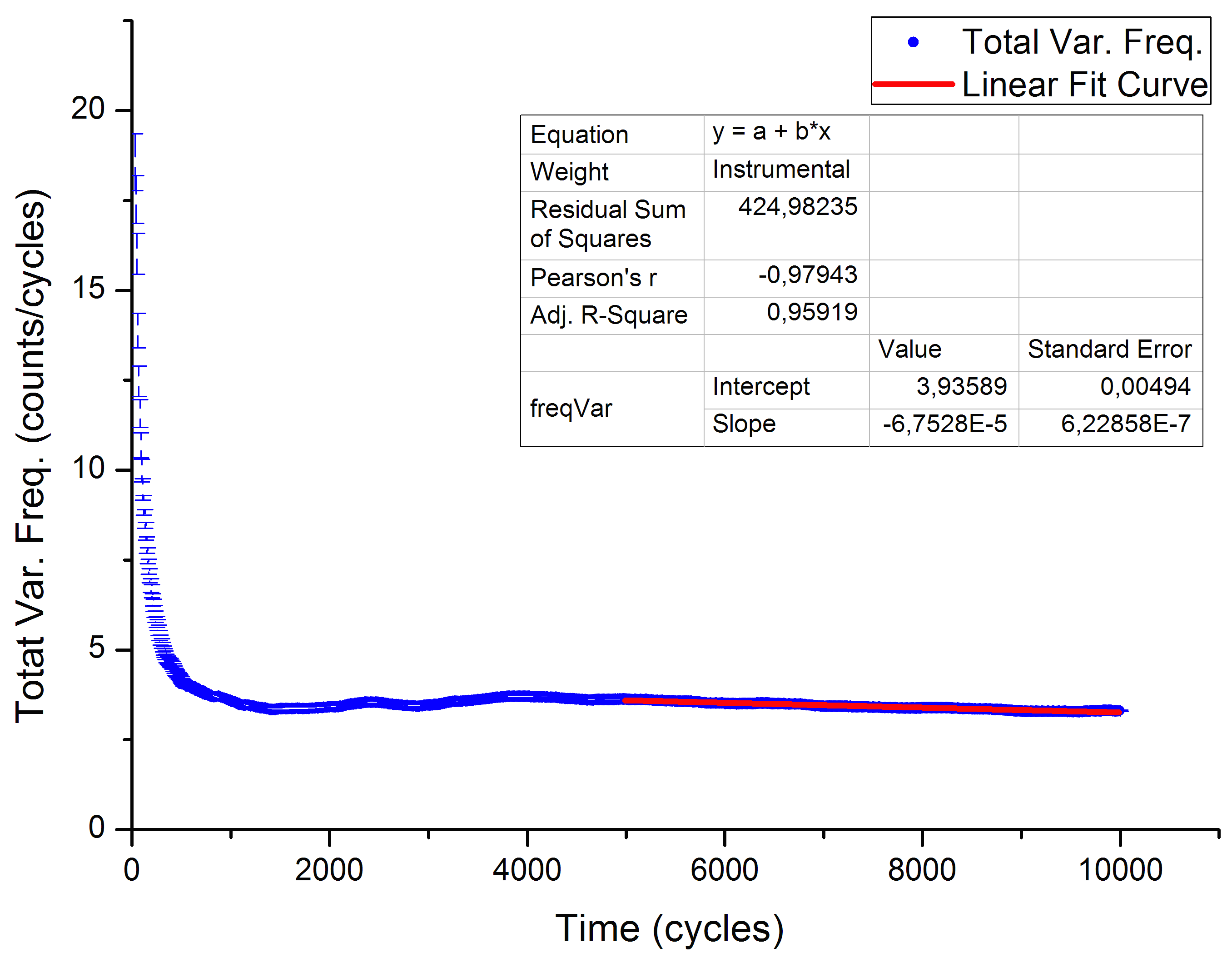}
\caption{Plot of total variations frequency.}
\label{totVarFreq}
%
\centering
\includegraphics[width=12cm]{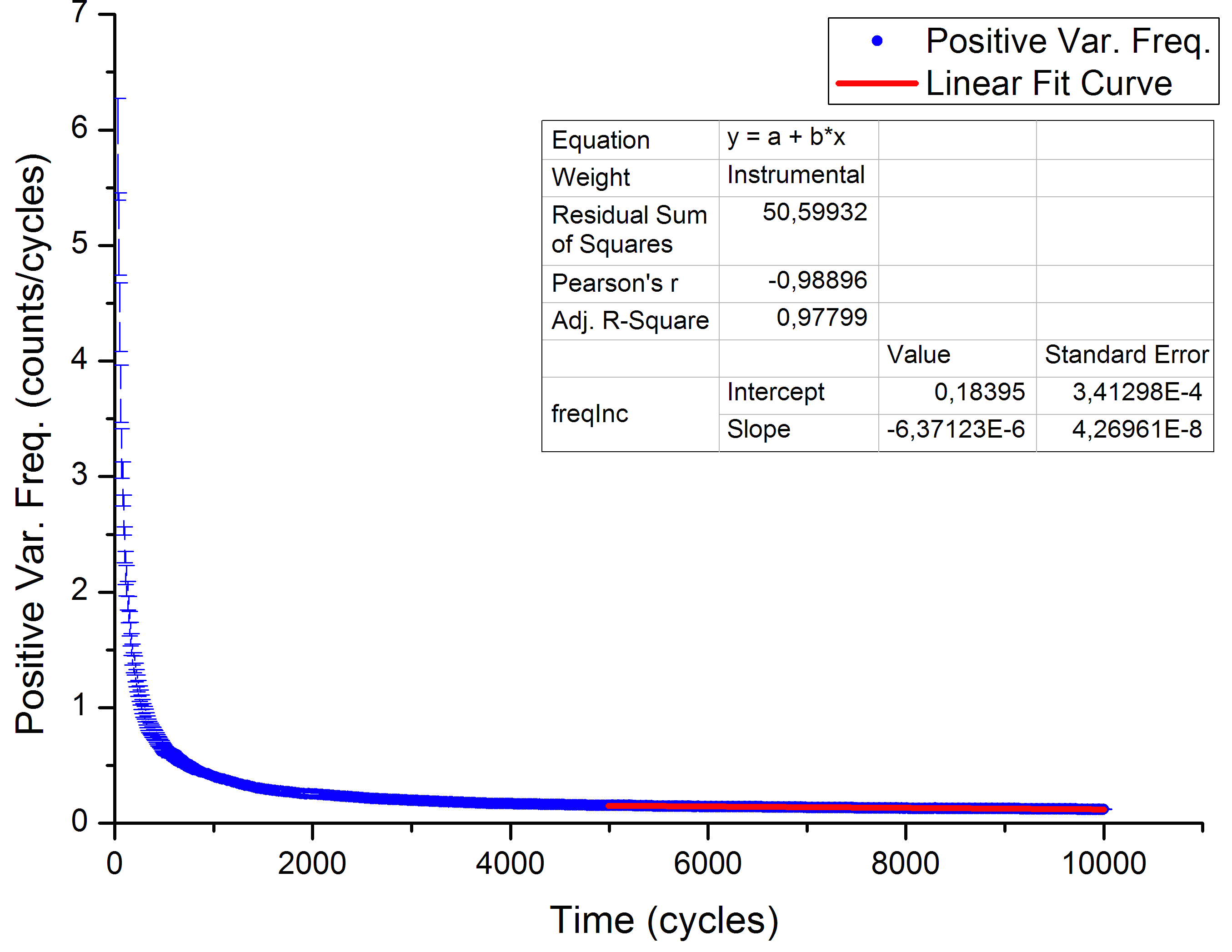}
\caption{Plot of positive variations frequency.}
\label{posVarFreq}
\end{figure}

We thus proceed considering the second region. After 5000 cycles we expect our system to stabilize and the trend to be linear. Figure \ref{totVarFreq} shows the linear fit concerning the frequency of total variations. The linearity is confirmed by the high absolute values of the Pearson's coefficient which equals 0.98. Moreover, since we measure the number of variations every 10 cycles, we are considering 500 data thus the reduced chi square value equals 0.85. 

We can now separately focus on the positive and negative variations. Figure \ref{posVarFreq} depicts the result of the linear fit of the positive variations. The absolute value of the Pearson's coefficient here equals 0.99 that indicates how the experimental trend matches with the linear one. Most of all, the reduced chi square value equals 0.10 which proves the goodness of the fit.

\begin{figure}[!h]
\centering
\includegraphics[width=12cm]{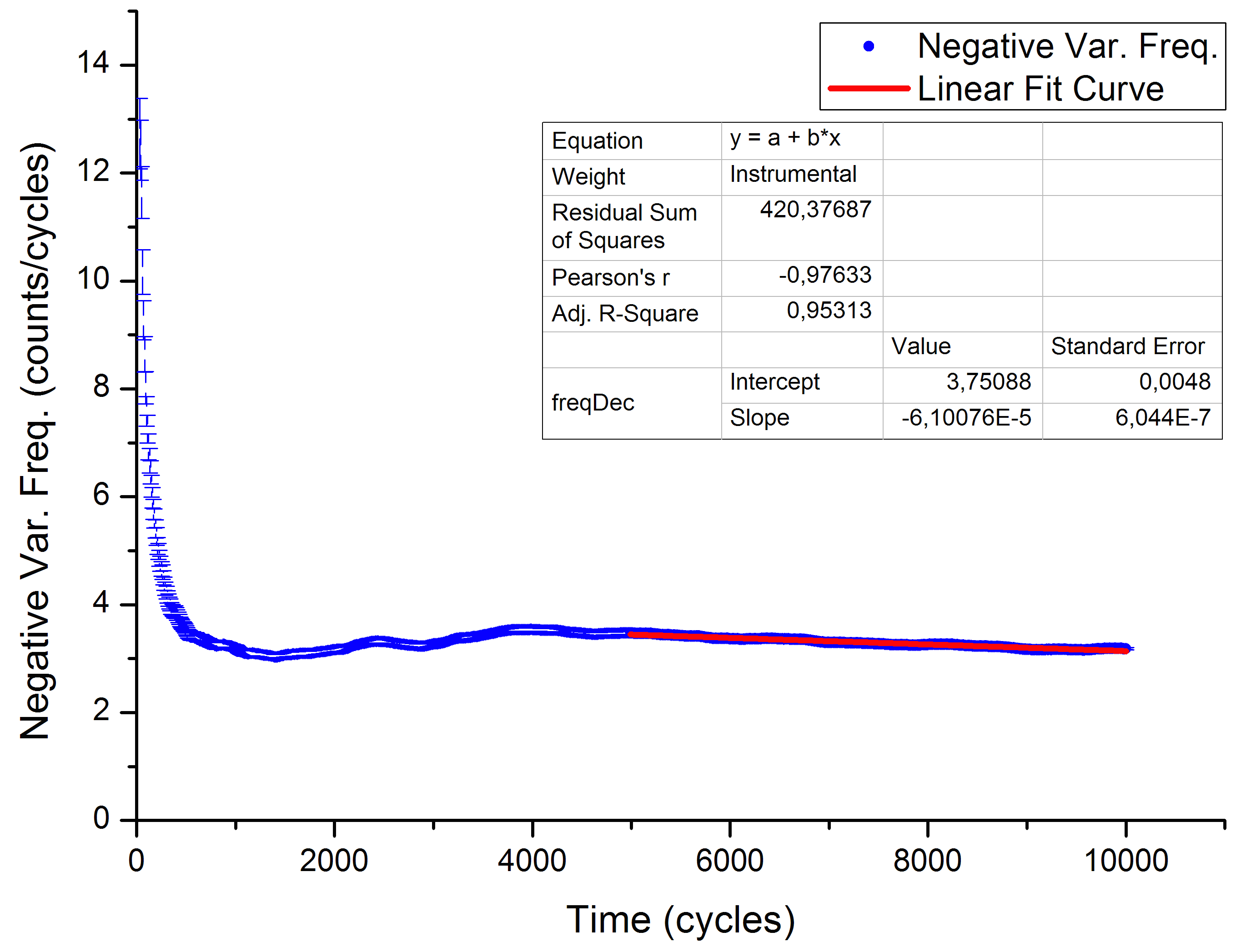}
\caption{Plot of negative variations frequency.}
\label{negVarFreq}
\end{figure}

As far as the negative variations are concerned, the linear fit is shown in Figure \ref{negVarFreq}. In this case the absolute value of the Pearson's coefficient equals 0.98 and the reduced chi square value is higher than the previous case but still below one. In particular it equals 0.84. We sum up the fit results in Table \ref{varFitTab}.

\begin{table}[!h]	
	\begin{center}
    \begin{tabular}{ |c|c|c|c|}
    \hline
    Fit Results  & Total & Positive & Negative  \\   
    \hline
	\hline	
	a ($\frac{counts}{cycles}$) & 3.9359 & $1.8395 \cdot 10^{-1}$ & 3.7509  \\
	\hline
	$\sigma_{a}$ ($\frac{counts}{cycles}$) & $4.9 \cdot 10^{-3}$ & $3.4 \cdot 10^{-4}$ &  $4.8 \cdot 10^{-3}$ \\
	\hline
	b ($\frac{counts}{cycles^2}$) & $-6.753 \cdot 10^{-5}$ & $-6.371 \cdot 10^{-6}$ & $-6.1008 \cdot 10^{-5}$  \\
	\hline
	$\sigma_{b}$ ($\frac{counts}{cycles^2}$) & $6.2 \cdot 10^{-7}$ & $4.3 \cdot 10^{-8}$ & $6.04 \cdot 10^{-7}$  \\
	\hline  
	$\rho$ & -0.98 & -0.99 & -0.98  \\
	\hline
	red. $\chi^2$ & 0.85 & 0.10 & 0.84  \\
	\hline            
    \end{tabular}
	\end{center}	
	\caption{Summary of fit results.}
	\label{varFitTab}
\end{table}

We also notice that the intercept value is one order of magnitude higher than the intercept of the positive case. That means that the system is effective in \-re\-co\-gni\-zing wrong suggestions and is conservative since all those suggestions that are not suitable enough for the agent are discarded. This considerations tell us that all the assumptions and the hypothesis made in the design process are meaningful. 

\clearpage

%% file: FreqTimeAnalysis.tex
\section{Frequency of ``Time-costing" Migrations}
\label{freqTimeAn}

Section \ref{stab_noTime} of this chapter contains an extensive analysis which proves the stability of the agent system. The stability has been tested under many conditions but with the assumption that migrations required only one cycle to be performed. As we already mentioned in Section \ref{migFreqTime} we need to introduce a temporal dependence and observe its effect on the system. In this section we analyse the frequency of migrations in a 10000 cycles simulation to check if the system is still stable.

At first we decided to fit the whole range of data fit the exponential function reported in \ref{exp}. The fit between 200 and 10000 cycles did not converge probably due to the evident linear trend for high number of cycles values. We thus split the data into two regions of interest: the first one involves data between 200 and 5000 cycles, the second one between 5000 and 10000. As far as the first region is concerned, we fit our data with the exponential function \ref{exp} adopted in Section \ref{stab_noTime} that we recall here:

\begin{equation}
f(t) = a \cdot \exp (-b \cdot \sqrt{t}) + c 
\label{exp}
\end{equation} 

\begin{figure}[!h]
\centering
\includegraphics[width=13cm]{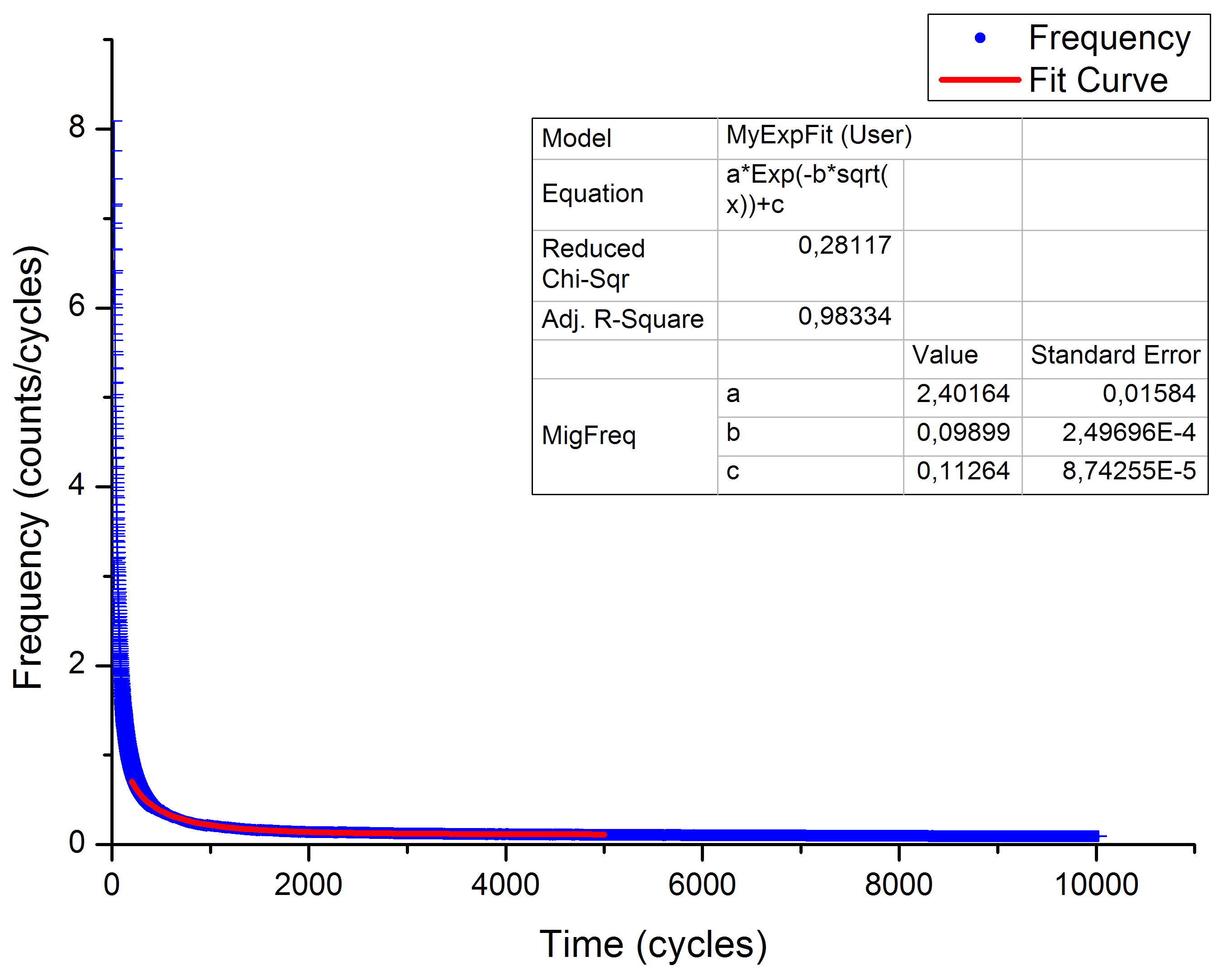}
\caption{Plot of exponential fit of the first ROI.}
\label{migFreqTimeExpFit}
\end{figure}

Figure \ref{migFreqTimeExpFit} shows how the experimental data are consistent with the fit function. The reduced chi square value equals 0.28 so the goodness of the fit is confirmed. This let us assess that, even though now migrations require a certain number of cycles to be performed, their frequency still faces an exponential decay. Therefore the agents need to perform less migration with increasing time because they solved most of their preservation issues by means of the solutions coming from the global interaction. The fit result is reported in Table \ref{expFreqTimeFitTable}.

\begin{table}[!h]	
	\begin{center}
    \begin{tabular}{ |c|c|c|c|}
    \hline
	$a \pm \sigma_a (\frac{counts}{cycles})$ & $b \pm \sigma_b (cycles^{-\frac{1}{2}})$ & $c \pm \sigma_c (\frac{counts}{cycles})$ & $red .\chi^2$\\ 
	\hline
	\hline
	$2.402 \pm 0.016 $ & $(9.899 \pm 0.025) \cdot 10^{-2}$ & $(1.12640 \pm 0.00087) \cdot 10^{-1}$ & 0.28 \\
  	\hline 
    \end{tabular}
	\end{center}	
	\caption{Exponential fit result.}
	\label{expFreqTimeFitTable}
\end{table}

\begin{figure}[!h]
\centering
\includegraphics[width=13cm]{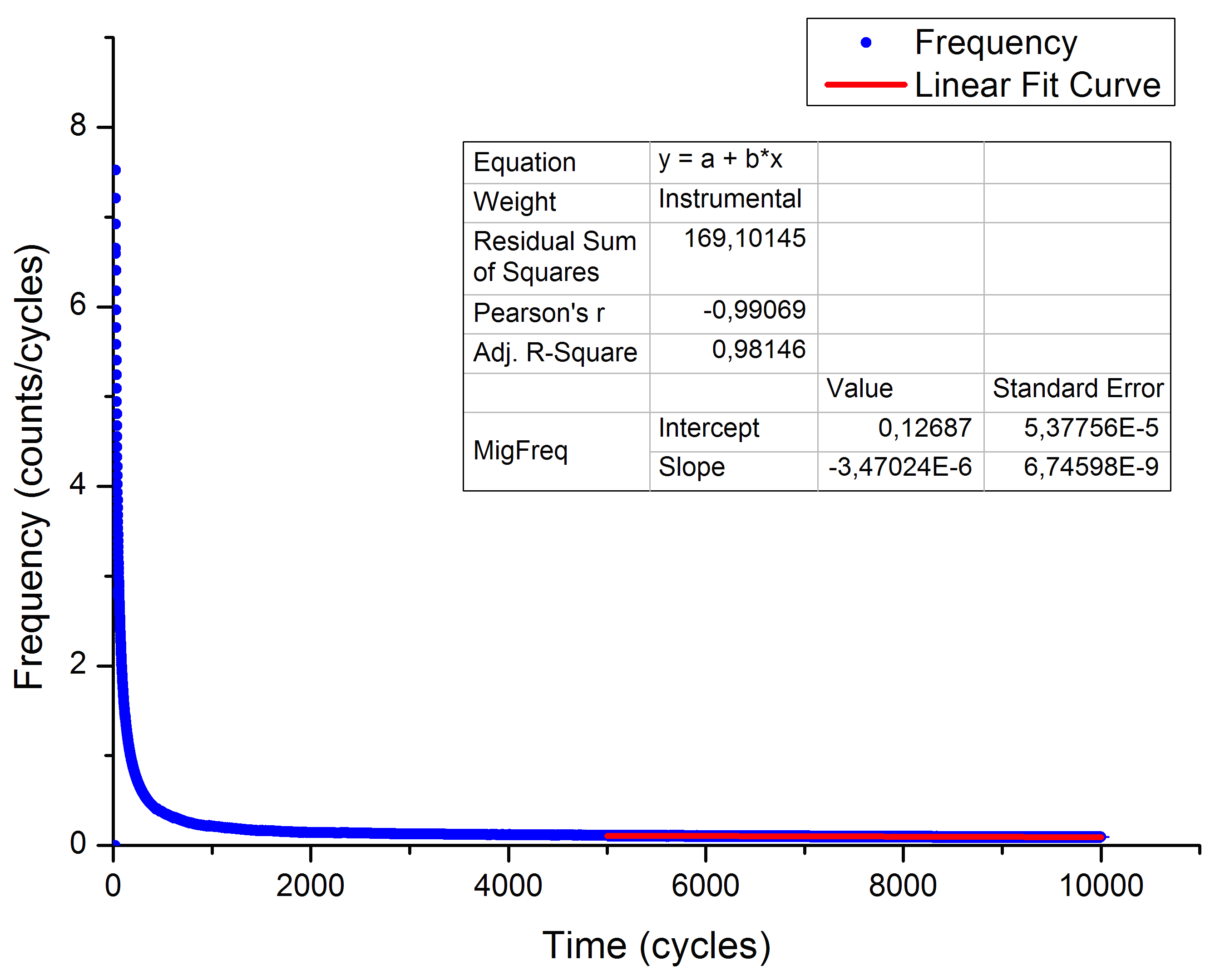}
\caption{Plot of exponential fit of the second ROI.}
\label{migFreqTimeLinFit}
\end{figure}

The second region of interest depicts the condition of the system after a large number of time steps. Under this conditions the system has already reached a state of stability where the frequency of migrations approaches its asymptotic values. The aim of the analysis of these data is to estimate the asymptotic value by means of a linear fit. Figure \ref{migFreqTimeLinFit} shows the linear fit and Table \ref{linFreqTimeFitTable} the result.

The high Pearson's coefficient, equal to -0.99, confirm that the trend of the data is linear. The fit parameters give us other interesting information: the intercept equals $1.26870 \cdot 10^{-1} \pm 5.4 \cdot 10^{-5} counts \cdot cycles^{-1}$ which is the asymptotic frequency of migrations value. 

The second parameter is the slope that equals $-3.4702 \cdot 10^{-6} \pm 6.9 \cdot 10^{-9} counts \cdot cycles^{-2}$. The slope indicates the speed with which the frequency of migrations increases, in particular its value is negative which means the the frequency decreases with increasing time. 

\begin{table}[!h]	
	\begin{center}
    \begin{tabular}{ |c|c|c|c|}
    \hline
	$a \pm \sigma_a (\frac{counts}{cycles})$ & $b \pm \sigma_b (\frac{counts}{cycles^2}) $ & $\chi ^2$ & $\rho$\\
	\hline
	\hline
	$(1.26870 \pm 0.00054) \cdot 10^{-1} $ & $(-3.4702 \pm 0.0067) \cdot 10^{-6}$ & 0.0038  & -0.99 \\
  	\hline 
    \end{tabular}
	\end{center}	
	\caption{Linear fit result.}
	\label{linFreqTimeFitTable}
\end{table}

As anticipated in Section \ref{migFreqTime} we run other simulations to verify that the stability observed also persists when various initial conditioned are applied. In particular we change the probability parameter in the range from 1 to 7\%, range in which the framework proved to be linear as explained in Section \ref{lin_noTime}. To do so, we follow the same procedure described in the previous lines performing an exponential fit between 200 and 5000 cycles and a linear fit between 5000 and 10000 cycles. For each probability value we run four simulations, we report in Table \ref{chiSquare} the reduced chi square values of both the exponential and the linear fit.

Since our goal is to determine the exponential and linear correlation and not a precise estimation of the fit parameters and their errors, we do not proceed to evaluate again the considered experimental point error bars.

\begin{table}[!h]	
	\begin{center}
    \begin{tabular}{ |c|c|c|c|}
    \hline
	Simulation nb. & Probability (\%) & Exp. $red \chi^2 $ & Lin. $red \chi^2 $ \\
	\hline
	\hline
1	&	1	&	0.22	&	0.016	\\
\hline							
2	&	2	&	0.08	&	0.038	\\
\hline							
3	&	3	&	0.07	&	0.041	\\
\hline							
4	&	5	&	0.11	&	0.023	\\
\hline							
5	&	7	&	0.03	&	0.027	\\
\hline							
6	&	1	&	0.19	&	0.024	\\
\hline							
7	&	2	&	0.04	&	0.024	\\
\hline							
8	&	3	&	0.10	&	0.018	\\
\hline							
9	&	5	&	0.05	&	0.015	\\
\hline							
10	&	7	&	0.06	&	0.016	\\
\hline							
11	&	1	&	0.10	&	0.031	\\
\hline							
12	&	2	&	0.16	&	0.015	\\
\hline							
13	&	3	&	0.11	&	0.023	\\
\hline							
14	&	5	&	0.11	&	0.040	\\
\hline							
15	&	7	&	0.04	&	0.043	\\
\hline							
16	&	1	&	0.11	&	0.048	\\
\hline							
17	&	2	&	0.06	&	0.023	\\
\hline							
18	&	3	&	0.03	&	0.036	\\
\hline							
19	&	5	&	0.07	&	0.016	\\
\hline							
20	&	7	&	0.05	&	0.070	\\
\hline							
				
    \end{tabular}
	\end{center}	
	\caption{Reduced $\chi^2$ values.}
	\label{chiSquare}
\end{table}

We notice how the reduced chi square values are always below 1 as far as the exponential fit is concerned. Moreover the values are even smaller for the linear fit which means that both the exponential and the linear fit are meaningful. The following Figures \ref{ExpChi2} and \ref{LinChi2} show the distribution of these values.

The functional dependences that we have considered, respectively exponential and linear, correctly interprete hence our data on a regular basis, and not for specific ``lucky" simulations performed with our model. 

\begin{figure}[!h]
\centering
\includegraphics[width=13cm]{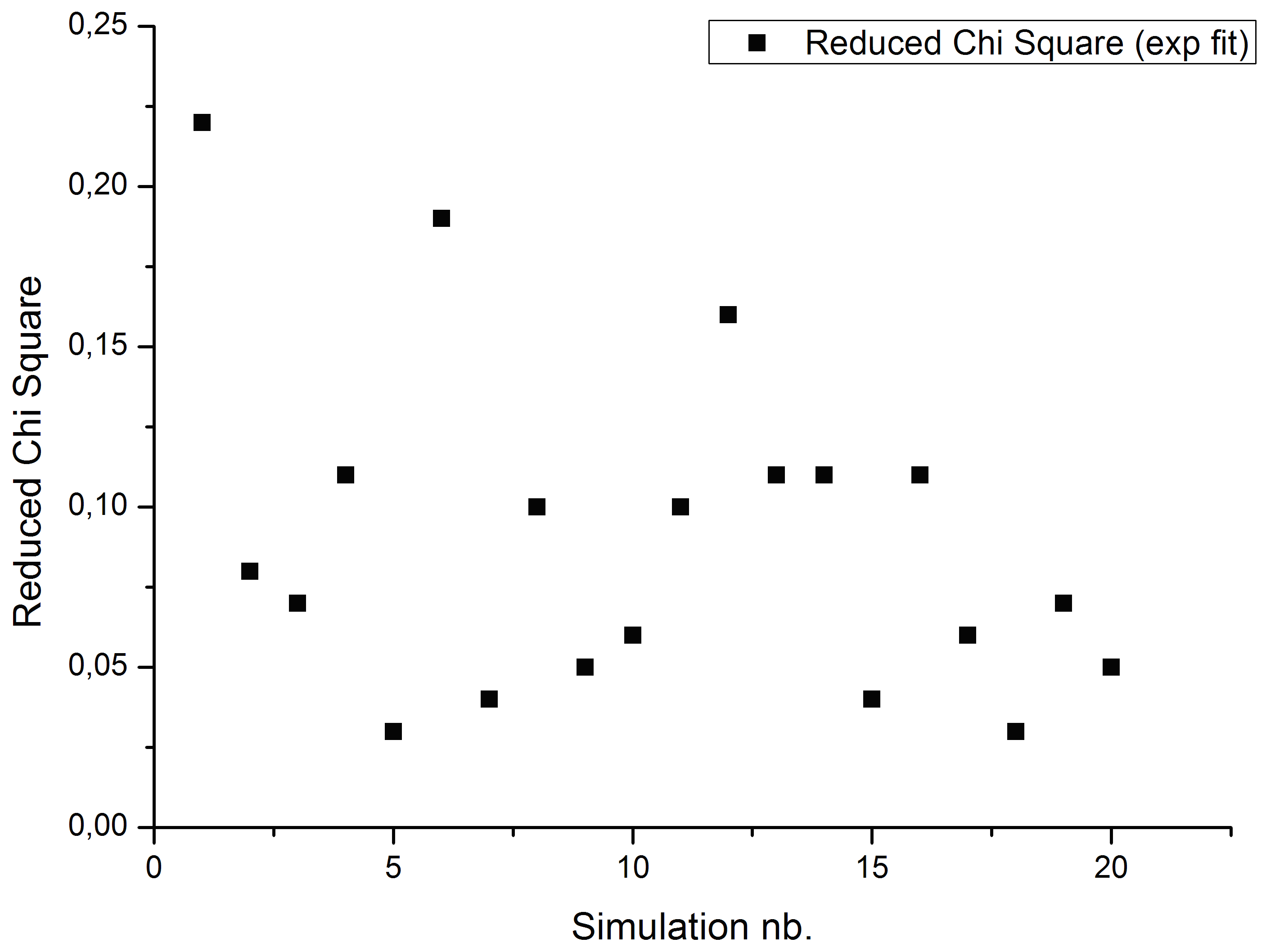}
\caption{Reduced $\chi^2$ values of the exponential fit.}
\label{ExpChi2}
%
%
%
\centering
\includegraphics[width=13cm]{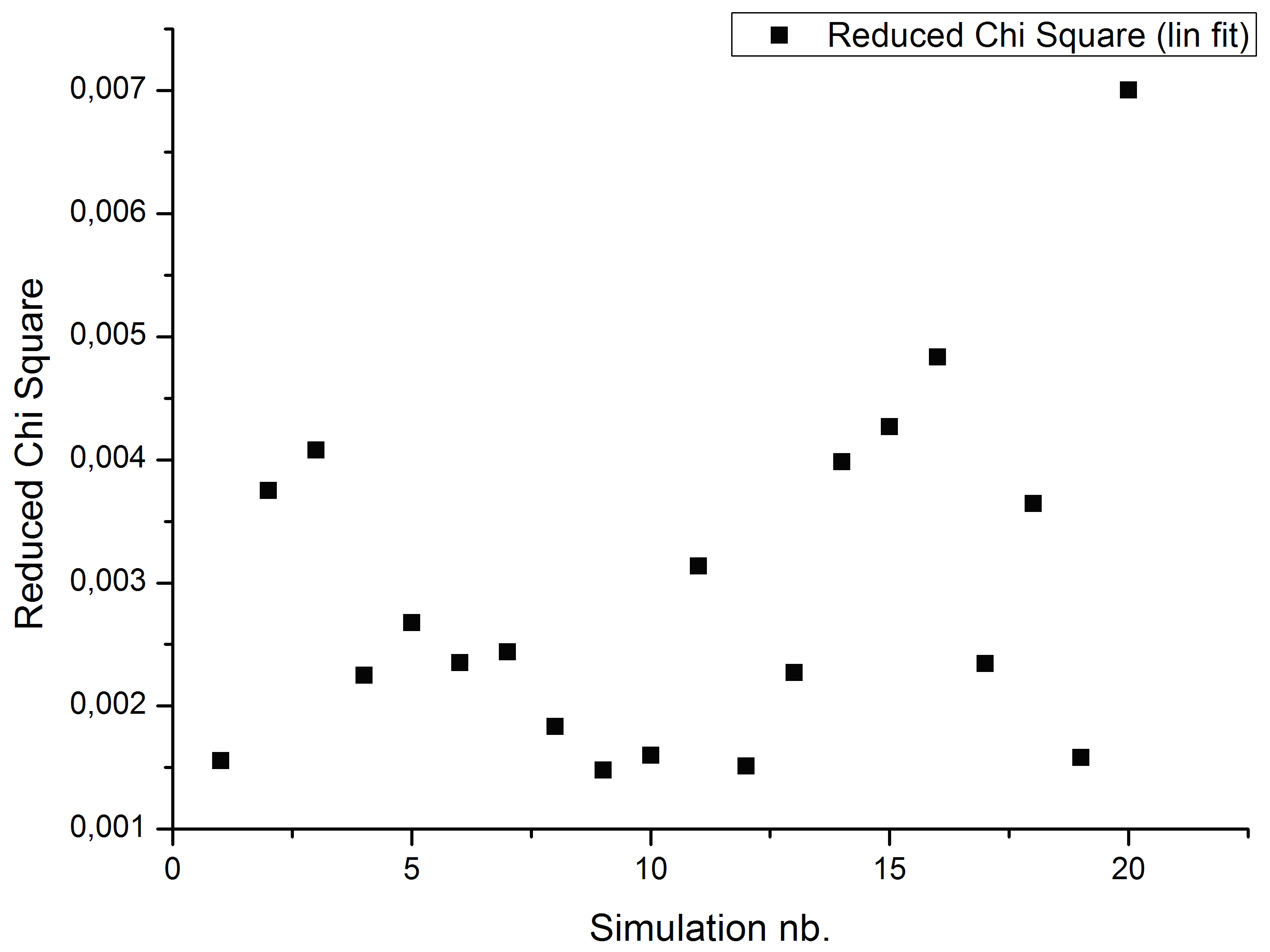}
\caption{Reduced $\chi^2$ values of the linear fit.}
\label{LinChi2}
\end{figure}

%% file: LinTimeAnalysis.tex
\section{Linear Response Range (``Time-costing" Migrations)}
\label{lin_resp_time}
\begin{figure}[!h]
\centering
\includegraphics[width=13cm]{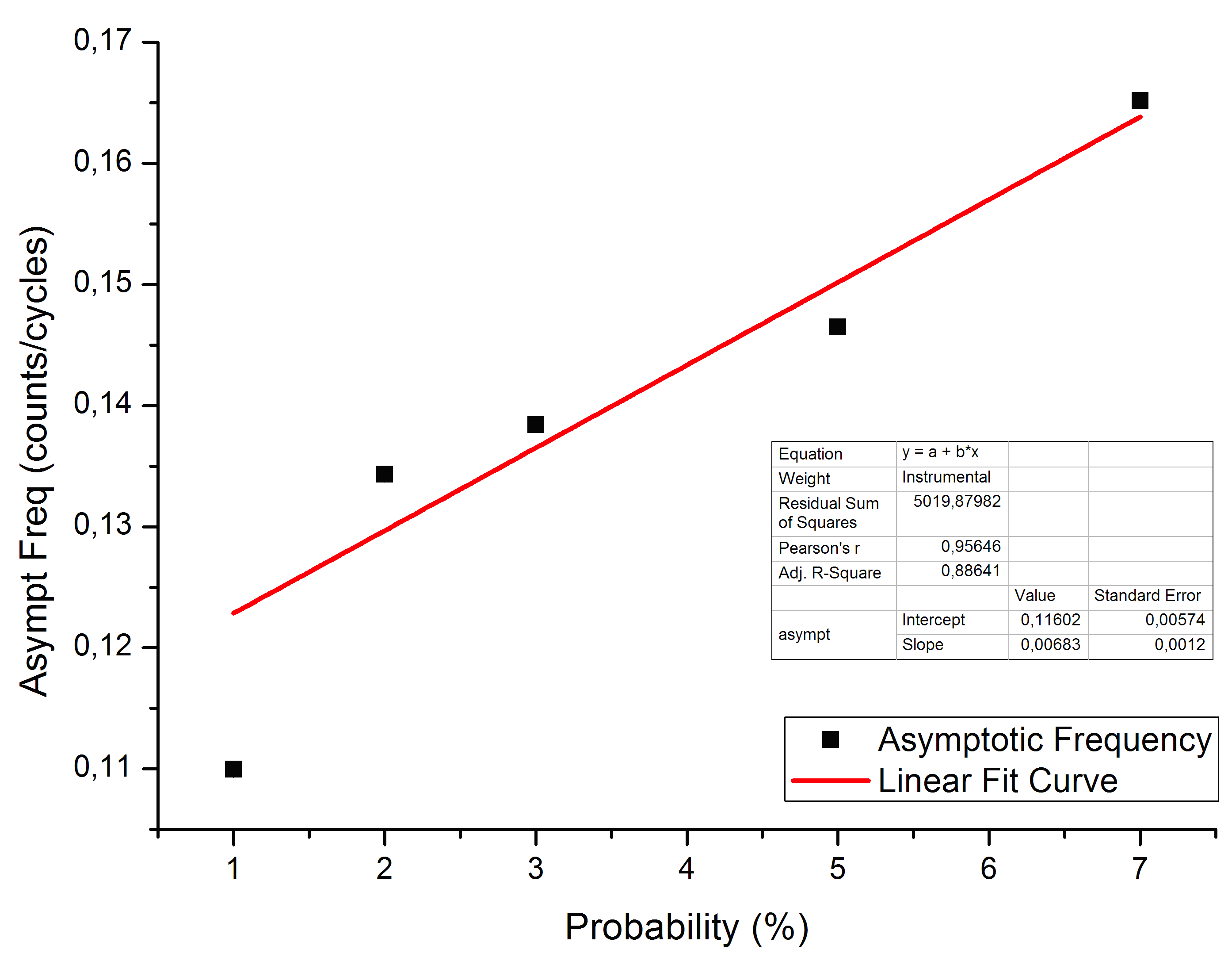}
\caption{Plot of asymptotic values of migrations frequency with low probability values.}
\label{linTime1-7}
\end{figure}

The introduction of time-costing migrations certainly is a significant change in the model. In this section we analyse the data acquired with the experiment discussed in Section \ref{lin_Time}. Following the procedure described in Section \ref{lin_resp_noTime} we fit the data concerning low creation and deletion probability values. Figure \ref{linTime1-7} reports the fitted data.

With respect to the fit depicted in Figure \ref{linNoTime1-7} we notice how the introduction of the temporal aspect causes a slight distortion from the quite perfect linear behaviour observed in Section \ref{lin_resp_noTime}. That is the effect of long lasting migrations that increase the inertia of the systems. The chi square value is again very high due to the small errors associated with our data. By the way the Pearson's coefficient equals 0.96, good enough to confirm the linear behaviour even if affected by the slight distortion just mentioned. Since our goal is to determine the linear correlation and not a precise estimation of the fit parameters and their errors, we do not proceed to revaluate the considered experimental point error bars. Table \ref{LinTimeFitTable} reports the fit result.

\begin{table}[!h]	
	\begin{center}
    \begin{tabular}{ |c|c|c|c|}
    \hline
	 $a \pm \sigma_a (\frac{counts}{cycles})$ & $b \pm \sigma_b (\frac{counts}{cycles \cdot probability}) $ & $\chi ^2$ & $\rho$\\
	\hline
	\hline
	$(1.160 \pm 0.057) \cdot 10^{-1} $ & $ (6.8 \pm 1.2) \cdot 10^{-3}$ & 5019.88 & 0.956 \\
  	\hline 
    \end{tabular}
	\end{center}	
	\caption{Linear fit result.}
	\label{LinTimeFitTable}
\end{table}

\begin{figure}[!h]
\centering
\includegraphics[width=13cm]{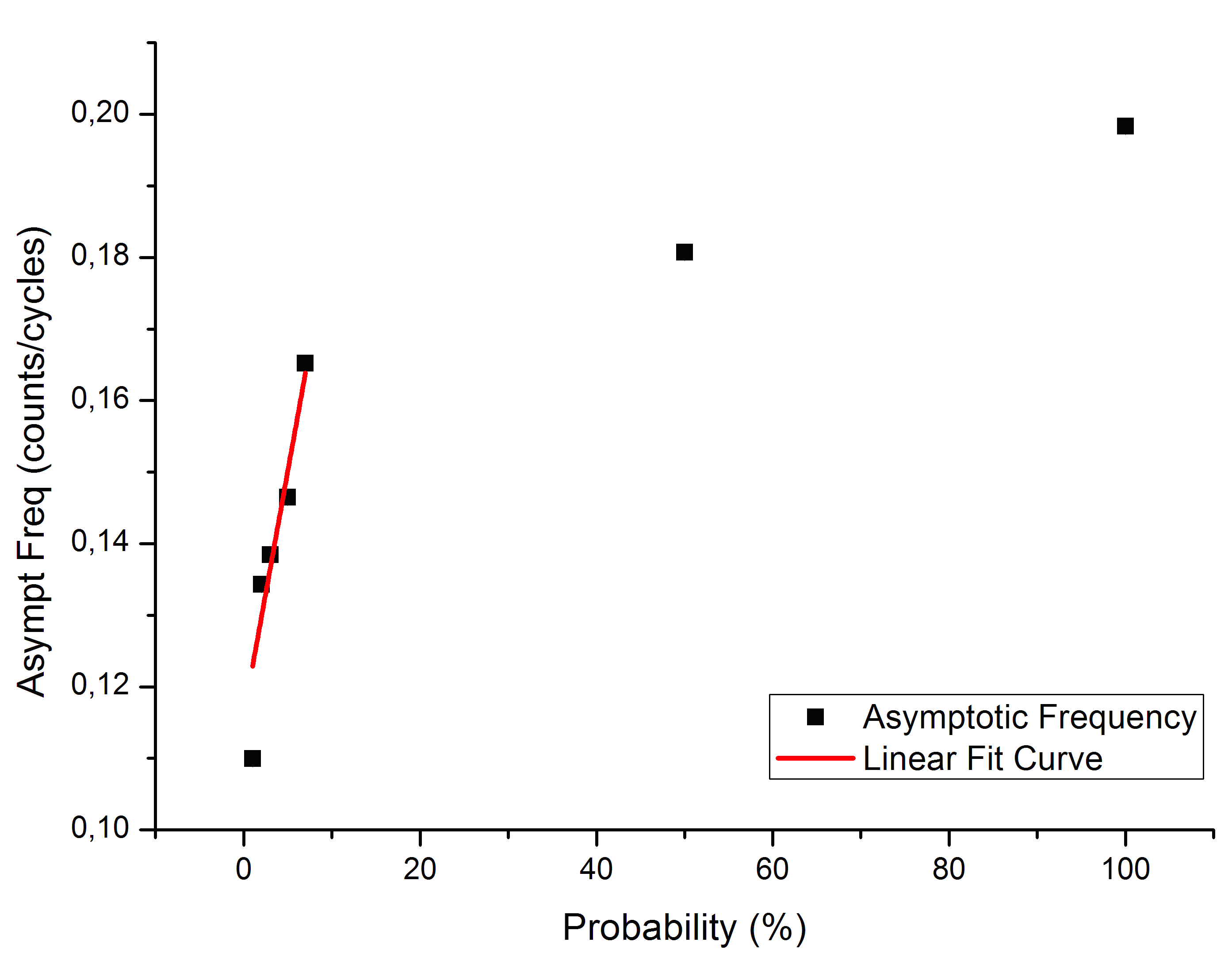}
\caption{Plot of asymptotic values of migrations frequency with high probability values.}
\label{linTime1-100}
\end{figure}

Let us now proceed with considering what happens for high probability values, in particular 50\% and 100\%. We remind that probability of 100\% means that every cycle each agent create or delete one format collection. Such a situation certainly is not very realistic, indeed the model does not behave linearly. Figure \ref{linTime1-100} shows how high probability data deviate from the linear behaviour of the low probability ones.

Here, besides the inertial distortion, we observe a clear saturation effect when the probability of creation or deletion of format collection becomes significant. This saturation effect can be fitted with the typical saturation exponential function, in particular \ref{expSat}: 

\begin{equation}
y(x) = y_0 + A \cdot \exp (R_0 \cdot x)
\label{expSat}
\end{equation} 

\begin{figure}[!h]
\centering
\includegraphics[width=13cm]{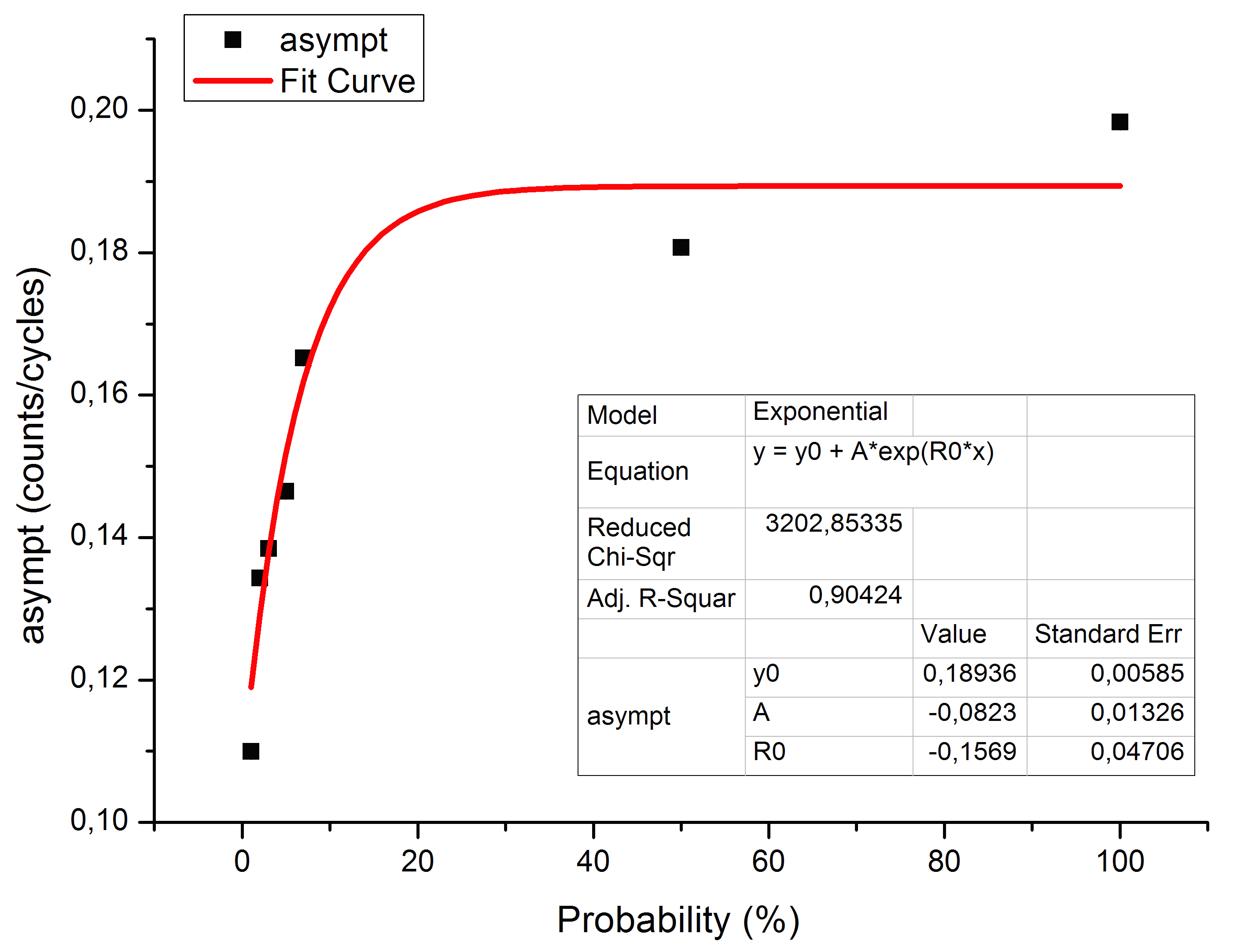}
\caption{Plot of exponential fit of asymptotic values of migrations frequency with high probability values.}
\label{linTime1-100Sat}
\end{figure}

The result of such a fit is reported in Table \ref{expSatFitTable}. Figure \ref{linTime1-100Sat} shows the result of this fit and we can observe how our data seem to follow this exponential trend despite of the very high chi square value obtained. Certainly we cannot confirm the exponential trend according to the fit but, here we are not interested in the specific numerical value.

\begin{table}[!h]	
	\begin{center}
    \begin{tabular}{ |c|c|c|c|}
    \hline
	$A \pm \sigma_A (\frac{counts}{cycles})$ & $R_0 \pm \sigma_{R_0} (cycles^{-\frac{1}{2}})$ & $y_0 \pm \sigma_{y_0} (\frac{counts}{cycles})$ & $red .\chi^2$\\ 
	\hline
	\hline
	$-0.082 \pm 0.013 $ & $-0.157 \pm 0.047$ & $0.1894 \pm 0.0059$ & 3202.85 \\
  	\hline 
    \end{tabular}
	\end{center}	
	\caption{Exponential fit result.}
	\label{expSatFitTable}
\end{table}

Thanks to the twenty simulations performed to prove the stable behaviour of the framework, we also have enough data to further analyse the linearity of the asymptotic frequency of migrations. Taking into account the set of data reported in \-Fi\-gu\-re \ref{linTime1-7} we have five asymptotic frequency values, with the corresponding errors, for each probability value from 1 to 7 \%. 

For each set of data we thus perform a linear fit to verify the linearity. In Table \ref{LinearFits} we report the result of each fit, including the one in Figure \ref{linTime1-7}. Even tough all the fit parameters are reported we would like to remind that our interest is focused on the Pearson's coefficient which indicated whether the distribution of our data can be considered linear or not.

\begin{table}[!h]	
	\begin{center}
    \begin{tabular}{ |c|c|c|c|}
    \hline
	 $a \pm \sigma_a (\frac{counts}{cycles})$ & $b \pm \sigma_b (\frac{counts}{cycles \cdot probability}) $ & red. $\chi ^2$ & $\rho$\\
	\hline
	\hline
	$(1.160 \pm 0.057) \cdot 10^{-1} $ & $ (6.8 \pm 1.2) \cdot 10^{-3}$ & 10.04 & 0.956 \\
  	\hline
  	$(1.179 \pm 0.027) \cdot 10^{-1} $ & $ (8.66 \pm 0.73) \cdot 10^{-3}$ & 2.66 & 0.989 \\
  	\hline  
  	$(1.196 \pm 0.068) \cdot 10^{-1} $ & $ (7.8 \pm 1.6) \cdot 10^{-3}$ & 17.05 & 0.941 \\
  	\hline
  	$(1.21 \pm 0.14) \cdot 10^{-1} $ & $ (7.6 \pm 4.1) \cdot 10^{-3}$ & 59.01 & 0.736 \\
  	\hline  
  	$(1.0845 \pm 0.1003) \cdot 10^{-1} $ & $ (1.04 \pm 0.27) \cdot 10^{-2}$ & 21.32 & 0.914 \\
  	\hline      
    \end{tabular}
	\end{center}	
	\caption{Linear fit result.}
	\label{LinearFits}
\end{table}

Despite of the reduced chi square values which are very high, we observe that the Pearson's coefficient remains very high through the various simulation. We can thus confirm the linear behaviour, and the distortion caused by the introduction of the migration time, as far as the given probability range is concerned.

%% file: InProgressAnalysis.tex
\section{Migrations ``in Progress"}
\label{ongoing_analysis}

\begin{figure}[!h]
\centering
\includegraphics[width=13cm]{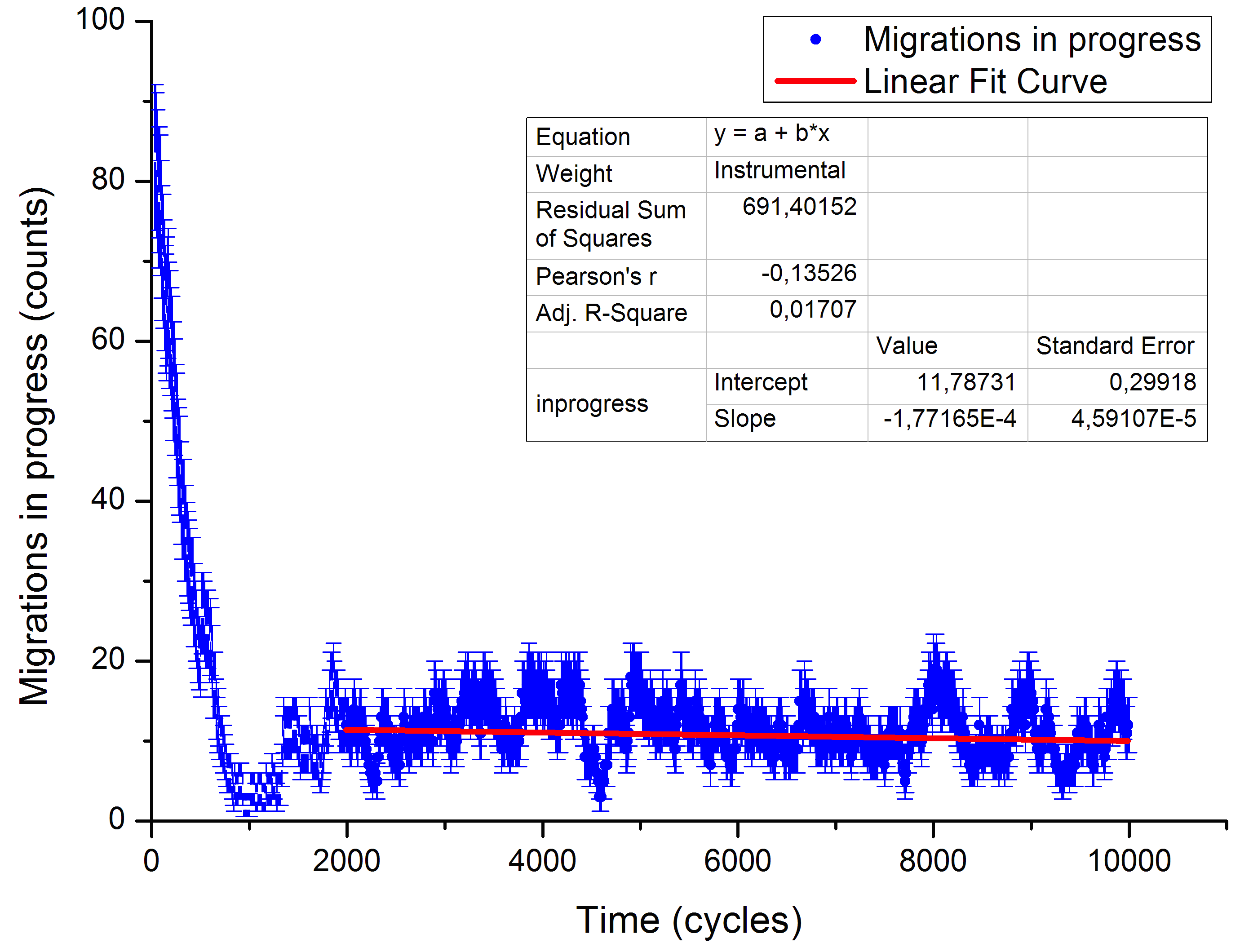}
\caption{Linear fit of the stable region.}
\label{linFitInProgress}
\end{figure}

The number of ongoing or ``in progress" migrations is another quantity that indicates the effectiveness of the presented model. We measured every 10 cycles the number of migration being performed by the agents. The Figure \ref{in_progress} of the \-pre\-vious chapter shows how the number of ongoing migrations decreases \-sig\-ni\-fi\-cantly in the first 1000 cycles. After that we see a little increase followed by a region of stability where the value seems to be stable a part from small but evident \-oscil\-la\-tions. Our analysis in focused on this stable region between 2000 and 10000 cycles. Figure \ref{linFitInProgress} depicts the linear fit of the data.

As usual, as far as the migrations are concerned, the error on the counts is assumed to be Poissonian. The Pearson's coefficient equals $-0.135$   which is not very high due to the evident oscillations of the variable. On the other hand, the reduced chi square value below 1, in particular $0.86$. More interesting results are the values of the fit parameters: the intercept, which equals $11.79 \pm 0.30$, may indicate the asymptotic value of the number of ongoing migrations. The slope equals $-1.77 \cdot 10^{-4} \pm 4.6 \cdot 10^{-5}$, that means that, as far as the fit can be considered valuable, the number of migrations in progress decreases slowly but linearly with increasing time. The fit result is reported in Table \ref{ProgressLinFitTable}.

\begin{table}[!h]	
	\begin{center}
    \begin{tabular}{ |c|c|c|c|}
    \hline
	 $a \pm \sigma_a (\frac{counts}{cycles})$ & $b \pm \sigma_b (\frac{counts}{cycles \cdot probability}) $ & $red. \chi ^2$ & $\rho$\\
	\hline
	\hline
	$11.79 \pm 0.30 $ & $ (-1.77 \pm 0.46) \cdot 10^{-4}$ & 0.86 & -0.135 \\
  	\hline 
    \end{tabular}
	\end{center}	
	\caption{Linear fit result.}
	\label{ProgressLinFitTable}
\end{table}

%% file: EvaluationAnalysis.tex
\section{Evaluation of Agents' Decisions}
\label{decisions}

This last experiment led to very interesting results. As anticipated in Section \ref{decisions_evaluation}, not only we require stability for our system, but also we would like our agents to perform only necessary and effective migrations. Figure \ref{class_plot} of the previous chapter shows a significant decrease of false positives and increase of correct actions.  We now proceed further analysing these results. First of all we consider the trend of the false positives percentage. These data are fitted with the following expression: 

\begin{equation}
f(t) = a \cdot \exp (-b \cdot t) + c 
\label{exp}
\end{equation} 

\begin{figure}[!h]
\centering
\includegraphics[width=13cm]{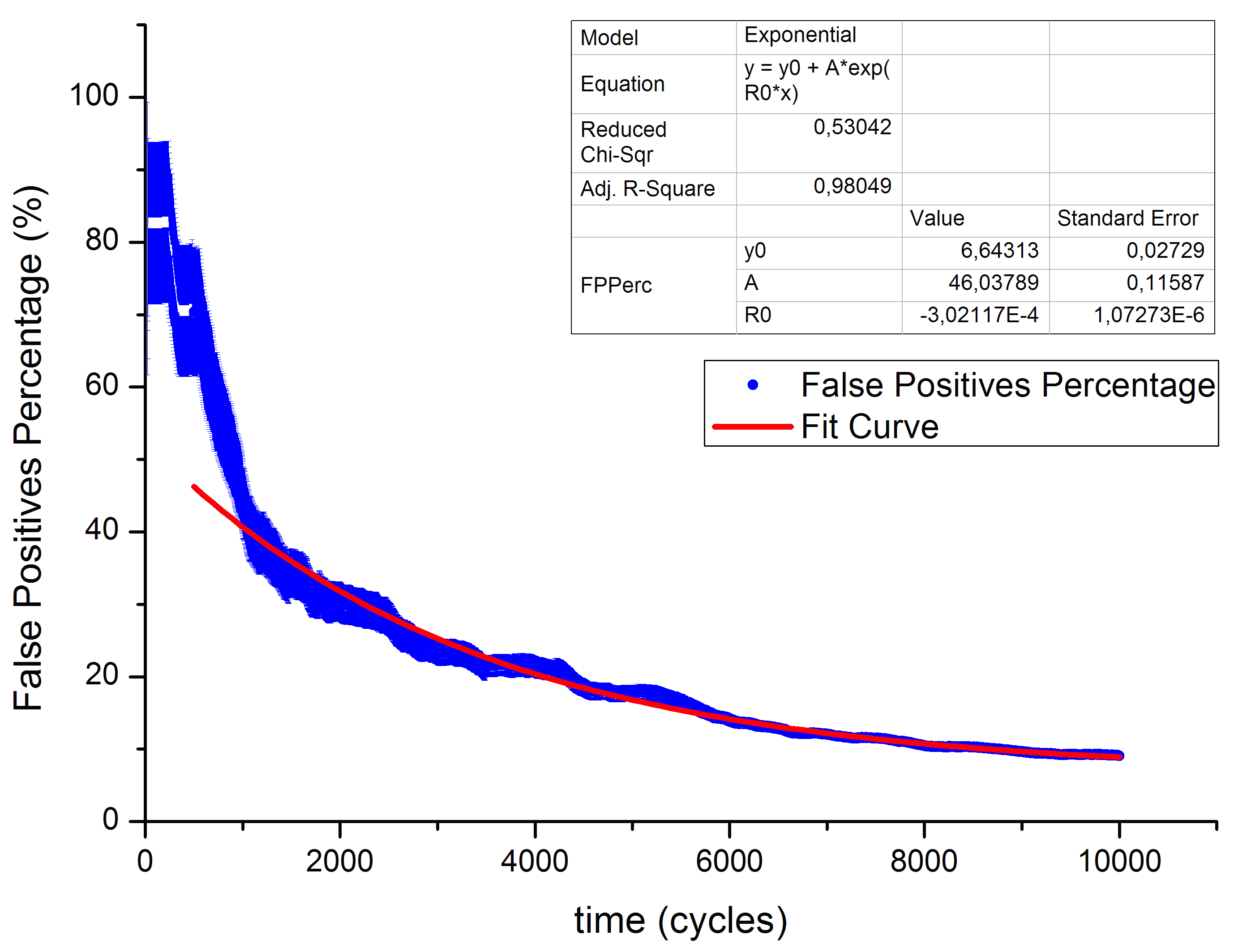}
\caption{False positives data and exponential fit.}
\label{fp_fit}
\end{figure}

The fit curve is evaluated excluding the first 200 cycles in order to avoid initial side effects. The reduced chi square value equals 0.53 thus this
result let us assert that the percentage of false positives faces an exponential decay with increasing time. The fit parameters are reported in Table \ref{fp_fit_table} while the fit of the data is shown in Figure \ref{fp_fit}.

\begin{table}[!h]	
	\begin{center}
    \begin{tabular}{ |c|c|c|c|}
    \hline
	$a \pm \sigma_a (\frac{counts}{cycles})$ & $b \pm \sigma_b (cycles^{-1})$ & $c \pm \sigma_c (\frac{counts}{cycles})$ & $red .\chi^2$\\ 
	\hline
	\hline
	$6.643 \pm 0.027 $ & $46.04 \pm 0.12$ & $(-3.021 \pm0.011) \cdot 10^{-4}$ & 0.53 \\
  	\hline 
    \end{tabular}
	\end{center}	
	\caption{False Positives exponential fit result.}
	\label{fp_fit_table}
\end{table}

The same procedure can be adopted to analyse the trend of the percentage of good actions. Table \ref{ga_fit_table} and the following Figure \ref{ga_fit} show the result of the fit. Again the very low reduced chi square value, equal to 0.36, confirm that the percentage of good actions grows exponentially with increasing time. 

\begin{table}[!h]	
	\begin{center}
    \begin{tabular}{ |c|c|c|c|}
    \hline
	$a \pm \sigma_a (\frac{counts}{cycles})$ & $b \pm \sigma_b (cycles^{-1})$ & $c \pm \sigma_c (\frac{counts}{cycles})$ & $red .\chi^2$\\ 
	\hline
	\hline
	$91.404 \pm 0.038 $ & $-56.97 \pm 0.18$ & $(-4.080 \pm 0.019) \cdot 10^{-4}$ & 0.36 \\
  	\hline 
    \end{tabular}
	\end{center}	
	\caption{Good Actions exponential fit result.}
	\label{ga_fit_table}
\end{table}

The results of the statistical analysis performed in this section are the the most interesting and well-promising ones. Here we clearly see how all the institutions benefit from the interaction: they became capable of choosing the correct action among the useful and risky migrations suggested. We also remind that not only good actions and false positives were taken into account, in the classification reported in Section \ref{institution} in the description of the \emph{analyse migration} action we considered false negatives and indifferent actions. By the way, as Figure \ref{class_plot} of the previous chapter depicts, we never encountered any migration classified as a false negative or an indifferent action. These behaviour is also very interesting since it means that agents may choose to perform an unnecessary migration (false positive) but they never skip a migration that could be relevant.

\clearpage

\begin{figure}[!h]
\centering
\includegraphics[width=13cm]{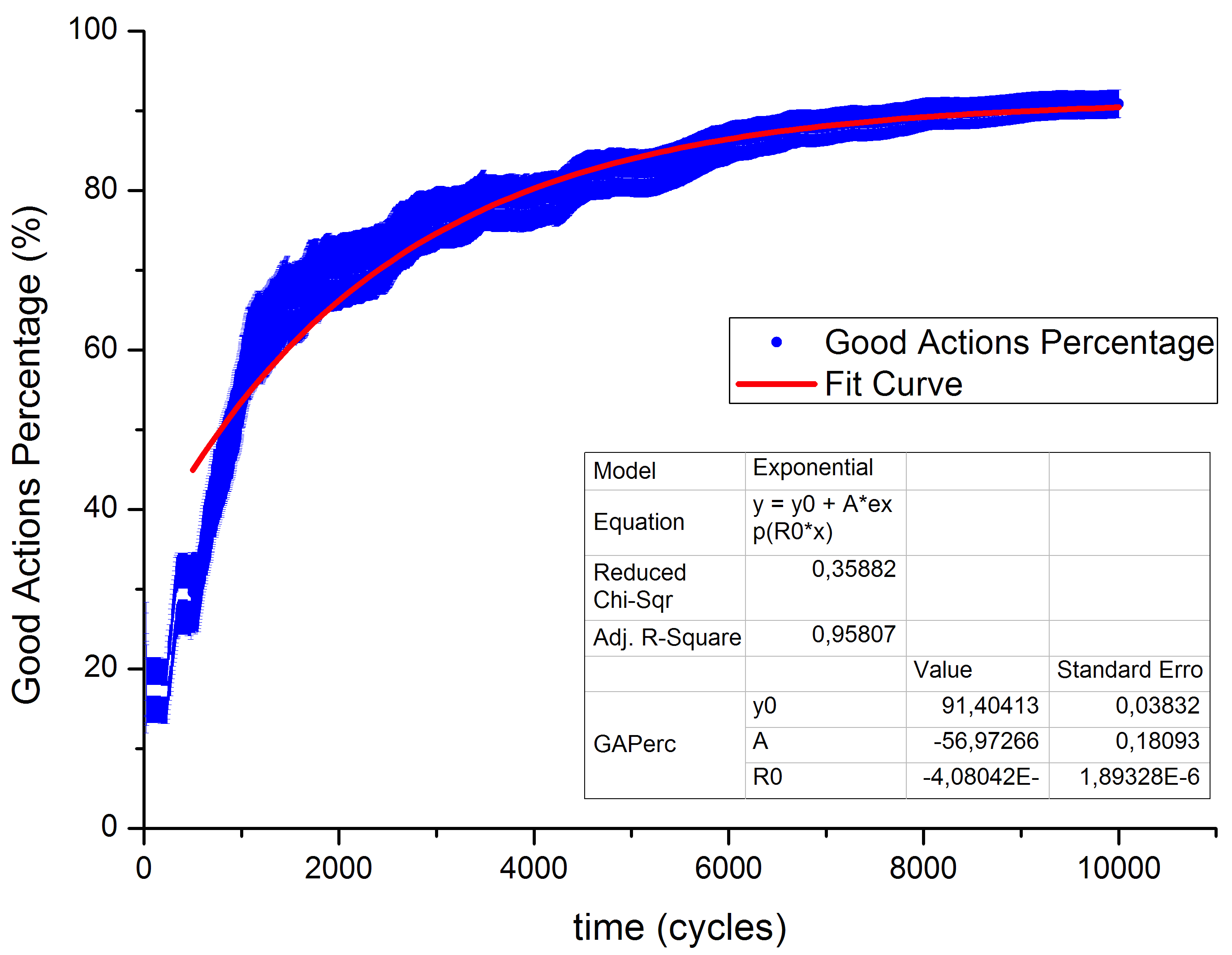}
\caption{Good actions data and exponential fit.}
\label{ga_fit}
\end{figure}